\documentclass[a4paper,11pt]{article}
\pdfoutput=1

\usepackage{jheppub}
\usepackage[dvipsnames]{xcolor}
\usepackage[T1]{fontenc}
\usepackage[english]{babel}
\usepackage{ulem}
\usepackage{xfrac}
\usepackage{amsmath}
\usepackage{tensor}
\usepackage{physics}
\usepackage{cancel}
\usepackage[utf8]{inputenc}
\usepackage{cleveref}
\usepackage{graphicx}
\usepackage{float}
\usepackage[caption=false]{subfig}
\MakeRobust{\overleftarrow}

\DeclareMathOperator*{\argmin}{arg\,min}

\title{EFT Validity and Truncation Uncertainty from few Nuisance Parameters}

\author{Beno\^it Assi$^1$,}
\author{Adam Martin$^2$,}
\author{and William Shepherd$^3$}

\emailAdd{assibt@ucmail.uc.edu}
\emailAdd{amarti41@nd.edu}
\emailAdd{shepherd@shsu.edu}

\affiliation{$^1$Department of Physics, University of Cincinnati, Cincinnati, Ohio 45221,USA\\$^2$Department of Physics and Astronomy, University of Notre Dame, South Bend, IN, 46556 USA\\$^3$Department of Physics and Astronomy, Sam Houston State University, Huntsville TX 77341, USA}

\abstract{
An observable in an Effective Field Theory (EFT) is an expansion in a set of small parameters, no different than any other perturbation series. Truncating such a series expansion leaves the leading dropped term as the dominant source of error, and that term itself contains a calculable portion. We explore the partial calculation of this next-order error, available at dim-6$^2$ in SMEFT with no additional tools needed to simulate it, and explore how to efficiently use that partial calculation to model the next-order uncertainty in full using a minimal number of nuisance parameters. This estimate of the uncertainty of the EFT signal rate naturally imposes EFT validity by ensuring that bounds are driven by kinematic regions where truncation uncertainties are parametrically smaller than the signal. We incorporate the calculable dim-6$^2$ piece into the signal and cover the remaining higher-dimension uncertainty with a small set of nuisance parameters derived from a scan over Wilson-coefficient values. Our algorithm to determine the relevant nuisance parameters and distributions is automation-friendly and applies to arbitrary truncation choices. We then provide multiple examples of its implementation in high-energy collider processes focused on SMEFT truncated at $\mathcal{O}(\Lambda^{-2})$, where the dim-6$^2$
piece is used to estimate the full $\mathcal O(\Lambda^{-4})$ dependence. In the examples considered here the reduction of nuisance parameters is appreciable, generically reducing the nuisance parameter count by an order of magnitude compared to na\"\i ve estimates.}

\begin{document}

\maketitle

\section{Introduction}
\label{sec:intro}

Precision measurements of Standard Model (SM) processes at the LHC are
increasingly interpreted within the Standard Model Effective Field Theory (SMEFT)
framework~\cite{Buchmuller:1985jz, Grzadkowski:2010es, Brivio:2017vri}, in which the effects of physics beyond the Standard Model at energies above a characteristic scale $\Lambda$ are encoded in Wilson coefficients of higher-dimensional operators. Because the SMEFT is an expansion in powers of $1/\Lambda$, any practical analysis must be truncated at finite order. The unknown contributions from the first neglected terms then constitute a theoretical uncertainty in the SMEFT interpretation of a given measurement. Concern for this uncertainty in the particular case of EFT analyses goes by the name of the EFT validity question. 

The standard perturbation-theory treatment of such an uncertainty is to estimate its size using a partial calculation of that next-order effect that is calculable from the previous-order result in perturbation theory. The best known example of this type is the QCD next-order uncertainty estimation achieved by variation of renormalisation and factorisation scales. These yield a partial next-order correction to the total cross section by dint of giving the effect of running couplings on the current-order calculation, which formally arises from additional one-loop divergences added onto the Feynman graphs of the current-order calculation. The modern form of this idea replaces the envelope over scale choices by theory nuisance parameters with theoretically motivated priors, profiled or marginalised like any other systematic~\cite{Tackmann:2024kci}. That work notes that the same strategy applies to the SMEFT expansion and cites for it, among others, the dijet analysis of Ref.~\cite{Alte:2017pme}. The use of the dim-6$^2$ piece to size the truncation uncertainty appears there and in Refs.~\cite{Alte:2018xgc,Keilmann:2019cbp,Horne:2020pot}. The present work shares that starting point and turns the estimate into a small set of nuisance parameters with calibrated priors.

The analogous partial next-order calculation in the SMEFT context is the portion of the cross-section that is calculable at next order which arises from the interference among amplitudes that are calculated to current-order accuracy. These terms are a priori comparable to all the other terms that arise at the same order in EFT perturbation theory, and therefore serve as the most reasonable ground available on which to build an estimate of the overall impact of the full next-order term.

Having a concrete estimate of the next-order effect in the EFT perturbation series will provide a handle on the question of EFT validity. Specifically, if the next-order effect is comparable to the would-be EFT signal at truncated order, the EFT expansion is no longer an appropriate perturbation series to estimate the signal and a more UV-complete model is needed. Said another way, including an error estimate in the interpretation of a given search will naturally ensure that only regions where the EFT series is well-understood meaningfully contribute to the constraint derived from experimental data.

Given the role higher order corrections play in a SMEFT analysis, how do we estimate them? To make the discussion more concrete, and in the interest of making contact with current experimental analysis practice and interest, we will focus our language below on cases where the observable is to be calculated to leading-order in the SMEFT~\footnote{We assume throughout that baryon and lepton number are conserved, so the leading SMEFT effects come at dimension-6, suppressed by $\mathcal O(1/\Lambda^2$)}, with the Lagrangian defined at dimension-6 (dim-6) and the signal cross section nominally truncated at $\mathcal{O}(\Lambda^{-2})$. The first neglected order, $\mathcal{O}(\Lambda^{-4})$, is then composed of the dim-6$^2$ correction, arising from the square of the dim-6 contribution to the scattering amplitude, the SM$\times$dim-8 interference,
and the SM interference with double insertions, where two dim-6 vertices contribute to a single
amplitude. Of these higher order terms, only the dim-6$^2$ term is fully calculable with the same tool set as the signal at $\mathcal{O}(\Lambda^{-2})$, namely e.g. Ref.~\cite{Brivio:2017btx, Aguilar-Saavedra:2018ksv, Corbett:2020bqv} within the Madgraph/UFO~\cite{Alwall:2014hca, Degrande:2011ua} framework. The remaining effects are dependent on choices about the dimension-8 basis definition~\cite{Li:2020gnx, Murphy:2020rsh}, and therefore beyond calculability for general processes with extant SMEFT simulation tools.

Many current experimental SMEFT interpretations routinely fold this dim-6$^2$ contribution into the signal prediction but assign that signal no theoretical uncertainty, leaving the genuinely uncalculable SM$\times$dim-8 and double-insertion remainders entirely unaccounted for. Clearly this is unsatisfactory. There are several other ideas in the literature on how to incorporate uncertainty from higher order terms, but they all have important limitations. Dropping
bins where the dim-6$^2$ correction is
large~\cite{Contino:2016jqw, Brivio:2022pyi} discards data and requires
process-specific tuning, while assigning a single overall scale
uncertainty~\cite{Trott:2021vqa} fails to capture the bin-to-bin shape
variation that the dim-6$^2$ correction actually produces. Using this dim-6$^2$ to set the scale of bin-by-bin errors on your distribution~\cite{Alte:2017pme,Alte:2018xgc,Keilmann:2019cbp,Horne:2020pot} gets the scaling of the error right as a function of kinematics but still drops correlations between bins. The recent proposal of Ref.~\cite{Chang:2025ohh} applies the theory-nuisance-parameter strategy of Ref.~\cite{Tackmann:2024kci} to EFT validity by rescaling the dim-6 coefficient inside each $2\to2$ amplitude with a polynomial in saturating form factors of the Mandelstam invariants over a stipulated cutoff $M$, so its nuisance parameters are the coefficients of the Mandelstam descendants of the operator and its bounds depend on $M$. A parametrisation built on an explicit dim-8 operator basis would in addition inherit the equation-of-motion and field-redefinition redundancies of that basis~\cite{Georgi:1991ch, Brivio:2017vri, Alonso:2025jvv}.

The procedure we advocate for and explore in this paper is as follows: the dim-6$^2$ piece, being fully calculable at the fitted Wilson coefficients, belongs in the signal model alongside the linear dim-6 piece, while the SM$\times$dim-8 and double-insertion remainders, which are not calculable with dim-6 simulation tools, are approximated by a set of mean-zero nuisance parameters. To determine these nuisance parameters, we reuse the same dimension six squared kernel, with coefficients drawn independently of the signal fit. Unlike Ref.~\cite{Chang:2025ohh} we parametrise the binned observable rather than the amplitude, the number of parameters follows from the rank of that observable's shape space, and no cutoff scale, form factor or dim-8 operator enters. The construction applies to any process and any binning, since it acts on the simulated observable and never on the amplitude, and two of the three examples below are $2\to3$ processes.

This approach would appear to suffer from a proliferation of correlated nuisance parameters. In a fit with signal parameters $\theta$ and nuisance parameters
$\boldsymbol\nu$, the likelihood
$\mathcal{L}(\mathcal{D}\,|\,\theta,\boldsymbol\nu)$ is reduced to an
effective likelihood over $\theta$ alone by marginalising over
$\boldsymbol\nu$,
$\int\!d\boldsymbol\nu\;
\mathcal{L}(\mathcal{D}\,|\,\theta,\boldsymbol\nu)\,
\pi(\boldsymbol\nu)$,
or by profiling,
$\max_{\boldsymbol\nu}
[\mathcal{L}(\mathcal{D}\,|\,\theta,\boldsymbol\nu)\,
\pi(\boldsymbol\nu)]$.
Either operation becomes computationally expensive when $\boldsymbol\nu$ is
high-dimensional and correlated. The (dim-6)$^2$ piece, used directly as an error, naively appears to introduce $\boldsymbol\nu \sim \mathcal O(N^2)$ parameters, where $N$ is the number of dim-6 operators that affect a given process. Profiling that many
correlated parameters through a detector-level fit, alongside the
many experimental systematics already present, is impractical. In this paper we show that the naive count of
$\mathcal{O}(N^2)$ vastly overcounts the necessary number of degrees of
freedom in this uncertainty, and we develop a method to compress the truncated correction
using $K \ll N$ representative Wilson coefficients whose joint prior follows from the
quadratic predictions already produced by existing signal
simulation tools. With the uncertainty reduced to a handful of uncorrelated parameters, profiling is no longer unwieldy.

 We emphasise at the outset that the object we are constructing is an
uncertainty on the truncated SMEFT cross section. The choice of truncation order is left to the analyser, and our method produces the
nuisance distribution that should accompany whichever choice is made.
The key observation is the fact that the kinematic shape space is low-rank, i.e.~multiple operators produce identically-shaped corrections to the observables being measured. The
correction's values across $B$ kinematic bins span a vector space whose
dimension $D_M$ is set by the number of distinct bin-by-bin shapes the
correction can produce, not by the number $N(N+1)/2$ of independent
quadratic products.
Operator combinations that produce identical shapes across bins are
supernumerary from the observable's point of view. Borrowing the
terminology of global SMEFT fits, where similar techniques are used
to identify poorly constrained parameter
combinations~\cite{Ellis:2020unq,Giani:2023gfq, Celada:2024mcf}, we call these
redundancies {flat directions}. A single representative parameter
suffices for each independent shape, so the number of nuisance parameters
needed is set by $D_M$ rather than $\mathcal{O}(N^2)$.

To take advantage of this fact in a way that does not require specialised knowledge of each collider process and observable, we develop an automation-friendly algorithm to select distributions of interest.
A singular-value decomposition (SVD) of the kinematic kernel identifies the independent shapes and guides the choice of $K$ representative operators. One representative Wilson
coefficient is assigned to each operator, and a large ensemble of random dim-6$^2$
corrections is generated by sampling all $N$ coefficients.
Fitting the $K$ representatives to each ensemble member yields
a joint distribution that is generically non-Gaussian and often multimodal, and yet faithfully reproduces the statistically-expected structure of the dim-6$^2$ correction with only $K$ parameters.

To extend coverage beyond the dim-6$^2$ piece, covering the same-order
SM$\times$dim-8 interference and double insertions, on the assumption, stated and tested in \cref{sec:algo}, that those terms are well approximated by the bin-by-bin shapes generated by the dim-6$^2$ piece, though they may differ from it in magnitude and sign, we decorrelate the fitted monomials via Principal Component Analysis
(PCA)~\cite{Jolliffe:2002}, allow each of the $K(K+1)/2$ representative-predicted kinematic shapes (rather than each representative operator) to vary its strength, and apply an $\alpha=\sqrt{2}$ scaling to the nuisance distributions. This factor accounts for the two uncalculable pieces at the same order, each of a size comparable to the dim-6$^2$ piece under naive dimensional analysis, adding in quadrature to twice its variance. Following this decorrelation, we are left with $D$ nuisance variables and their distributions, with no correlations between them. By default we therefore take $D=K(K+1)/2$, though it is possible that in some cases the truly necessary number of parameters is less than this. In such cases our algorithm outputs a distribution for the unneeded nuisance parameter which has negligible variance from zero.

We apply this algorithm to three collider processes as proof of principle, selected for their differing kinematic properties and for the fact that all three already have tools to calculate the full $\mathcal O(\Lambda^{-4})$ correction for comparison purposes. All of these processes are corrected by large numbers of operators, but using our nuisance reduction algorithm has significant impacts. The needed number of representative operators is $K=2$ for high-$p_T$ Drell-Yan, $K=3$ to $5$ depending on the observable of interest for
$Zh$ production, and $K=2$ to $4$ for vector boson fusion (VBF) Higgs production. In all these cases, the representatives reproduce the full dim-6$^2$ contribution, and, after our decorrelation and scaling procedure, generate an uncertainty band that covers the full $\mathcal O(\Lambda^{-4})$ SMEFT contribution. This serves as an explicit test of the ability to approximate error sizes due to full next-order EFT effects using the dim-6$^2$ distributions and this algorithm.

We recommend that searches both with and without these errors be performed, to provide reliable LHC constraints on new physics that matches onto the SMEFT in the first case and an estimation of the possible sensitivity of the LHC to such models in the second case. If a model matched onto the SMEFT is detectable by an errorless analysis but not by one which carefully treats this perturbation theory uncertainty using the technique presented here then an understanding of its actual detectability at the LHC requires a specific study of that UV model at the LHC.

The remainder of this article is organised as follows: in the next section we detail the calculation of the partial next-order distribution that will be the input to our estimation of the full next-order uncertainty on SMEFT cross sections, then in \cref{sec:algo} we present the algorithm to produce that uncertainty and step through how to use it in a SMEFT analysis. In \cref{sec:examples} we apply that algorithm to three LHC processes, deriving estimated next-order uncertainty bands and comparing them to the full range of next-order effects. We then summarize the essential results of that set of examples in \cref{sec:summary} and draw conclusions in \cref{sec:conclusions}. We discuss some of the parametric details of our implementation of this algorithm in \cref{app:implementation}, work through a pedagogical toy example in \cref{app:mwe}, and collect additional figures from the examples considered in \cref{app:figs}.

\section{EFT observables and flat directions}
\label{sec:eft}

Here, we discuss the general features of next-order EFT uncertainties. While we specialise in the discussion below to focus on the case of $\mathcal O\left(\Lambda^{-4}\right)$ corrections to a signal calculated at linear order in $\Lambda^{-2}$, the general features identified will carry over to arbitrary orders.

\subsection{The quadratic dim-6$^2$ correction}
\label{subsec:qb}

Consider a differential cross-section with $B$ kinematic bins (in an observable space of arbitrary dimensionality) in an
EFT process involving $N$ Wilson coefficients $\mathbf{c} = (c_1,\dots,c_N)$. The dim-6$^2$ partial next-order result to be used in estimating the full truncation error is a quadratic form in these coefficients,
\begin{equation}
\label{eq:qb}
 q_b(\mathbf{c}) = \sum_{i,j=1}^{N} A_{b,ij}\, c_i c_j
 = \mathbf{c}^\top A_b\, \mathbf{c},
\end{equation}
where $A_b$ is a symmetric matrix encoding the squared and
interference contributions of operators $i$ and $j$ in bin $b$. In
words, $q_b$ is the dim-6$^2$ piece of the predicted differential
cross-section in bin $b$, the part quadratic in the Wilson coefficients
that the truncated series adds beyond the SM and its linear dim-6
interference without ambiguities introduced by dim-8 basis definitions. It is the per-bin quantity on which the rest of this paper builds its estimate of the full next-order uncertainty due to truncation of the SMEFT perturbation series.

In
practice, $A_b$ is extracted from signal simulation tools (e.g.\
MadGraph~\cite{Alwall:2014hca} with SMEFTSim~\cite{Brivio:2017btx}). Each entry can be calculated directly in such tools by applying appropriate coupling restrictions to the simulation to insist on the appearance of the required Wilson coefficients. If this approach is undesired or technically challenging due to the structure of a particular simulation it is also possible to extract each matrix entry by systematically setting couplings to ones and zeroes. The diagonal entry $A_{b,ii}$ is the
squared-insertion cross-section in bin $b$ with $c_i=1$ and all other
coefficients set to zero.
The off-diagonal entries follow from expanding eq.~\eqref{eq:qb}
with $c_i = c_j = 1$ (all others zero):
$q_b = A_{b,ii} + 2A_{b,ij} + A_{b,jj}$, so
\begin{equation}
\label{eq:subtract}
 A_{b,ij} = \tfrac{1}{2}\bigl[\sigma_b(c_i{=}1,c_j{=}1) - \sigma_b(c_i{=}1) - \sigma_b(c_j{=}1)\bigr]
 \quad (i\neq j),
\end{equation}
where $\sigma_b(\cdot)$ denotes the dim-6$^2$ cross section in bin $b$ simulated with the indicated coefficients set to one and all others to zero.

Note that it is entirely possible for two operators to have
non-vanishing diagonal entries $A_{b,ii},\,A_{b,jj}\neq 0$ while their
interference entry $A_{b,ij}$ vanishes in every bin -- as happens when
their amplitudes carry different quantum numbers and cannot interfere. This subtraction-based approach will then produce statistical noise, while the direct simulation of each term independently will correctly identify the analytic zero.

The full
$\mathcal{O}(\Lambda^{-4})$ observable also receives SM$\times$dim-8
interference and double-insertion corrections:
\begin{equation}
\label{eq:lam4}
 q_b^{\rm full}(\mathbf{c}, \mathbf{d}) =
 \mathbf{c}^\top A_b\,\mathbf{c} + \mathcal{B}_b\cdot\mathbf{d}
 + \Delta_b,
\end{equation}
where $\mathbf{d} = (d_1,\dots)$ collects the Wilson coefficients of
the next-order operators (for SMEFT these are the dim-8 operators of
Refs.~\cite{Li:2020gnx, Murphy:2020rsh}) and $\mathcal{B}_{bk}$ encodes their
linear interference with the SM. The remainder $\Delta_b$ collects
double insertions of dim-6 operators in a single amplitude, which sit at the same
$\mathcal{O}(\Lambda^{-4})$ as the explicit terms but depend on basis choices at the dim-8 level and are therefore not well-defined
in standard dim-6 signal simulations. We focus here on the dim-6$^2$ piece encoded in $A_b$ because it is the only term which is reliably calculable from the signal-level tools, while all remaining terms (which we aim to estimate using the dim-6$^2$ piece as input) require additional calculations on a process-by-process basis.

Once the $A_b$ matrix is extracted from simulation, the entire downstream pipeline (detailed in \cref{sec:algo}) is pure reweighting of pre-computed kernels and runs in seconds to minutes on a laptop. The Monte Carlo cost is paid once per process, and analyses that update binning, prior, or representative choices need only rerun the cheap steps.

A word on what looks at first like a circularity. The dim-6$^2$ piece is built from the same $c_i$ that appear linearly in the dim-6 signal, so at the fitted $\mathbf{c}$ it is fully calculable. We include it in the signal model. The bin-level prediction reads
\begin{equation}
\label{eq:signal_with_dim6sq}
 \mu_b(\mathbf{c},\mathbf{s})
 = \sigma^{\rm SM}_b
 + \sum_i c_i\,\sigma^{\rm dim\text{-}6}_{b,i}/\Lambda^2
 + \Bigl(\mathbf{c}^\top A_b\,\mathbf{c}
 +\varepsilon_b
 \Bigr)\!\big/\Lambda^4,
\end{equation}
where $\varepsilon_b$ is the error distribution we construct below, dependent on the nuisance parameters $\mathbf{s}$ (the PCA scores introduced in Section~\ref{subsec:decorr}) that we algorithmically determine based on the physics of the dim-6$^2$ kernel $A_b$ by the use of an artificial independent draw of Wilson coefficients $\mathbf c'$ as described in \cref{sec:algo} to come.

\subsection{Flat directions}
\label{subsec:flatdirections}

 To count the distinct kinematic shapes summed over in eq.~\eqref{eq:qb},
note that the quadratic form $q_b = \mathbf{c}^\top A_b\,\mathbf{c}$
is not linear in the Wilson coefficients $c_i$ themselves but is
linear in the {monomials} $\{c_i^2,\, c_i c_j\}$, the
quadratic products formed from pairs of coefficients.\footnote{For a signal function truncated at a higher power of $\Lambda^{-2}$ the result would still be linear in monomials and the following technique still applies, but the monomials would not all be quadratic in Wilson coefficients.}
Define the monomial feature matrix $M$ of dimension $B \times P$, with
$B$ the number of kinematic bins (as in eq.~\eqref{eq:qb}) and
$P \equiv N(N+1)/2$ independent monomials,
\begin{equation}
\label{eq:monomial}
 M_{b,k} =
 \begin{cases}
 A_{b,ii} & \text{if monomial } k = c_i^2,\\
 2 A_{b,ij} & \text{if monomial } k = c_ic_j\ (i<j).
 \end{cases}
\end{equation}
In terms of $M$, the observable vector is $\mathbf{q} = M\,\mathbf{m}$, where
$\mathbf{m} = (c_1^2,\, c_1 c_2,\, \dots)^\top$ is the monomial vector.
Every possible realisation of the quadratic correction lies in the
column space of $M$, so its rank $D_M = \mathrm{rank}(M) \leq \min(B, P)$
counts the number of distinct bin-by-bin shapes the dim-6$^2$ correction can
produce. In practice $D_M$ is smaller than $N$, let alone
$P = \mathcal{O}(N^2)$: the many operator combinations that produce the same
shape across bins are `flat directions' that need not be independently parametrised. In this way flat directions, rather than being a hindrance to strong constraints as they are when they occur in the space of constraints on signal parameters, are actually beneficial in uncertainty space, because they reduce the amount of work needed to adequately model the truncation error.

\section{Algorithm}
\label{sec:algo}

We now develop an algorithm to efficiently estimate the uncertainty that is only partially calculable using our signal simulation tools. This algorithm is constructed in the limit of large simulation statistics, where the shapes of simulation in the experimental binning of interest are fully reliable.

In plain terms the algorithm has four steps. It first finds how many genuinely different bin-by-bin shapes the dim-6$^2$ correction can take in the chosen observable, which is a matrix rank. It then picks one operator to stand for each shape. It next learns, by fitting those few operators to a large random sample of full dim-6$^2$ predictions, how large each shape can be and how the shapes are correlated. Finally it loosens those correlations and widens the result so that the same shapes can also stand in for the terms at the same order that the simulation does not compute.

\paragraph{How to apply this error estimate.} The optimal use of errors constructed in this way depends on what one assumes about the full EFT truncation uncertainty. If the uncalculated $\mathcal O(\Lambda^{-4})$ terms, the SM$\times$dim-8 interference and the double insertions, are taken to populate the same bin-by-bin shapes as the dim-6$^2$ piece, differing from it in magnitude and sign but not in kinematic form, then these nuisance parameters can be taken as a complete model of the EFT uncertainty. This allows for marginalisation or profiling over these parameters in any straightforward fashion. The physics support for such an assumption would be that these terms enter at the same power of $E/\Lambda$ and, in every process studied here except one topology in VBF, through the same vertices and diagrams, so at leading power in the energy they may be the same functions of the kinematics.

This assumption does not hold generally. A next-order operator that opens a topology absent at dim-6 can produce a shape outside the dim-6$^2$ span, and whether it does is a question for each process and observable. Examples are also known where new shapes are introduced in fairly straightforward processes like Drell-Yan when carefully chosen observables are measured~\cite{Alioli:2020kez}. In the language of nuisance parameters, the scores $s_i$ introduced in \cref{subsec:decorr} enter the prediction as $\varepsilon_b=\alpha\sum_i W_{b,i}s_i$, so their true value is the projection of the true remainder onto the shape vectors $W_{b,i}$, and whatever part of the remainder lies outside that span is a modelling error that no value of the scores reproduces. Imperfect modelling is to be expected when we are explicitly discarding an infinite tower of operators and Wilson coefficients, as is the essential nature of truncation error estimation in the EFT. The check we can offer is empirical, checking the error band size and correlations between bins for our processes. In the three processes of \cref{sec:examples} a full $\mathcal O(\Lambda^{-4})$ calculation is available, and the per-bin coverage of the calibrated nuisance against that truth is reported there for every observable. The double-insertion part of that assumption can be tested directly in one case, which we do in \cref{subsec:decorr} where it bears on the inflation factor $\alpha$. These checks show that the error estimates are correct in magnitude for the three processes, which were chosen because a full result exists to compare against and not because the technique looked promising there. Beyond them we can give no guarantee that the estimate holds for processes or observables we have not examined. Given that higher-order effects can have different shapes, applying these error estimates to a complete analysis would benefit from careful treatment of these nuisance parameters that allows for the possibility of mismodelling, such as the Sandwich Variance Estimator technique~\cite{huber1967behavior,white1980heteroskedasticity,white1982maximum}.

Despite not producing a complete parametrised model of the truncation error in the SMEFT perturbation series, these error estimates do encapsulate all knowledge that is unambiguously available about higher-order EFT effects with a fixed-order expansion of the Lagrangian of the theory, and they do so as compactly as the dim-6$^2$ kernel allows. Combined with the $\alpha$-inflation, carefully checked below, they give a reasonable, conservative error band with minimal added computational cost to any given analysis.

\subsection{Minimal parametrisation of partial next order distribution}

The goal, given $M$ defined in \cref{eq:monomial}, is to find both the rank $D_M$ of $M$ and also to extract the corresponding kinematic shapes and the normalisation of the associated nuisance parameters that will best estimate the truncation error in the SMEFT series. Achieving this in an automation-friendly and computationally tractable way proceeds through multiple steps. We utilize the monomial matrix $M$ to identify promising candidates to serve as representative operators, adding one at a time to the list of representatives and exploring how well we can fit the full dim-6$^2$ distribution with the current list. We stop when our fits are no longer improved by the addition of the next would-be representative. The process is explained in detail below, and Appendix~\ref{app:mwe} walks through a minimal toy example end to end. \Cref{tab:terminology} collects the notation used in constructing our uncertainty estimates.

\begin{table}[H]
 \centering
 \small
 \renewcommand{\arraystretch}{1.4}
 \begin{tabular}{llp{0.50\textwidth}}
 \hline
 \textbf{Term} & \textbf{Symbol} & \textbf{Role} \\
 \hline
 Wilson coefficient & $c_i$ &
 $N$ parameters of a SMEFT operator \\
 Monomial & $m_{i,j}\equiv c_i c_j$ &
 $P=N(N+1)/2$ products of two Wilson coefficients \\
 Representative coefficient & $c_{r_k}$ &
 $K \ll N$ Wilson coefficients spanning independent shapes \\
 Monomial matrix, rank & $M$, $D_M$ &$B\times P$ table of bin-shapes, and the number of distinct shapes in it \\
 Shape weight, score & $W_{b,i}$, $s_i$ &how much of shape $i$ enters bin $b$, and the nuisance parameter that scales it \\
 Nuisance count, inflation & $D$, $\alpha$ &$D=K(K{+}1)/2$ scores, each inflated by $\alpha=\sqrt2$ \\
 Coverage & $C_{95}$ &fraction of full $\mathcal O(\Lambda^{-4})$ throws inside the 95\% band of a bin \\
 \hline
 \end{tabular}
 \caption{ Notation used by the algorithm. The first three rows are the objects it manipulates, the remaining rows are defined where they first appear and collected here for reference.}
 \label{tab:terminology}
\end{table}

\subsubsection{SVD and representative selection}
\label{subsec:svd}

To identify the kinematic shapes of interest and rank their importance, we perform a
singular value decomposition (SVD)~\cite{GolubVanLoan:2013} of the full monomial
feature matrix, $M = U S V^\top$. An SVD writes a matrix as a sum of rank-one pieces, each a single bin-shape times a single monomial pattern times a positive weight, ordered by that weight, so it answers at once how many independent shapes the matrix contains and which of them matter. The columns of $U$ are orthogonal bin-shape
patterns, the diagonal entries of $S$ (the singular values
$\sigma_1 \geq \sigma_2 \geq \dots \geq 0$) measure the magnitude of each pattern's
contribution to the observable, and the columns of $V$ identify which monomial
combinations produce each pattern. A singular value $\sigma_k$ that is much smaller
than $\sigma_1$ corresponds to a shape that contributes negligibly to the total
correction. The number of significant singular values therefore gives
the effective rank $D_M$ of $M$.

The shapes identified by the SVD live in the abstract space of monomials $c_ic_j$, but the representatives we actually report and feed into the algorithm are individual Wilson coefficients $c_i$, so we need a translation. Each $V_{\cdot k}$ (which has $P=N(N{+}1)/2$ entries, one per monomial) tells us which monomials carry shape $k$. We then map that weight back onto operators by summing the $V^2$-weight of every monomial that touches each $c_i$. Specifically, we assign a score to each operator $c_i$ by summing:
\begin{equation}
\label{eq:score}
 \mathrm{score}_k(i) \;=\; V^2_{c_i^2,\,k} \;+\; \sum_{j\neq i} V^2_{c_i c_j,\,k}.
\end{equation}
This is a heuristic, but it is the natural one, summing all the impacts that an operator can have on the final dim-6$^2$ correction. By contrast, considering only the diagonal term or only interference terms has clear failure modes when one is negligible or vanishes but the other is important to the fit.

The score is evaluated one shape at a time: we take the SVD directions $k$ in order of decreasing singular value, so that at each step the shape under consideration is the leading one that does not yet have an assigned representative. This ordering is where the singular-value hierarchy enters the selection. The score itself need not be weighted by the singular value, since dominant shapes are simply resolved first and sub-dominant ones are pruned by the $R^2$ ladder below.
The candidate representative $r_k$ is the highest-scoring operator not already included in the list of representatives whose diagonal $A_{b,r_kr_k}$ is
non-vanishing in at least some bins (to avoid runaway directions in any fits). In the analyses in Sec.~\ref{sec:examples}, this filter is active: it removes two operators from the DY set ($c_{HB}, c_{HW}$), five from the $Zh$ set ($c_H, c_{HG}, c_{H\Box}, c_{Hud}, c_W$), and six from the VBF set ($c_H, c_{HL}^{(1)}, c_{HL}^{(3)}, c_{H\Box}, c_{He}, c_W$), whose self-interference is suppressed by quark Yukawas or vanishes for the chosen final state. The off-diagonal entries $A_{b,ij}$ of these operators with the remaining set are also negligible, so they contribute neither to the diagonal scoring nor through cross-coupling and are effectively absent from the dim-6$^2$ structure of the process at hand.

\subsubsection{Ensemble fitting}
\label{subsec:fitting}

Given the $K$ representative Wilson coefficients $\mathcal{R} = \{c_{r_1},\dots,c_{r_K}\}$,
we need to determine what values they must take to reproduce the
quadratic correction for an arbitrary point in Wilson coefficient
space.
We draw $T$ samples
$\mathbf{c}^{(t)}$ with all $N$ coefficients sampled uniformly from $[-p, +p]$, where the
prior scale is set by naive dimensional analysis
(NDA)~\cite{Manohar:1983md, Gavela:2016bzc}. NDA estimates the maximum
natural size of a Wilson coefficient by requiring that loop corrections
do not exceed tree-level contributions,
\begin{equation}
\label{eq:prior}
 c_i\le p = 4\pi.
\end{equation}
The Wilson coefficients $\mathbf{c}$ here are the dimensionless objects that enter the Lagrangian as $\mathbf{c}/\Lambda^2$ multiplying each dim-6 operator (the same objects fed directly to MadGraph). NDA bounds them by $|c_i|\le 4\pi$ regardless of $\Lambda$. Because $p=4\pi$, the calibration draws and the representative-coefficient distributions they produce routinely reach values of $\mathcal{O}(10)$, so the coefficient axes of the figures below extend well beyond the $\mathcal{O}(1)$ range a reader might expect. These are the extremes of the NDA-allowed band being scanned to size the uncertainty, not a claim that any physical coefficient sits at that value. The signal model of \cref{eq:signal_with_dim6sq} carries the same $\mathbf{c}$, and for self-consistency the values being fit must lie in the same range $|c_i|\le p$ used for the uncertainty calibration scan.

The combination that actually enters any cross section is $\mathbf{c}/\Lambda^2$, so the NDA bound $|c_i|\le 4\pi$ together with the explicit $1/\Lambda^2$ factor sets the physical scale at which the dim-6 operators contribute. Throughout this
paper we set $\Lambda = 3~\text{TeV}$.
With $p=4\pi$ itself $\Lambda$-independent, the prior, the fitted-coefficient distributions, the $R^2$ ladder and the coverage check all live in coefficient space and are unaffected by $\Lambda$. What $\Lambda$ does affect is the relative size of the dim-6 linear signal and the calibrated error estimate constructed below: the linear signal scales as $1/\Lambda^2$ and the error as $1/\Lambda^4$, so smaller $\Lambda$ makes the truncation error larger relative to the signal and thus more important to consider in the construction of a given collider analysis.

For each truth point we compute the
observable $q_b^{(t)} = \mathbf{c}^{(t)\top} A_b\, \mathbf{c}^{(t)}$ using the full $A_b$
matrix, then fit the $K$ representative values to minimise
\begin{equation}
\label{eq:lsq}
 \hat{\mathbf{c}}_{\mathcal{R}}^{(t)} = \argmin_{\mathbf{c}_{\mathcal{R}}}
 \sum_b \bigl(q_b^{(t)} - \mathbf{c}_{\mathcal{R}}^\top \tilde{A}_b\, \mathbf{c}_{\mathcal{R}}\bigr)^2
 + \lambda_{\rm reg} \|\mathbf{c}_{\mathcal{R}}\|^2,
\end{equation}
where $\tilde{A}_b$ is the $K\times K$ submatrix of $A_b$ restricted to the representative
operators and the redundancy of the quadratic form under $\mathbf{c}_{\mathcal{R}}\to-\mathbf{c}_{\mathcal{R}}$ is removed by folding each fitted tuple to $\hat c_{r_1}\ge0$ after convergence. The $\lambda_{\rm reg}\|\mathbf{c}_{\mathcal{R}}\|^2$ term keeps the fit well-posed
when multiple representative tuples reproduce the truth equally well by selecting the smallest-norm tuple. Our selection process for representatives is constructed to avoid this instance, and in the examples explored in this article that is the case, but the term is important for algorithmic future-proofing.
The value of $\lambda_{\rm reg}$ is given in Appendix~\ref{app:implementation}.

The quality of each fit is measured by the uncentered coefficient of
determination~\cite{ParticleDataGroup:2024cfk}
\begin{equation}
\label{eq:r2}
 R^2 = 1 - \frac{\|\mathbf{q}^{(t)} - \tilde{\mathbf{q}}^{(t)}\|^2}{\|\mathbf{q}^{(t)}\|^2},
\end{equation}
where $\tilde{q}_b^{(t)} = \hat{\mathbf{c}}_{\mathcal{R}}^{(t)\top}\tilde{A}_b\,\hat{\mathbf{c}}_{\mathcal{R}}^{(t)}$. Stated simply, $R^2$ quantifies how well the distribution generated with $K$ representative coefficients can fit the full $N$ coefficient distribution.
The denominator is $\|\mathbf{q}\|^2$ rather than the variance
$\|\mathbf{q} - \bar{q}\|^2$ used in the standard (centered) $R^2$. This is
the natural choice here because the null model is $\tilde{q}_b = 0$
(no quadratic correction at all), not the bin-averaged correction
$\bar{q}$, and because the distributions are either calculated analytically or with large Monte Carlo simulation samples such that there is no appreciable uncertainty on the truth or representative distributions.
A consistently high $R^2$ across the ensemble confirms that $K$ representatives genuinely
span the observable space.

In a procedure we call the {$R^2$ ladder}, we iteratively select one new representative following the approach of \cref{subsec:svd} and re-fit with the expanded list of representatives until the median $R^2$ value is not improved by the addition of the latest representative. At that point, we remove the last candidate representative, which did not improve our fit, from our signal model and proceed with $K$ fixed representatives to model the uncertainty. This ladder condition climbs the median $R^2$ with as few representatives as needed, stopping when a further representative no longer improves it. To prevent a candidate that is partly degenerate with an already-accepted representative from prematurely halting the ladder, we allow a look-ahead of two further candidates and report the representative list of length $K$ that achieves the best median $R^2$.

With the representative list fixed, the ensemble of fitted values $\{\hat{c}_{r_k}^{(t)}\}_{t=1}^T$ for each
representative encodes the
range and shape of values the representative Wilson coefficients must take to reproduce
the quadratic correction across the full EFT prior. No parametric model is imposed on these distributions.

\subsection{Estimating uncalculated terms: the decorrelated monomial prescription}
\label{subsec:decorr}

The ensemble of fitted tuples
$(\hat c_{r_1}^{(t)},\dots,\hat c_{r_K}^{(t)})$, ($t \in 1\cdots T$ throws) from
Section~\ref{subsec:fitting} is already a usable nuisance distribution:
resample one tuple per pseudo-experiment and plug it into the
likelihood. By construction this reproduces the full dim-6$^2$ distribution using $K<N$ parameters, though they are still correlated. Appendix~\ref{app:mwe} walks through this construction on a six-bin, three-operator toy where $K=2$ representatives capture the rank-2 shape space exactly.

Given this compressed understanding of the partial next-order term in terms of $K$ representative operators, we aim to appropriately broaden the estimated full $\mathcal O(\Lambda^{-4})$ uncertainty to account for the presence of other terms at the same order in EFT perturbation theory, i.e.~the $\mathcal{B}_b\cdot\mathbf{d}$ and $\Delta_b$ terms of \cref{eq:lam4}. The broadening adds no new shapes. It changes the magnitude and the sign pattern of the correlations between the shapes the dim-6$^2$ kernel already spans.

In order to add freedom to the uncertainty estimate and account for these new terms we must break some correlations that are enforced by the dim-6$^2$ quadratic structure. We therefore work in monomial space, where the observable is linear,
rather than in coefficient space where it is quadratic. We call the
resulting procedure the {decorrelated monomial prescription}. The first step is to
reparametrise. In analogy to the $M$ construction from \cref{eq:monomial}, we form the monomial vector for each fitted tuple of representation Wilson coefficients,
\begin{equation}
 m_{kl}^{(t)} = \hat c_{r_k}^{(t)}\,\hat c_{r_l}^{(t)},
\end{equation}
for $k\leq l$,
of length $K(K{+}1)/2$ (note the contrast with eq.~\eqref{eq:monomial}, where the count $N(N{+}1)/2$ ran over the full operator set, while here it runs over the $K$ representatives only, which is the source of the dimensional reduction). The observable is linear in these monomials,
\begin{equation}
\tilde q_b = \sum_{k\leq l} w_{b,kl}\,m_{kl},
\end{equation}
with the weight matrix
$w_{b,kk} = \tilde A_{b,kk}$ and $w_{b,kl} = 2\tilde A_{b,kl}$ for
$k<l$ being the representation-space analogue of $M$.

The second step is a principal component analysis (PCA) of the monomial
ensemble. A PCA finds the directions along which a cloud of points spreads most, in decreasing order of spread, and the score of a point along a direction is its coordinate along it. Here the cloud is the ensemble of fitted monomial vectors, one point per throw. We center by subtracting the ensemble mean $\bar m$ and apply
SVD to the $T \times K(K{+}1)/2$ matrix of centered monomial vectors for each throw $t$.
Centering plays two roles. First, the diagonals
$c_i^2$ are non-negative and have a positive mean while the cross
monomials $c_i c_j$ change sign with the relative sign of the fitted
coefficients, so subtracting the empirical mean decouples their
statistical behaviour. Without that step the natural correlation
between the always-positive diagonals and the sign-symmetric crosses
would dominate the principal directions. Second, the leading principal
direction would otherwise point from the origin toward the mean of the
monomial ensemble, an offset that is nonzero because the diagonals are
non-negative, and so produce a ``nuisance parameter'' that describes the average of the dim-6$^2$ contribution rather than the dispersion of possibilities, which is what we are aiming to model. This centering shifts our nuisance parameters to naturally produce corrections with either sign to the cross section in a given bin, which is the expected behaviour of all the uncalculated terms we aim to model. The mean that is subtracted away here corresponds to the necessarily-positive total dim-6$^2$ term alone, which we shift into the signal function to account for its known sign.

Upon SVD, the centered ensemble matrix $X$, with one row per throw, $X_{t,kl} = m^{(t)}_{kl} - \bar m_{kl}$, can be decomposed as $X = \tilde U \tilde S\tilde V^\top$, where $\tilde S$ is diagonal and holds the reduced singular values $\tilde\sigma_i$. As in Sec.~\ref{subsec:svd}, $\tilde V_{m,i}$ denotes the entry of
the matrix whose columns are the right singular vectors, so that the
principal directions of variation in monomial space are the columns
$\tilde V_{\cdot,i}$ and the $\tilde\sigma_i$ set the natural scale along each. The per-throw scores $s_i$ are
\begin{equation}
 s_i^{(t)} = \sum_{k\le l} \tilde V_{kl,i}\,(m^{(t)}_{kl} - \bar m_{kl}).
\end{equation}
In plain terms, the columns of $\tilde V$ are the eigenvectors of the (representative) monomial
covariance matrix, ordered by variance, and each score $s_i^{(t)}$ tells
us how far throw $t$ sits along the $i$-th eigendirection. The leading
direction is the one along which the ensemble spreads most. The
sub-leading directions account for progressively smaller spreads. We then decorrelate the scores by force, resampling each score variable from different throws and therefore mixing deviations that were tied together in the original fit. This widens the resulting
nuisance distribution to cover patterns the dim-6$^2$-only fit cannot
produce.

The third step is an independent bootstrap with a scaling
factor $\alpha \ge 1$ to account for additional sources broadening the distribution, not just allowing shape deviations. Our physics motivation for this factor is the two next-order contributions, not calculable with the dim-6 tool set, arising at the same order in $1/\Lambda$ as the
quadratic piece: in our examples these are the SM$\times$dim-8 interference and the
dim-6 double insertions. Under NDA each carries a variance comparable to the dim-6$^2$ scale, so their sum, which is the residual we model, has approximately twice that variance. Setting $\alpha=\sqrt 2$ matches this. We stress what $\alpha$ is and is not. It is a multiplicative inflation of the variance of each decorrelated score, entering the fit through the prior of that score, and the total uncertainty then follows from profiling or marginalising the scores jointly with the Wilson coefficients. It is not the endpoint of a range over which an unphysical parameter is varied, and no envelope over variations is ever taken. The coverage scan of \cref{subsec:alpha} shows that this choice sits at the knee of the coverage curve.

Putting the pieces together, for each nuisance sample we draw each score $s_i$
independently from its empirical distribution over throws, and
reconstruct the per-bin full $\Lambda^{-4}$ contribution as
\begin{equation}
\label{eq:decorr_reconstruct}
 q^{\rm dec}_b = \bar q_b + \sum_{i=1}^{D} W_{b,i}\,\alpha\, s_i.
\end{equation}
Here $D = K(K{+}1)/2$ is the number of principal components, all of which we retain by default. Components with negligible variance contribute negligibly to $q^{\rm dec}_b$ and can be dropped without measurable effect on coverage. $\bar q_b = \sum_{k\le l} w_{b,kl}\,\bar m_{kl}$ is the mean
quadratic-form prediction (which is the only portion of this term that has nonzero mean value), and $W_{b,i} = \sum_{k\le l} w_{b,kl}\,\tilde V_{kl,i}$
rotates the weight matrix $w$ into the PCA basis. The matrix $\tilde V$ is
the same in both places: in $s_i$ it rotates monomials forward into
scores, and in $W_{b,i}$ it rotates the weights into the space in
which those scores live (the two $\tilde V$'s differ only in whether one
reads $\tilde V_{kl,i}$ as an entry of $\tilde V$ or of $\tilde V^\top$).
The decorrelation of scores for the monomial PCA directions from one another releases the
constraint that the monomials must arise from a single Wilson
coefficient vector, allowing $q^{\rm dec}_b$ to take negative values and
explore regions inaccessible to the correlated quadratic form. We never
fit a Gaussian or any other functional form to the scores. The marginal
of each PCA score is preserved exactly from the fits.

The resulting nuisance distribution preserves the per-direction scales
learned from the fits while conservatively covering the full next-order
space. The prescription applies uniformly
to all processes, including the $K=1$ case in which there is nothing to
decorrelate but the $\alpha$-inflation still widens the single monomial's
distribution.

The NDA premise behind $\alpha=\sqrt 2$ can be tested for the double-insertion half. Isolating the $c_{HQ}^{(3)}$ self-term of $pp\to jjh$, with every dim-8 coefficient set to zero so that $\Delta_b$ alone is measured (\cref{app:implementation}), gives a double insertion equal to $-0.119\pm0.007$ times that operator's dim-6$^2$ kernel $A_{b,ii}$, the same factor in all ten $m_{jj}$ bins of \cref{sec:vbf}. One constant fits the ten ratios with $\chi^2/\text{ndf}=0.49$ against per-bin statistical errors of $2.3$ to $2.5\%$, and a linear trend sits $1.5\sigma$ from zero. The double insertion reproduces the dim-6$^2$ shape and only rescales it, which is what the broadening assumes, and it does so at a tenth of the dim-6$^2$ scale rather than the comparable size NDA would give. A change of dim-8 basis moves content between $\Delta_b$ and $\mathcal{B}_b\cdot\mathbf{d}$, so the ratio is defined only at fixed basis, and one operator in one process settles neither the double insertions in general nor the SM$\times$dim-8 interference. We conclude only that $\alpha=\sqrt 2$ is conservative here.

\begin{figure}[t!]
 \centering
 \includegraphics[width=\textwidth]{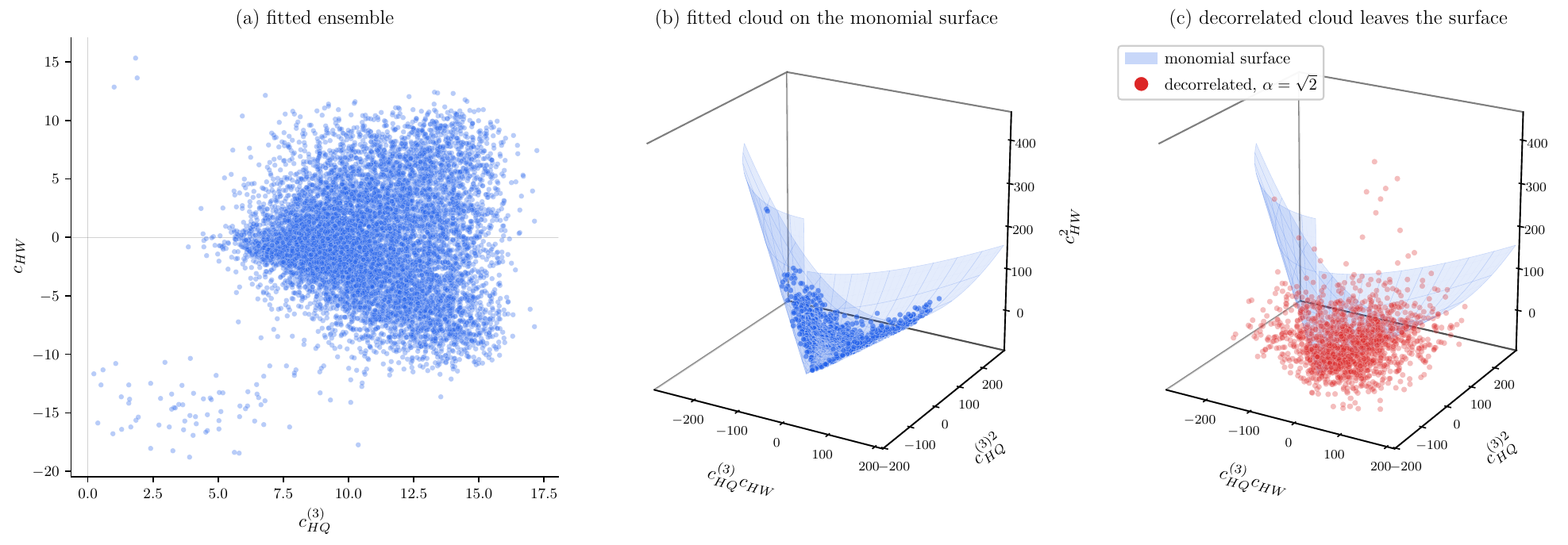}
 \caption{Pedagogical illustration of the decorrelation prescription,
 using the VBF $\Delta\phi_{jj}$ $K=2$ ensemble.
 (a)~Fitted ensemble in coefficient space. The dominant representative
 $c_{HQ}^{(3)}$ is pinned positive by convention and carries most of
 the variance, while the sub-leading representative $c_{HW}$ scatters
 around zero, so the ensemble forms a single cloud. (b)~In monomial space the fitted points
 lie on the curved surface $m_{12}^2 = m_{11}m_{22}$ (translucent
 sheet), the constraint satisfied whenever the monomials
 $(c_{HQ}^{(3)2},\, c_{HW}^2,\, c_{HQ}^{(3)}c_{HW})$ derive from a
 single coefficient pair. The sheet is two-dimensional but curved,
 so the linear PCA spans it with all three monomial directions
 (variance fractions $45\%/41\%/14\%$). (c)~The decorrelated
 nuisance (red), built by resampling the PCA scores independently
 and inflating by $\alpha=\sqrt 2$, no longer satisfies the
 constraint and lifts off the surface, reaching configurations the
 constrained fit cannot.}
 \label{fig:decorr_cartoon}
\end{figure}

Figure~\ref{fig:decorr_cartoon} illustrates the construction on the VBF
$\Delta\phi_{jj}$ ensemble, selected from our later examples in \cref{sec:examples} as a $K=2,~D=3$ case whose three
monomial principal directions all carry non-trivial variance
(45\%, 41\%, 14\%). The coefficient-space distribution forms a single
cloud: $c_{HQ}^{(3)}$ is the dominant direction and is held positive by
convention, while $c_{HW}$ is sub-leading and scatters around zero
(panel a). In monomial space the fitted points
are confined to the curved surface $m_{12}^2 = m_{11}\,m_{22}$, where the three independent monomials are $m_{11} = (c_{HQ}^{(3)})^2$, $m_{22} = c_{HW}^2$, and $m_{12} = c_{HQ}^{(3)}\,c_{HW}$ (panel b).
The sheet is two-dimensional but curved, so a linear PCA spans it with
all three monomial directions $(m_{11}, m_{22}, m_{12})$, which is why $K=2$
yields $D=K(K{+}1)/2 = 3$ principal components. The third step of
the prescription, illustrated in panel (c), resamples each principal
score independently from its empirical distribution inflated by
$\alpha=\sqrt 2$. The key step here is the {forced} decorrelation: the constrained fit (blue) populates a (curved) two-dimensional surface in the monomial space, while the decorrelated distribution (red) fills a volume by allowing each principal direction to vary independently, explicitly taking the samples off the curved surface they would otherwise be confined to. The resulting nuisance distribution is also visibly
broader than the original constrained fit due to the $\alpha$ scaling.

\subsection{Using the calibrated band in an analysis}
\label{subsec:howtouse}

The output of the algorithm is, per process and per binning, a set of $K$ representative directions, the shape weights $W_{b,i}$, and the empirical marginal of each PCA score $s_i$ (post-decorrelation, with $\alpha=\sqrt 2$ as set in \cref{subsec:decorr}). Plugging this into an analysis requires three steps.

\paragraph{Step 1: extend the signal model to include dim-6$^2$.} Standard SMEFT pipelines compute $\sigma^{\rm SM}_b$ and the linear dim-6 vector $\sigma^{\rm dim\text{-}6}_{b,i}$. The dim-6$^2$ kernel $A_b$ is already available from the same simulation. The bin-level prediction is \cref{eq:signal_with_dim6sq}: SM, linear-in-$\mathbf{c}$, the deterministic $\mathbf{c}^\top A_b\,\mathbf{c}/\Lambda^4$ at the fitted $\mathbf{c}$ adding the known positive shift from this positive-definite term, and the calibrated nuisance contribution $\varepsilon_b=\sum_{i=1}^{D} W_{b,i}\,\alpha\,s_i$, entering through the overall $1/\Lambda^4$ of \cref{eq:signal_with_dim6sq} and centered at zero change, as the positive dim-6$^2$ term is already present in the signal modelling.

\paragraph{Step 2: treat the $D=K(K{+}1)/2$ PCA scores as nuisance parameters.} The residual enters every bin through the same set of $D$ scores $s_i$ introduced in eq.~\eqref{eq:decorr_reconstruct}, so the nuisance contribution in bin $b$ is $\varepsilon_b = \sum_{i=1}^{D} W_{b,i}\,\alpha\, s_i$, with the $W_{b,i}$ and $\alpha=\sqrt 2$ fixed by the calibration scan and the same across all bins of a given process. Each $s_i$ is profiled or marginalised over jointly with $\mathbf{c}$ under its calibrated marginal prior, which is mean-zero by PCA construction (though skewed, hence not strictly symmetric) and decorrelated from the other $s_j$, so no cross-bin or cross-component constraint is needed. The number of nuisance parameters is therefore $D$, not the number of bins, and bin-to-bin correlations of $\varepsilon_b$ are inherited from the shared $W_{b,i}$.

\paragraph{Step 3: read off the EFT-validity behaviour.}

With this approach, EFT validity at parton energies approaching $\Lambda$ is handled automatically by the structure of \cref{eq:signal_with_dim6sq}. Any bin in which the linear signal $\sum_i c_i\,\sigma^{\rm dim\text{-}6}_{b,i}/\Lambda^2$ at the fitted $\mathbf{c}$ falls below the calibrated nuisance scale built from the joint PCA-score draw $\sum_{i=1}^{D} W_{b,i}\,\alpha\, s_i / \Lambda^4$ of \cref{eq:decorr_reconstruct} is one in which the next-order remainder can be as large as the signal it would correct, so the bin's likelihood weight collapses and it contributes no information to bounds on $\mathbf{c}$. Because the nuisance scale is set by the independent NDA-prior scan over $\mathbf{c}'$ rather than by the fitted $\mathbf{c}$, the attenuation persists even when the fit pulls $\mathbf{c}$ toward zero. Self-consistency only requires applying the same NDA bound $|c_i|\le p$ to the fitted signal coefficients that was applied to the calibration draws. This constraint on the range of the Wilson coefficients to fit in the analysis is the only bound that needs to be enforced by hand in such an analysis. There is no need for specialised phase-space cuts or externally imposed energy thresholds.

\section{Examples}
\label{sec:examples}

We now present three distinct examples of this technique. The processes studied here (DY, associated $Zh$ production, VBF Higgs) are chosen for their kinematic distinctions from each other and because their full
$\Lambda^{-4}$ distribution, including the dim-8 contribution, {has been studied in the literature, analytically or with simulation/numerically~\cite{Hays:2018zze, Dawson:2021xei, Boughezal:2021tih, Kim:2022amu, Corbett:2023yhk, Assi:2024zap, Araz:2020zyh}}. The comparison serves as a cross-check demonstrating
that the dim-6-only nuisance prescription covers the full correction even without
an explicit dim-8 model.
The {decorrelated nuisance distribution} is generated by
applying the full algorithm of Section~\ref{sec:algo}, importantly including the decorrelation and expansion of \cref{subsec:decorr} with
$\alpha = \sqrt{2}$. In the binwise comparison plots throughout the examples below it is shown in orange, alongside the full $\Lambda^{-4}$ truth (including the dim-8 contribution and double insertions) in green.

The pure dim-6$^2$ observable used as input for this algorithm is extracted from dedicated
dim-6-only simulations, which contain only the $|\mathcal{M}_6|^2$
amplitude-squared terms without double-insertion interference.
The dim-6$^2$ piece is moreover the unique
basis-change-invariant contribution at $\mathcal{O}(\Lambda^{-4})$ from
dim-6 operators alone~\cite{Helset:2020yio, Brivio:2017vri}, and thus
the only well-defined piece available without dedicated dim-8 simulation,
which makes it the natural basis for an estimator for the truncation
uncertainty.

\subsection{Validation of estimates}
\label{subsec:validation}

The primary validation of our technique in these examples is a per-bin comparison of two distributions: the full $\Lambda^{-4}$ prior (green) and the decorrelated nuisance band (orange).
The full $\Lambda^{-4}$ prior is obtained by
drawing all $N$ dim-6 coefficients from $\mathrm{U}[-p,+p]$ as in \cref{eq:prior} and all
dim-8 coefficients from $\mathrm{U}[-p_8,+p_8]$ with $p_8 = p^2$, consistent with NDA estimation. We then evaluate
$q_b^{\rm full}(\mathbf{c},\mathbf{d})$ through eq.~\eqref{eq:lam4}, with all terms calculated explicitly to this order. This full $\Lambda^{-4}$ result contains both dim-8 interference
and double-insertion contributions consistently included in the full simulation.

Agreement is quantified by the {coverage} -- the fraction of
full-prior throws that fall within the $X\%$ credible interval of
the nuisance distribution. For an uncertainty estimate this is the
relevant metric: a nuisance distribution that is wider than the truth has high
coverage and is conservative and safe, while one that is narrower
misses real effects. Per-bin comparison figures throughout this
paper annotate the per-bin $95\%$ coverage $C_{95}$.

\subsection{Example I: High-$p_T$ Drell-Yan}
\label{sec:dy}

\subsubsection{Setup}
\label{subsec:dy_setup}

\begin{figure}[H]
 \centering
 \includegraphics[width=0.7\textwidth]{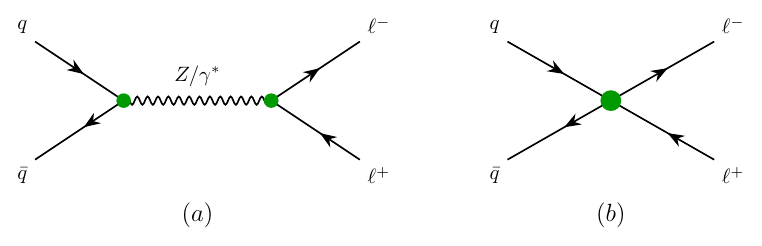}
 \caption{Tree-level topologies for $q\bar{q}\to\ell^+\ell^-$ with dim-6 SMEFT
 insertions (green blobs).
 (\textit{a})~$s$-channel: $q\bar{q}\to Z/\gamma^*\to\ell^+\ell^-$, where SMEFT
 operators modify the $\bar{q}qV$ and $V\ell\bar\ell$ vertices.
 (\textit{b})~Four-fermion contact interaction from $\psi^4$ operators.
 At dim-6$^2$ the full observable receives contributions from the squares
 of each topology and their interference.}
 \label{fig:dy_diagrams}
\end{figure}

As a first example we apply the algorithm to high-$p_T$ Drell-Yan production
$pp\to\ell^+\ell^-$ in the invariant-mass distribution
$m_{\ell\ell}$~\cite{Allwicher:2022gkm, Boughezal:2022nof}. This process provides a
clean test case: it receives
contributions from two distinct topologies (Figure~\ref{fig:dy_diagrams}) involving $N=14$
dim-6 operators with nonzero contribution at tree level and $N_8=40$ dim-8 operators
(Table~\ref{tab:dy_ops}) under the
assumption of $U(3)^5$ flavour symmetry and CP conservation\footnote{Furthermore, we take all fermions to be massless, so mixed-chirality combinations (e.g.\ LR initial or final states) do not interfere with the SM at the dim-6 level.}. Relaxing these assumptions would enlarge $N$ substantially, adding flavour-non-universal and CP-odd partners of the operators above (the flavour-general dim-6 basis has $2499$ baryon-number-conserving parameters), but leaves the rank-reduction procedure unchanged. Analytical
studies of full $\Lambda^{-4}$ contributions to dilepton production at high $p_T$ have
been carried out in
Refs.~\cite{Alioli:2020kez, Boughezal:2021tih, Boughezal:2022nof},
providing independent cross-checks of the kinematic structure used here.

The $s$-channel topology receives corrections from $\psi^2H^2D$ operators ($c_{Hu}$, $c_{Hd}$, $c_{He}$, $c_{HL}^{(1,3)}$,
$c_{HQ}^{(1,3)}$) that modify the $Z/\gamma^*$ couplings, while the
contact topology arises from four-fermion operators ($c_{LQ}^{(1,3)}$,
$c_{ed}$, $c_{Ld}$, $c_{Lu}$, $c_{eu}$, $c_{Qe}$). Operators that enter only as shifts to the electroweak input parameters ($c_{HWB}$, $c_{HD}$, and the four-lepton $c_{LL}$ that rescales $G_F$) are absorbed into the input scheme and not tracked separately, consistent with the $Vh$ and VBF examples below. The bosonic operators $c_{HW}$ and $c_{HB}$ modify $hVV$ vertices that this final state does not contain, so they contribute exactly zero at tree level and are removed by the diagonal filter of Section~\ref{subsec:svd}.

\begin{table}[t]
 \centering
 \small
 \renewcommand{\arraystretch}{1.2}
 \begin{tabular}{lll}
 \hline
 \textbf{Class} & \textbf{Operators} & \textbf{Topology} \\
 \hline
 \multicolumn{3}{l}{\textbf{Dim-6 operators ($N=14$)}} \\
 \hline
 $\psi^4$ (four-fermion) &
 $c_{LQ}^{(1)},\ c_{LQ}^{(3)},\ c_{ed},\ c_{eu},\ c_{Ld},\ c_{Lu}$ &
 contact \\
 &
 $c_{Qe}$ &
 contact \\
 \hline
 $\psi^2 H^2 D$ &
 $c_{HL}^{(1)},\ c_{HL}^{(3)},\ c_{HQ}^{(1)},\ c_{HQ}^{(3)}$ &
 $s$-channel \\
 &
 $c_{Hu},\ c_{Hd},\ c_{He}$ &
 $s$-channel \\
 \hline
 \multicolumn{3}{l}{\textbf{Dim-8 operators ($N_8=40$)}} \\
 \hline
 $\psi^4 D^2,\ \psi^4 HD,\ \psi^4 X$ &
 $c^{(8)}_{LQ^{(1\text{--}4)}},\ c^{(8)}_{Ld^{(1,3)}},\ c^{(8)}_{Lu^{(1,3)}},\ c^{(8)}_{Qe^{(1,3)}}$ &
 contact \\
 &
 $c^{(8)}_{LQ^{(1)},s/t},\ c^{(8)}_{LQ^{(3)},s/t}$ &
 contact \\
 &
 $c^{(8)}_{Ld,s/t},\ c^{(8)}_{Lu,s/t},\ c^{(8)}_{eQ,s/t}$ &
 contact \\
 &
 $c^{(8)}_{ed},\ c^{(8)}_{ed,s/t},\ c^{(8)}_{eu},\ c^{(8)}_{eu,s/t}$ &
 contact \\
 \hline
 $\psi^2 H^4 D$ &
 $c^{(8,a)}_{HL}$ ($a=1,2,3$), $c^{(8,a)}_{HQ}$ ($a=1,2,3$) &
 $s$-channel \\
 &
 $c^{(8)}_{Hu},\ c^{(8)}_{Hd},\ c^{(8)}_{He},\ \delta G_F^{(8)}$ &
 $s$-channel \\
 \hline
 $H^2 X^2 D^2$ &
 $c^{(8)}_{HD},\ c^{(8)}_{HD^2},\ c^{(8)}_{HW^2},\ c^{(8)}_{HWB}$ &
 $s$-channel \\
 \hline
 \end{tabular}
 \caption{Operators contributing to high-$p_T$ Drell-Yan at $\mathcal{O}(\Lambda^{-4})$,
 grouped by operator class and the topology through which they enter.
 The dim-6 operators produce the dim-6$^2$ quadratic observable. The dim-8
 operators enter through linear SM$\times$dim-8 interference and are included
 in the full $\Lambda^{-4}$ validation. Dim-8 operators follow the basis of
 Refs.~\cite{Li:2020gnx, Murphy:2020rsh}. Superscripts $s,t$ distinguish
 Mandelstam-channel variants within a given four-fermion class.}
 \label{tab:dy_ops}
\end{table}

The dim-6$^2$ matrix $A_b$ is computed for the inclusive $m_{\ell\ell}$
distribution from $300~\text{GeV}$ to $2400~\text{GeV}$ in $B=21$ bins.

\subsubsection{Rank and representatives}
\label{subsec:dy_rank}

The monomial feature matrix has dimensions $21 \times 105$, since
$N=14$ active operators give $N(N{+}1)/2 = 105$ unique monomials. The
median-$R^2$ ladder (Section~\ref{subsec:fitting})
accepts the first two directions and rejects the rest, giving $K=2$. The
leading direction carries $99.3\%$ of the variance and the
sub-leading direction a further $0.70\%$.

The leading representative is $c_{LQ}^{(3)}$, a four-fermion contact
operator. It represents the flat direction containing the set of operators $c_{LQ}^{(1)},\ c_{ed},\ c_{eu},\ c_{Ld},\ c_{Lu},\ c_{Qe}$. The sub-leading representative is $c_{HL}^{(3)}$, an $s$-channel
$\psi^2H^2D$ operator. Its flat direction contains the remaining
$s$-channel operators
$c_{HL}^{(1)},\ c_{HQ}^{(1,3)},\ c_{Hu},\ c_{Hd},\ c_{He}$. Adding a third representative
no longer improves the median $R^2$.

The fact that two representatives, one per topology, suffice is consistent with
the energy-enhanced expansion of Ref.~\cite{Assi:2025zmp}, which introduced the parameter $\lambda$ to track different powers of $E/\Lambda$ vs. $v/\Lambda$ for different SMEFT operator classes. In the regime $\Lambda \gg E \gg v$, $\lambda$ scales as $(\Lambda, E, v)\sim(\lambda^{-3}, \lambda^{-2}, \lambda^{-1})$, with $\lambda \ll 1$. The net $\lambda$ scaling for an amplitude can be derived from the product of the scaling of each vertex involved along with factors of $1/E^2 \sim \lambda^4$ from each propagator. Following this approach, all SMEFT corrections show up in the vertex factors as higher powers of $\lambda$ accompanied by the Wilson coefficients from different operator classes. In the expansions that follow we use operator-class notation: for example, $c_{\psi^2H^2D}$ stands collectively for any operator in
that class ($c_{HQ}^{(3)}$, $c_{HQ}^{(1)}$, $c_{HL}^{(3)}$, etc.,
enumerated in Table~\ref{tab:dy_ops}), and $c_{\psi^4}$ stands for any
four-fermion operator. The $\lambda$-weight depends only on the class,
not on which specific operator within the class is active.

The $\lambda$-powers tabulated below are amplitude-level. Observable-level scalings are products of amplitudes: a SM$\times$dim-6 linear interference contributes at $\lambda^{a_{\rm SM}+a_X}$ (linear in $c_X$). The dim-6$^2$ self-interference of operator $X$ contributes at $\lambda^{2 a_X}$ and its cross with operator $Y$ at $\lambda^{a_X+a_Y}$, both quadratic in dim-6 coefficients. The SM$\times$dim-8 piece of the $\Lambda^{-4}$ truth scales as $\lambda^{a_{\rm SM}+a_d}$ (linear in a dim-8 coefficient $d$ with amplitude power $a_d$). In every case the relative ordering of contributions at the observable level matches the amplitude ordering.

Working through the $\lambda$ scaling for the $s$-channel Drell-Yan contribution, the on-shell $\bar{q}qV$ vertex scales as
\begin{align}
\label{eq:dy_qqv}
g_{\bar{q}qV} &= \frac{g^{\rm SM}_{\bar{q}qV}}{\lambda^2}
+ \frac{\lambda^2}{\hat\Lambda^2}\, c_{\psi^2 H^2 D}
+ \frac{\lambda^5}{\hat\Lambda^4}\, c^{(8)}_{\psi^2 H^4 D}
+ \cdots\,,
\end{align}
where $\hat\Lambda$ is the cutoff scale with its power-counting of $\lambda^{-3}$ explicitly removed and the factor $1/\lambda^2$ comes from the two external spinor
wave-function normalisations, each contributing $\sqrt{E}\sim\lambda^{-1}$.
The $V\ell\bar\ell$ vertex has the same structure with lepton-sector
Wilson coefficients.

Multiplying the $g_{\bar{q}qV}$ and $g_{V\ell\bar\ell}$ expansions
with a propagator factor, the $s$-channel
topology gives (to $\mathcal{O}(\lambda^5)$)
\begin{align}
\label{eq:dy_amp1}
\mathcal{A}^{(1)}_{\rm DY} &= g^{\rm SM}_{\bar{q}qV}\, g^{\rm SM}_{V\ell\bar\ell}
+ \frac{\lambda^4}{\hat\Lambda^2}\,
 \bigl(g^{\rm SM}_{V\ell\bar\ell}\, c_{\psi^2 H^2 D}
 + g^{\rm SM}_{\bar{q}qV}\, c_{\psi^2 H^2 D}\bigr)
+ \cdots\,.
\end{align}
Similar logic applied to the contact topology gives
\begin{align}
\label{eq:dy_contact}
\mathcal{A}^{(2)}_{\rm DY} = g_{\bar{q}q\ell\bar\ell} &= \frac{\lambda^2}{\hat\Lambda^2}\, c_{\psi^4}
+ \frac{\lambda^4}{\hat\Lambda^4}\, c^{(8)}_{\psi^4 D^2}
+ \frac{\lambda^5}{\hat\Lambda^4}\, c^{(8)}_{\psi^4 HD}
+ \cdots\,.
\end{align}

We see that the SM contribution enters the amplitude at $\lambda^0$ (two vertices at
$\lambda^{-2}$ each, propagator at $\lambda^4$), the contact $\psi^4$
enters at $\lambda^2$, and the $s$-channel $\psi^2 H^2 D$ modification
enters at $\lambda^4$.
The contact operators are therefore the most energy-enhanced,
dominating by $\lambda^{-2}$ over the $s$-channel modifications. At the dim-6$^2$ observable level this translates to a $c_{LQ}^{(3)}$ self-interference at $\lambda^4$, a $c_{LQ}^{(3)}c_{HL}^{(3)}$ cross at $\lambda^6$, and a $c_{HL}^{(3)}$ self-interference at $\lambda^8$.
This hierarchy is reflected in the fit: the leading representative
$c_{LQ}^{(3)}$ belongs to the contact topology and carries $99.3\%$ of
the variance, while the sub-leading representative $c_{HL}^{(3)}$ from
the $\psi^2H^2D$ class of the $s$-channel topology carries the
remaining $\sim 0.7\%$. The two topologies produce distinct dim-6$^2$
bin-by-bin shapes, so both directions survive the $R^2$ ladder.

\subsubsection{Nuisance parameter distributions}
\label{subsec:dy_dists}

\begin{figure}[t!]
 \centering
 \includegraphics[width=0.5\textwidth]{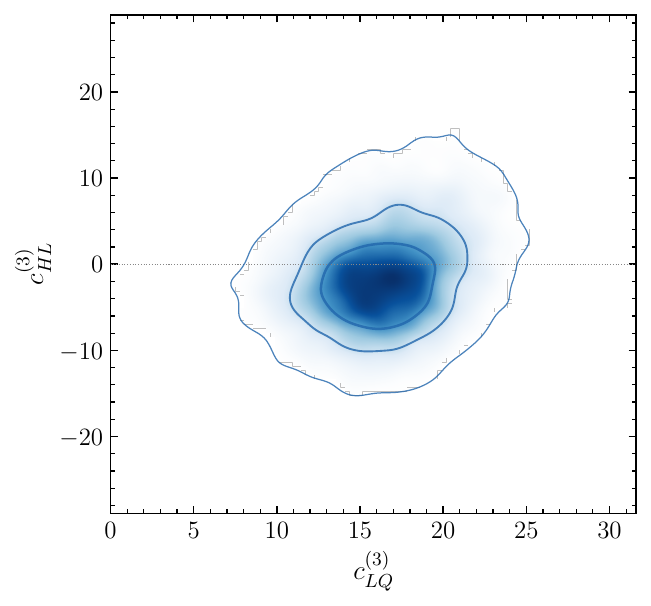}
 \caption{Joint distribution of the two DY representatives
 $c_{LQ}^{(3)}$ and $c_{HL}^{(3)}$ from $T=10{,}000$ throws
 at $\Lambda=3~\text{TeV}$. Blue KDE contours enclose $38\%$,
 $68\%$, and $95\%$ of the density. The takeaway is that the fitted coefficient distribution is a single skewed cloud rather than a simple Gaussian (broad and unimodal in $c_{LQ}^{(3)}$, which is pinned positive by convention and dominant, with $c_{HL}^{(3)}$ sub-leading and scattered around zero), which is what the decorrelated empirical prescription of Sec~\ref{subsec:decorr} resamples, with no parametric approximation imposed.}
 \label{fig:dy_2d}
\end{figure}

Figure~\ref{fig:dy_2d} shows the joint fitted distribution of the two
DY representatives from $T=10{,}000$ throws at
$\Lambda=3~\text{TeV}$. For visualisation, kernel density estimation (KDE)
is used to overlay smooth density curves on the marginal histograms and to
produce 2D density contours for pairwise distributions.
With the prior $|c_i|\le p=4\pi$ independent of $\Lambda$, the fitted-coefficient distributions in \cref{fig:dy_2d} are themselves $\Lambda$-independent in coefficient space: $\Lambda$ enters only through the explicit $1/\Lambda^4$ that converts the dim-6$^2$ kernel output $\mathbf{c}^\top A_b\,\mathbf{c}$ to a cross-section contribution at $\mathcal{O}(\Lambda^{-4})$ (Section~\ref{subsec:fitting}).

Refitting the two representatives directly to the full
$\Lambda^{-4}$ truth, including the SM$\times$dim-8 interference, gives
a fit quality indistinguishable from the dim-6$^2$-only fit, with
median $R^2 = 0.9999$ and $99.9\%$ of throws above $0.95$: the dim-8
operators of Table~\ref{tab:dy_ops} enter through the same two
topologies as the dim-6 set, so in this observable they produce no new
bin-by-bin shapes for the representatives to miss. This is a statement
about DY, not about the expansion, and it does not remove the need for
the truncation error: the dim-8 piece still shifts the observable
within those shapes, and in processes where dim-8 opens new topologies
the same comparison retains a visible residual
(Figure~\ref{fig:r2_vbf} in Section~\ref{sec:vbf}). That uncalculable
remainder is what the $\alpha=\sqrt 2$ inflation of the decorrelated
prescription is calibrated to cover, as the coverage scan of
Section~\ref{subsec:alpha} confirms. Recall that this $R^2$ measures how
well the representative coefficients reconstruct the ensemble of
simulated truth throws, not how well they fit experimental data.


\subsubsection{$\Lambda^{-4}$ comparison}
\label{subsec:dy_lam4}

\begin{figure}[H]
 \centering
 \includegraphics[width=\textwidth]{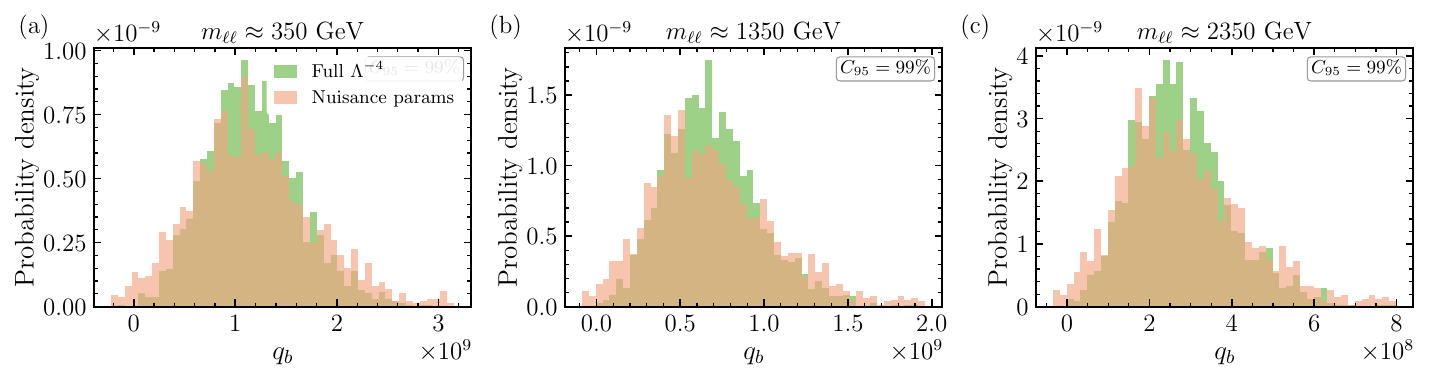}
 \caption{Per-bin comparison of the full $\Lambda^{-4}$ distribution (green) and the
 $K=2$ decorrelated nuisance distribution (orange) for representative
 $m_{\ell\ell}$ bins, for high-$p_T$ Drell-Yan ($K=2$).
 Resampling the $D=K(K{+}1)/2$ decorrelated PCA scores $s_i$ (the
 principal directions of the representative monomial ensemble),
 inflated by $\alpha=\sqrt 2$ and reconstructed per bin through the
 fixed shape weights $W_{b,i}$ as in eq.~\eqref{eq:decorr_reconstruct},
 reproduces the full $\mathcal{O}(\Lambda^{-4})$ uncertainty budget bin
 by bin. Per-bin $95\%$ coverage $C_{95}$ annotated per panel. }
 \label{fig:dy_lam4}
\end{figure}

Figure~\ref{fig:dy_lam4} compares the per-bin distribution of $q_b$
from the full prior (green) against the decorrelated two-representative
nuisance (orange) with $\alpha=\sqrt 2$, which achieves $C_{95}\geq 0.99$ across all $m_{\ell\ell}$ bins. The correlated
bootstrap without $\alpha$-inflation (pure dim-6$^2$ reconstruction)
already covers most of the $\Lambda^{-4}$ variation by itself. The
$\alpha=\sqrt 2$ step widens the distribution to conservatively
cover the remaining dim-8 and double-insertion contributions. Three nuisance parameters, one per monomial of the two representatives, cover everything the forty dim-8 operators can do to the $m_{\ell\ell}$ spectrum at this order.

\subsubsection{Forward and backward $m_{\ell\ell}$ distributions}
\label{subsec:dy_fwdbwd}

Next, we analyze the forward ($\sigma^F$) and backward ($\sigma^B$)
distributions in $\cos\theta^*$ separately. Each hemisphere yields $K=2$
with representatives $(c_{LQ}^{(3)}, c_{HL}^{(3)})$, the same pair as the
inclusive result, with median $R^2 = 0.9999$ in each hemisphere. Analogous figures to \cref{fig:dy_2d,fig:dy_lam4} for the
forward and backward distributions are presented in \cref{app:figs}.

The row labelled $(F{+}B)$ in \cref{tab:summary} is the bin-by-bin sum of the two hemispheres, which is the inclusive $m_{\ell\ell}$ spectrum of the previous subsection and carries no forward-backward information. It gives $K=2$ with the same representatives and $95\%$ coverage of $0.993$ at worst across its $N_{\rm bin}$ bins. What the hemisphere split shows is that the reduction is stable under it, since each hemisphere separately returns the same $K$ and the same representatives as the sum.

\subsection{Example II: $Zh$ production}
\label{sec:vh}

As our second example, we explore associated Higgs production where we decay the vector boson $Z$ into a fermion-anti-fermion pair, $pp \to Zh \to f\bar f h$. Associated Higgs production is an interesting test case for three reasons. First,
the three-body final state exposes a
richer kinematic structure than Drell-Yan. The 2$\to$3 phase space has
five independent invariants. We focus on three observables commonly
studied in $Zh$ EFT analyses ($p_{T,H}$, $p_{T,\ell}$, and
$\cos\theta^*$), together with the joint 2D distribution
$p_{T,\ell}\times\cos\theta^*$ in Section~\ref{subsec:vh_2d}. Second, the
dim-8 operators contributing to $Zh$ all flow through topologies already
present at dim-6 (this will not be the case for our third example, VBF). Finally, a
complete basis of dim-8 operators for $Zh$ has been enumerated and encoded in a simulation framework in
Ref.~\cite{Corbett:2023yhk}, allowing a quantitative validation of the method
against the full $\Lambda^{-4}$ truth rather than just the pure
dim-6$^2$ piece.

\subsubsection{Setup}
\label{subsec:vh_setup}

\begin{figure}[H]
 \centering
 \includegraphics[width=0.85\textwidth]{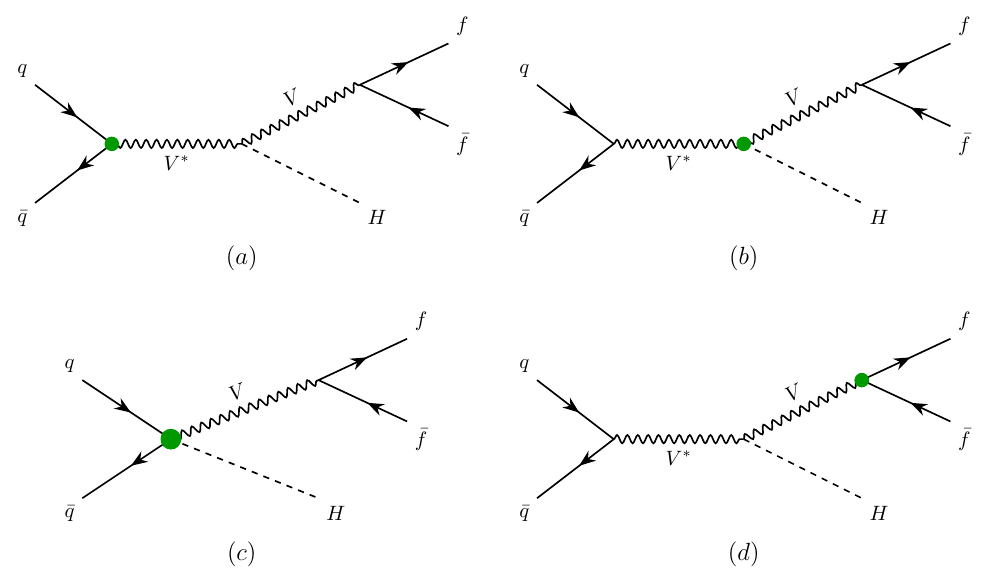}
 \caption{Tree-level topologies for $q\bar{q}\to f\bar{f}H$ ($Zh$ production
 with $Z\to f\bar{f}$ decay) with dim-6 SMEFT insertions (green blobs).
 (\textit{a})~$s$-channel with modified $\bar{q}qZ/\gamma$ vertex,
 (\textit{b})~$s$-channel with modified $ZZH$ vertex,
 (\textit{c})~contact $\bar{q}qZh$ interaction from $\psi^2 H^2 D$ operators,
 and (\textit{d})~$s$-channel with modified $Zf\bar{f}$ decay vertex.
 The operators contributing to each vertex are listed in Table~\ref{tab:vh_ops}.}
 \label{fig:vh_diagrams}
\end{figure}

$Zh$ production~\cite{Corbett:2023yhk, Bishara:2022vsc} receives
tree-level contributions from Higgs-strahlung,
$q\bar q\to Z^*\to Zh$ with $Z\to f\bar f$
(Figure~\ref{fig:vh_diagrams}), involving $N=14$ operators with
nonzero contribution at tree level under an assumption of $U(3)^5$
flavour symmetry, acting through several vertex
structures. Unlike the Drell-Yan example, the CP-odd bosonic
operators are retained here, and two of them are in fact resolved as
independent representatives below. We restrict to tree level and do not include the
loop-induced channel $gg\to Zh$. As a result, the operator $c_{HG}$ is absent from our
analysis. The full set of $N=14$ active dim-6 and $N_8=29$ dim-8 operators is listed in
Table~\ref{tab:vh_ops}.

\begin{table}[t]
 \centering
 \small
 \renewcommand{\arraystretch}{1.2}
 \begin{tabular}{lll}
 \hline
 \textbf{Class} & \textbf{Operators} & \textbf{Vertex} \\
 \hline
 \multicolumn{3}{l}{\textbf{Dim-6 operators ($N=14$)}} \\
 \hline
 $\psi^2 H^2 D$ &
 $c_{HL}^{(1)},\ c_{HL}^{(3)},\ c_{HQ}^{(1)},\ c_{HQ}^{(3)}$ &
 $\bar{q}qZ$, $Zf\bar{f}$, contact \\
 &
 $c_{Hu},\ c_{Hd},\ c_{He}$ &
 $\bar{q}qZ$, $Zf\bar{f}$, contact \\
 \hline
 $H^4 D^2$ &
 $c_{HD}$ &
 $ZZH$ \\
 \hline
 $H^2 X^2,\ H^2 X D$ &
 $c_{HW},\ c_{H\tilde{W}},\ c_{HB},\ c_{H\tilde{B}}$ &
 $ZZH$ \\
 &
 $c_{HWB},\ c_{H\widetilde{WB}}$ &
 $ZZH$ \\
 \hline
 \multicolumn{3}{l}{\textbf{Dim-8 operators ($N_8=29$)}} \\
 \hline
 $\psi^2 H^4 D$ &
 $c^{(8)}_{HL^{(1,2,3)}},\ c^{(8)}_{HQ^{(1,2,3)}}$ &
 $\bar{q}qZ$, $Zf\bar{f}$ \\
 &
 $c^{(8)}_{Hu},\ c^{(8)}_{Hd},\ c^{(8)}_{He}$ &
 $\bar{q}qZ$, $Zf\bar{f}$ \\
 \hline
 $H^4 D^4,\ H^2 X^2 D^2$ &
 $c^{(8)}_{HD},\ c^{(8)}_{HD^2}$ &
 $ZZH$ \\
 &
 $c^{(8)}_{HB},\ c^{(8)}_{HW},\ c^{(8)}_{HW^2},\ c^{(8)}_{HWB}$ &
 $ZZH$ \\
 &
 $c^{(8)}_{HD\cdot HB},\ c^{(8)}_{HD\cdot HW}$ &
 $ZZH$ \\
 \hline
 $\psi^2 H^2 X D^2,\ \psi^2 H^2 D^3$ &
 $c^{(8)}_{Q^2 B H^2 D^{(1,3)}},\ c^{(8)}_{Q^2 W H^2 D^{(1,3)}}$ &
 contact $\bar{q}qZh$ \\
 &
 $c^{(8)}_{Q^2 H^2 D^{3,(1,3)}},\ c^{(8)}_{u^2 B H^2 D^{(1)}},\ c^{(8)}_{u^2 W H^2 D^{(1)}}$ &
 contact $\bar{q}qZh$ \\
 &
 $c^{(8)}_{u^2 H^2 D^{3,(1)}},\ c^{(8)}_{d^2 B H^2 D^{(1)}},\ c^{(8)}_{d^2 W H^2 D^{(1)}},\ c^{(8)}_{d^2 H^2 D^{3,(1)}}$ &
 contact $\bar{q}qZh$ \\
 \hline
 \end{tabular}
 \caption{Operators contributing to $Zh$ production at $\mathcal{O}(\Lambda^{-4})$,
 grouped by operator class and the vertex they modify.
 The dim-6 $\psi^2 H^2 D$ operators modify the $\bar{q}qZ$ and $Zf\bar{f}$
 vertices and contribute a contact interaction. Dim-8 operators enter
 through linear SM$\times$dim-8 interference and are included in the
 full $\Lambda^{-4}$ validation, following the basis of
 Refs.~\cite{Li:2020gnx, Murphy:2020rsh}.}
 \label{tab:vh_ops}
\end{table}

\subsubsection{Rank and representatives}
\label{subsec:vh_rank}

The first kinematic variable we look at is the Higgs transverse momentum $p_{T,H}$, split into $B=16$ bins spanning $25$ GeV to $775$ GeV. Multiplying the bins times the dim-6$^2$ possibilities, we begin with a matrix of $16\times 105$ entries. Running through the median-$R^2$ ladder procedure described in Section~\ref{subsec:fitting}, we find this reduces to five
representatives, giving $K=5$. The variance is distributed across
the five as follows: the leading direction carries $97.6\%$ of the variance
and is represented by $c_{HQ}^{(3)}$, with the flat directions
$\{c_{HQ}^{(1)},\,c_{Hd},\,c_{Hu}\}$.
The second direction ($2.0\%$) is represented by $c_{HW}$, absorbing
$c_{HB}$ and $c_{HWB}$. The third ($0.14\%$) is $c_{H\widetilde{W}}$
alone. The fourth ($0.08\%$) is represented by $c_{H\widetilde{B}}$,
absorbing $c_{H\widetilde{WB}}$. Finally, the fifth ($0.05\%$) is $c_{HL}^{(3)}$,
representing $\{c_{HD},\,c_{HL}^{(1)},\,c_{He}\}$.

This hierarchy is consistent with the $\lambda$-counting
of Ref.~\cite{Assi:2025zmp}. At tree level in the SM, $q\bar{q}\to Zh$
proceeds via an $s$-channel diagram connecting a $\bar{q}qZ$ vertex to
a $hZZ$ vertex by a vector boson propagator. Beyond the SM, a
four-point $\bar{q}qZh$ contact vertex also contributes.

The on-shell $\bar{q}qZ$ vertex scales as
\begin{align}
\label{eq:vh_qqv}
g_{\bar{q}qZ} &= \frac{g^{\rm SM}_{\bar{q}qZ}}{\lambda^2}
+ \frac{\lambda^2}{\hat\Lambda^2}\, c_{\psi^2 H^2 D}
+ \frac{\lambda^5}{\hat\Lambda^4}\, c^{(8)}_{\psi^2 H^4 D}
+ \cdots\,,
\end{align}
the $hZZ$ vertex as
\begin{align}
\label{eq:vh_hvv}
g_{hZZ} &= \frac{g^{\rm SM}_{hZZ}}{\lambda}
+ \frac{\lambda}{\hat\Lambda^2}\, c_{H^2 X^2}
+ \frac{\lambda^3}{\hat\Lambda^2}\, c_{H^4 D^2}
+ \frac{\lambda^5}{\hat\Lambda^4}\, c^{(8)}_{H^4 X^2}
+ \cdots\,,
\end{align}
and the four-point contact vertex (no SM counterpart) as
\begin{align}
\label{eq:vh_qqvh}
g_{\bar{q}qZh} &= \frac{\lambda^3}{\hat\Lambda^2}\, c_{\psi^2 H^2 D}
+ \frac{\lambda^5}{\hat\Lambda^4}\, c^{(8)}_{\psi^2 H^2 D^3}
+ \frac{\lambda^5}{\hat\Lambda^4}\, c^{(8)}_{\psi^2 H^2 X D}
+ \cdots\,.
\end{align}
Here $c_{\psi^2 H^2 D}$ includes $c_{HQ}^{(3)}$ (the leading
representative), $c_{H^2 X^2}$ includes $c_{HW}$, $c_{HB}$ and their
CP-odd counterparts, and $c_{H^4 D^2}$ includes $c_{HD}$ and $c_{H\Box}$.

Multiplying the $g_{\bar{q}qZ}$ and $g_{hZZ}$ expansions with a
propagator factor $E^{-2}\sim\lambda^4$, the $s$-channel topology gives
(to $\mathcal{O}(\lambda^5)$)
\begin{align}
\label{eq:vh_amp1}
\mathcal{A}^{(1)}_{Vh} &= \lambda\, g^{\rm SM}_{\bar{q}qZ}\, g^{\rm SM}_{hZZ}
+ \frac{\lambda^3}{\hat\Lambda^2}\, g^{\rm SM}_{\bar{q}qZ}\, c_{H^2 X^2}
\nonumber\\
&\quad + \frac{\lambda^5}{\hat\Lambda^2}\,\bigl(
 g^{\rm SM}_{hZZ}\, c_{\psi^2 H^2 D}
 + g^{\rm SM}_{\bar{q}qZ}\, c_{H^4 D^2}\bigr)
+ \cdots\,,
\end{align}
while the contact topology gives
$\mathcal{A}^{(2)}_{Zh} = g_{\bar{q}qZh}$.
The SM enters at $\lambda^1$ (vertices at $\lambda^{-2}$ and
$\lambda^{-1}$, propagator at $\lambda^4$). In the helicity-blind
counting the $\psi^2 H^2 D$ contact and the $s$-channel $H^2 X^2$
both enter at $\lambda^3$, and the $s$-channel $\psi^2 H^2 D$ and
$H^4 D^2$ at $\lambda^5$. High-$p_T$ $Zh$ production is dominated by a
longitudinal $Z$, whose polarisation vector carries
$\varepsilon_L \sim E/m_Z \sim E/v$. This factor enhances the
$\psi^2 H^2 D$ contact by one power to an effective $\lambda^2$,
making it the leading SMEFT correction. At the dim-6$^2$ level, writing each class through its SVD-selected representative ($c_{HQ}^{(3)}$ for $\psi^2 H^2 D$, $c_{HW}$ for $H^2 X^2$), this maps to a $c_{HQ}^{(3)}$ contact-squared at $\lambda^4$, a $c_{HQ}^{(3)}c_{HW}$ cross-interference at $\lambda^5$, and a $c_{HW}^2$ self-term at $\lambda^6$.

The $K=5$ independent shapes arise from
five distinct operators spanning these classes: the leading quark coupling
$c_{HQ}^{(3)}$ and lepton coupling $c_{HL}^{(3)}$ from $\psi^2 H^2 D$,
together with three bosonic operators $c_{HW}$, $c_{H\widetilde{W}}$,
$c_{H\widetilde{B}}$ from $H^2 X^2$.
This explains why $c_{HQ}^{(3)}$ ($\psi^2 H^2 D$) is the leading SVD
direction: through the longitudinally-enhanced contact vertex it enters at
the lowest effective power $\lambda^2$, so $c_{HQ}^{(3)}$ carries the largest independent
shape at high~$p_T$.
The bosonic operators appear as sub-leading SVD directions that
resolve finer kinematic structure.

\subsubsection{Nuisance parameter distributions}
\label{subsec:vh_dists}

The pairwise joint distributions (Figure~\ref{fig:vh_theory_2d} in Appendix~\ref{app:figs}) are
largely uncorrelated and centered at zero, except $c_{HQ}^{(3)}$ which
carries the majority of the variance in the fit. The four subleading
representatives are likewise largely uncorrelated, with small and
comparable variances.
The $R^2$ distribution gives median $R^2 = 0.999$.

\subsubsection{$\Lambda^{-4}$ comparison}
\label{subsec:vh_lam4}

\begin{figure}[H]
 \centering
 \includegraphics[width=\textwidth]{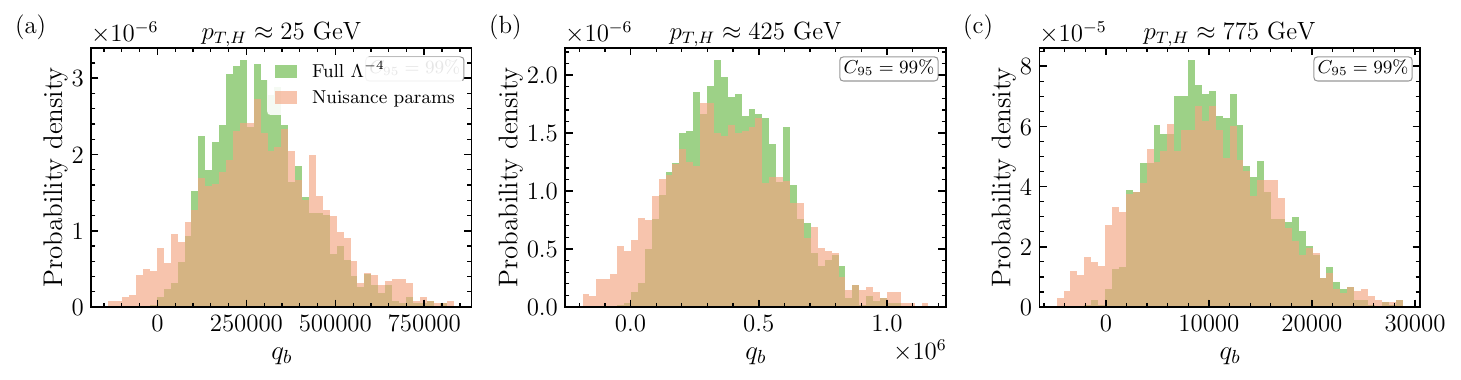}
 \caption{Per-bin comparison of the full $\Lambda^{-4}$ distribution (green) and the
 $K=5$ decorrelated nuisance distribution (orange) for $p_{T,H}$ bins in $Zh$ production.
 Per-bin $95\%$ coverage $C_{95}$ annotated per panel.}
 \label{fig:vh_theory_lam4}
\end{figure}

The $K=5$ comparison (Figure~\ref{fig:vh_theory_lam4}) gives $95\%$
coverage $\geq 0.9$ across all bins. The $R^2$ ladder
progressively improves the fit quality: $K=1$ gives median
$R^2 = 0.994$, $K=2$ adds $\Delta R^2 = +0.005$, and the three
sub-leading directions contribute a further $\Delta R^2 = +0.0006$.
These $R^2$ values measure the fit used to select representatives. The
decorrelation and $\alpha=\sqrt 2$ inflation are applied afterward to
build the nuisance distribution shown in the figure. Fifteen nuisance parameters bracket the full next-order correction to the $p_{T,H}$ spectrum in every bin at better than the 90\% level.

\subsubsection{Additional $Zh$ observables}
\label{subsec:vh_other}

We apply the same algorithm to two additional $Zh$ observables, the lepton
$p_T$ from the $Z$ decay ($B=12$ bins, $25$--$575~\text{GeV}$) and the
production angle $\cos\theta^*$ ($B=20$ bins). In both cases the $R^2$ ladder procedure selects $K=3$. The lepton $p_T$ fit selects representatives
$c_{HQ}^{(3)}$, $c_{HW}$, $c_{HD}$ with median $R^2 = 1.000$ and
$95\%$ coverage above $0.9$, while the $\cos\theta^*$ fit selects
$c_{HQ}^{(3)}$, $c_{HW}$, $c_{He}$ with median $R^2 = 0.998$ and
$95\%$ coverage above $0.9$. Both three-representative fits achieve
excellent $R^2 > 0.99$.

Physically, $p_{T,H}$ directly probes the high-energy regime where the
bosonic operators $c_{HW}$, $c_{H\widetilde W}$, $c_{H\widetilde B}$
generate distinct $E^2/\Lambda^2$ shapes, so each appears as its own
resolvable direction. The lepton $p_T$ inherits only the parent $Z$'s
boost and is therefore less energy-sensitive, while $\cos\theta^*$ is
an angular variable that does not select high energies at all. Both
observables collapse the bosonic-operator contributions into fewer
independent shapes, leaving $K=3$ rather than the $K=5$ resolved by
$p_{T,H}$. Distributions of representative Wilson coefficients and comparisons between error estimates and full $\mathcal O(\Lambda^{-4})$ distributions are presented in \cref{app:figs}.

\subsubsection{Two-dimensional $p_{T,\ell} \times \cos\theta^*$ analysis}
\label{subsec:vh_2d}

The 1D analyses of $p_{T,\ell}$ and $\cos\theta^*$ each find $K=3$
independent directions, but these marginal projections may hide
additional operator structure that is only visible in the joint
two-dimensional distribution. To test this, we apply the algorithm to the
$p_{T,\ell}\times\cos\theta^*$ double-differential distribution.

The 2D grid has $24 \times 22 = 528$ bins in
$(p_{T,\ell},\,\cos\theta^*)$. Because the double-differential distribution
has a large dynamic range, a tighter MC statistics cut of $5\%$ (rather
than the $1\%$ used for 1D observables) is applied to remove sparsely
populated tail bins. $B = 292$ bins survive, and the $R^2$ ladder accepts $K = 4$ independent representatives,
$c_{HQ}^{(3)}$, $c_{HW}$, $c_{H\widetilde{W}}$, and $c_{He}$
(median $R^2 = 0.992$, with $>98\%$ of throws above $R^2 = 0.95$).

The 2D observable resolves more independent operator behaviours than
either 1D marginal alone ($K=4$ in 2D vs $K=3$ for the $p_{T,\ell}$ or
$\cos\theta^*$ marginal). The CP-odd direction $c_{H\widetilde{W}}$ in
particular is resolved by neither of these 1D marginals, only by their joint distribution: its
distinguishing shape is a joint modulation in $(p_{T,\ell},\cos\theta^*)$
that averages away when the distribution is projected onto either
variable alone, so neither 1D-marginal ladder selects it.

\begin{figure}[H]
 \centering
 \includegraphics[width=\textwidth]{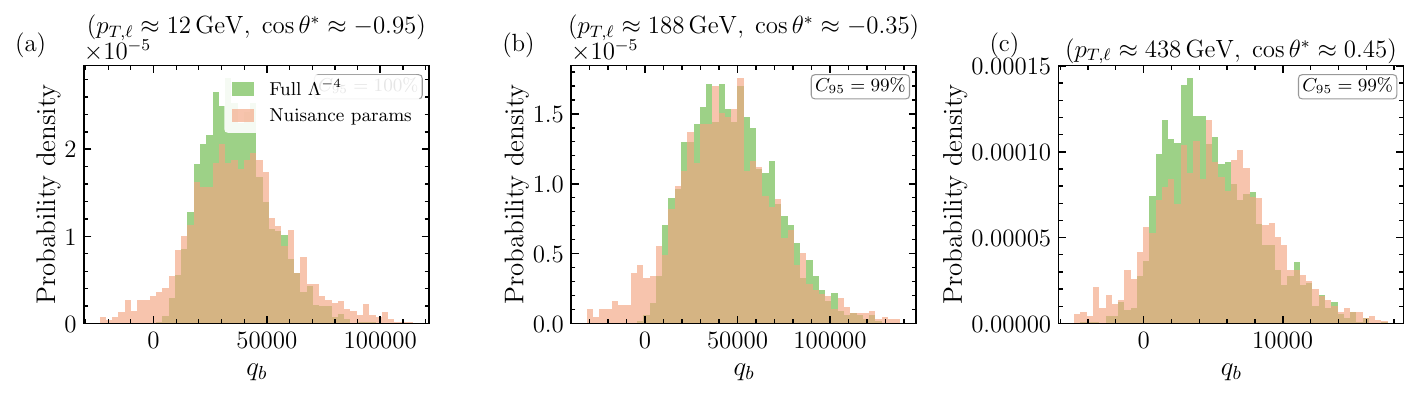}
 \caption{Per-bin comparison of the full $\Lambda^{-4}$ distribution (green)
 and the $K=4$ decorrelated nuisance (orange) for three representative
 $(p_{T,\ell},\,\cos\theta^*)$ bins in the 2D $Vh$ analysis.
 Per-bin $95\%$ coverage $C_{95}$ annotated per panel. }
 \label{fig:vh_2d_lam4}
\end{figure}

Figure~\ref{fig:vh_2d_lam4} shows the per-bin comparison for three
representative $(p_{T,\ell},\,\cos\theta^*)$ bins spanning the kinematic
plane. The decorrelated nuisance with $\alpha=\sqrt{2}$ achieves median
$C_{95}=0.994$ across all 292 bins, with minimum $C_{95}=0.921$ at
$(p_{T,\ell},\cos\theta^*)\approx(412~\text{GeV},-0.15)$. The
two-dimensional binning demands more Monte Carlo per bin than the
one-dimensional analyses, so the worst-bin floor will tighten further
with additional simulation.

\subsection{Example III: Vector boson fusion}
\label{sec:vbf}

\begin{figure}[H]
 \centering
 \includegraphics[width=0.85\textwidth]{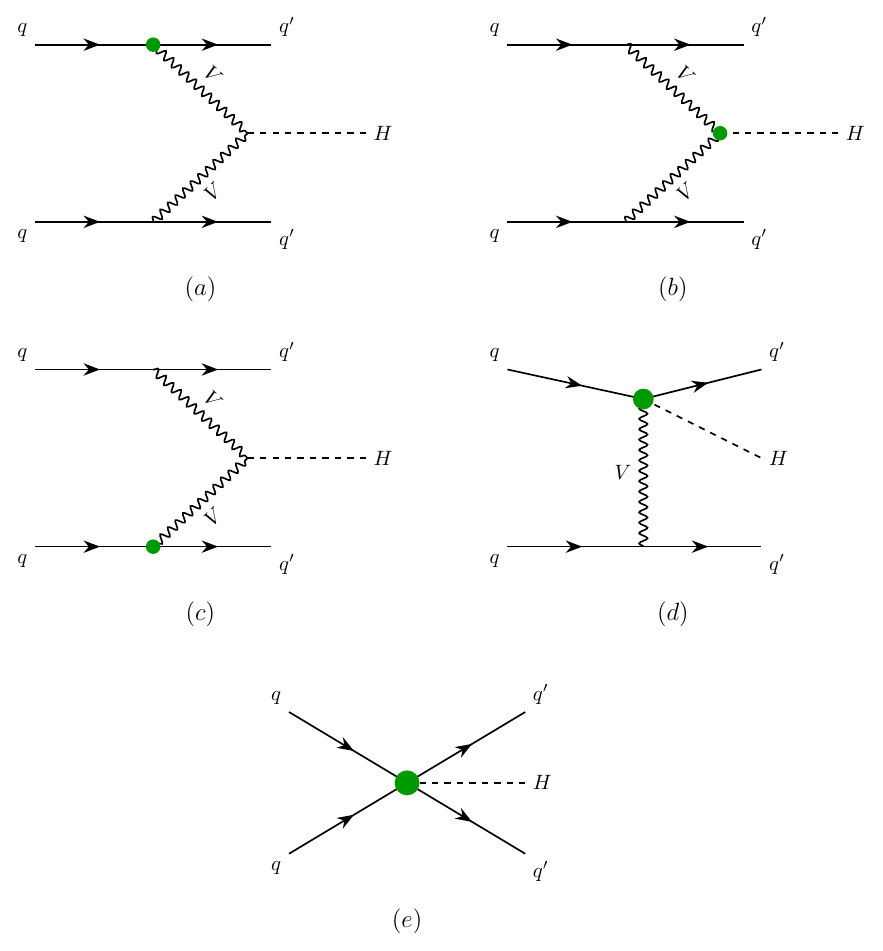}
 \caption{Feynman diagram topologies contributing to VBF at
 $\mathcal{O}(\Lambda^{-4})$. Green blobs indicate SMEFT vertex
 modifications: (a)~upper $qqV$, (b)~$VVH$, (c)~lower $qqV$,
 (d)~contact $qqVh$ from dim-6 operators,
 (e)~five-point $qqqqH$ contact from dim-8 operators.}
 \label{fig:vbf_diagrams}
\end{figure}

As a third application we consider Higgs production via vector boson
fusion, $qq'\to qq'H$, at $\mathcal{O}(\Lambda^{-4})$. The relevant
Feynman topologies (Figure~\ref{fig:vbf_diagrams}) are $t$-channel
electroweak exchange with SMEFT modifications to the $qqV$ and $VVH$
vertices, a $qqVh$ contact vertex generated by the dim-6 operators, and
a five-point $qq\bar q\bar qH$ contact vertex that first appears at
dim-8.

The dim-6$^2$ matrix involves $N=12$ operators with nonzero contribution
at tree level (the $Vh$ set of Table~\ref{tab:vh_ops} with the lepton
couplings $c_{HL}^{(1,3)}, c_{He}$ removed, since no leptons appear in
this final state, and the charged-current coupling $c_{Hud}$ added,
entering through $W$ fusion),
and the full $\Lambda^{-4}$ comparison uses $N_8=38$ dim-8 operators
(Table~\ref{tab:vbf_ops}). Among the dim-8 operators, ten belong to
the $\psi^4 H^2$ class generating the five-point $\bar qq\bar q'q'H$
topology of Figure~\ref{fig:vbf_diagrams}e, a topology with no dim-6
analogue.
We analyse four observables: the dijet invariant mass $m_{jj}$ ($B=10$ bins,
$580$--$1080~\text{GeV}$), the Higgs $p_T$ ($B=8$ bins), the dijet rapidity
separation $\Delta\eta_{jj}$ ($B=12$ bins after a 1\% MC statistics cut), and the dijet azimuthal separation
$\Delta\phi_{jj}$ ($B=5$ bins). We present the $m_{jj}$ analysis in detail
and summarise the remaining three in Section~\ref{subsec:vbf_other}.

\begin{table}[t]
 \centering
 \small
 \renewcommand{\arraystretch}{1.2}
 \begin{tabular}{lll}
 \hline
 \textbf{Class} & \textbf{Operators} & \textbf{Vertex} \\
 \hline
 \multicolumn{3}{l}{\textbf{Dim-6 operators ($N=12$)}} \\
 \hline
 $\psi^2 H^2 D$ &
 $c_{HQ}^{(1)},\ c_{HQ}^{(3)}$ &
 $\bar{q}qV$, contact \\
 &
 $c_{Hu},\ c_{Hd},\ c_{Hud}$ &
 $\bar{q}qV$, contact \\
 \hline
 $H^4 D^2$ &
 $c_{HD}$ &
 $VVH$ \\
 \hline
 $H^2 X^2,\ H^2 X D$ &
 $c_{HW},\ c_{H\tilde W},\ c_{HB},\ c_{H\tilde B},\ c_{HWB},\ c_{H\widetilde{WB}}$ &
 $VVH$ \\
 \hline
 \multicolumn{3}{l}{\textbf{Dim-8 operators ($N_8=38$)}} \\
 \hline
 $\psi^4 H^2$ (five-point) &
 $c^{(8)}_{Q^2 H^4,(1,2,3)},\ c^{(8)}_{u^2 H^4},\ c^{(8)}_{d^2 H^4}$ &
 $\bar q q\bar q'q'H$ \\
 &
 $c^{(8)}_{Q^2 H^2 u^2,(1,2)},\ c^{(8)}_{Q^2 H^2 d^2,(1,2)},\ c^{(8)}_{u^2 H^2 d^2}$ &
 $\bar q q\bar q'q'H$ \\
 \hline
 $\psi^2 H^4 D$ &
 $c^{(8)}_{HQ^{(1,2,3)}},\ c^{(8)}_{Hu},\ c^{(8)}_{Hd}$ &
 $\bar{q}qV$ \\
 \hline
 $H^4 D^4,\ H^2 X^2 D^2$ &
 $c^{(8)}_{HD},\ c^{(8)}_{HD^2}$ &
 $VVH$ \\
 &
 $c^{(8)}_{HB},\ c^{(8)}_{HW},\ c^{(8)}_{HW^2},\ c^{(8)}_{HWB}$ &
 $VVH$ \\
 &
 $c^{(8)}_{HD\cdot HB},\ c^{(8)}_{HD\cdot HW},\ c^{(8)}_{HD\cdot HW^2}$ &
 $VVH$ \\
 \hline
 $\psi^2 H^2 X D^2,\ \psi^2 H^2 D^3$ &
 $c^{(8)}_{Q^2 B H^2 D^{(1,3)}},\ c^{(8)}_{Q^2 W H^2 D^{(1,3,5)}}$ &
 contact $\bar{q}qVh$ \\
 &
 $c^{(8)}_{Q^2 H^2 D^{3,(1,3,4)}}$ &
 contact $\bar{q}qVh$ \\
 &
 $c^{(8)}_{u^2 B H^2 D^{(1)}},\ c^{(8)}_{u^2 W H^2 D^{(1)}},\ c^{(8)}_{u^2 H^2 D^{3,(1)}}$ &
 contact $\bar{q}qVh$ \\
 &
 $c^{(8)}_{d^2 B H^2 D^{(1)}},\ c^{(8)}_{d^2 W H^2 D^{(1)}},\ c^{(8)}_{d^2 H^2 D^{3,(1)}}$ &
 contact $\bar{q}qVh$ \\
 \hline
 \end{tabular}
 \caption{Operators contributing to VBF at $\mathcal{O}(\Lambda^{-4})$,
 grouped by operator class and the vertex they modify.
 VBF carries the tree-level dim-6 set of $Vh$ without the lepton
 couplings and with $c_{Hud}$ added through $W$ fusion.
 At dim-8 it additionally receives the $\psi^4 H^2$ class that
 generates the five-point $\bar qq\bar q'q'H$ contact topology
 (Figure~\ref{fig:vbf_diagrams}e), a topology with no dim-6
 analogue. Here, the dimension-8 operators follow the basis of
 Refs.~\cite{Li:2020gnx, Murphy:2020rsh} and are included in the
 full $\Lambda^{-4}$ validation.}
 \label{tab:vbf_ops}
\end{table}

\subsubsection{Rank and representatives}
\label{subsec:vbf_rank}

In this example, the $R^2$ ladder approach finds optimal fitting with $K=2$. The leading direction carries $97.7\%$ of the variance and is
represented by $c_{HQ}^{(3)}$, with the flat directions
$\{c_{HQ}^{(1)},\,c_{Hd},\,c_{Hu},\,c_{Hud}\}$.
The second direction ($1.3\%$) is represented by $c_{HB}$ and absorbs
$\{c_{HD},\,c_{HW},\,c_{H\widetilde{B}},\,c_{H\widetilde{W}},\,c_{HWB},\,c_{H\widetilde{WB}}\}$.

This hierarchy again follows from the $\lambda$-counting of
Ref.~\cite{Assi:2025zmp,Assi:2024zap}. To see this, recall that the $\bar{q}qV$ and $hVV$ vertex expansions are given in
eqs.~\eqref{eq:vh_qqv}--\eqref{eq:vh_hvv}.
The four-point $\bar{q}qVh$ contact vertex scales as in
eq.~\eqref{eq:vh_qqvh}, while the five-point contact vertex
(no SM or dim-6 counterpart) is
\begin{align}
\label{eq:vbf_5pt}
g_{\bar{q}q\bar{q}'q'H} &= \frac{\lambda^7}{\hat\Lambda^4}\, c^{(8)}_{\psi^4 H^2}
+ \cdots\,.
\end{align}

Multiplying two $g_{\bar{q}qV}$ expansions, the $g_{hVV}$ expansion,
and two propagator factors $E^{-2}\sim\lambda^4$ each, the $t$-channel
topology gives (to $\mathcal{O}(\lambda^7)$)
\begin{align}
\label{eq:vbf_amp1}
\mathcal{A}^{(1)}_{\rm VBF} &= \lambda^3\,
 (g^{\rm SM}_{\bar{q}qV})^2\, g^{\rm SM}_{hVV}
\nonumber\\
&\quad + \frac{\lambda^5}{\hat\Lambda^2}\,
 (g^{\rm SM}_{\bar{q}qV})^2\, c_{H^2 X^2}
\nonumber\\
&\quad + \frac{\lambda^7}{\hat\Lambda^2}\,
 g^{\rm SM}_{\bar{q}qV}\, g^{\rm SM}_{hVV}\, c_{\psi^2 H^2 D}
+ \cdots\,.
\end{align}
The second topology (one $t$-channel exchange + one contact vertex)
gives
\begin{align}
\label{eq:vbf_amp2}
\mathcal{A}^{(2)}_{\rm VBF} &=
 \frac{\lambda^5}{\hat\Lambda^2}\, g^{\rm SM}_{\bar{q}qV}\, c_{\psi^2 H^2 D}
+ \cdots\,,
\end{align}
and the five-point contact topology gives
$\mathcal{A}^{(3)}_{\rm VBF} = g_{\bar{q}q\bar{q}'q'H}$.
The SM VBF amplitude enters at $\lambda^3$. The leading SMEFT contributions are the contact $\mathcal{A}^{(2)}$ and the $H^2X^2$ $t$-channel modification $\mathcal{A}^{(1)}$, both first entering at $\lambda^5$. The five-point contact topology $\mathcal{A}^{(3)}$ (a pure dim-8 effect from the $\psi^4 H^2$ class), enters at $\lambda^7$.

\begin{figure}[t!]
 \centering
 \includegraphics[width=0.62\textwidth]{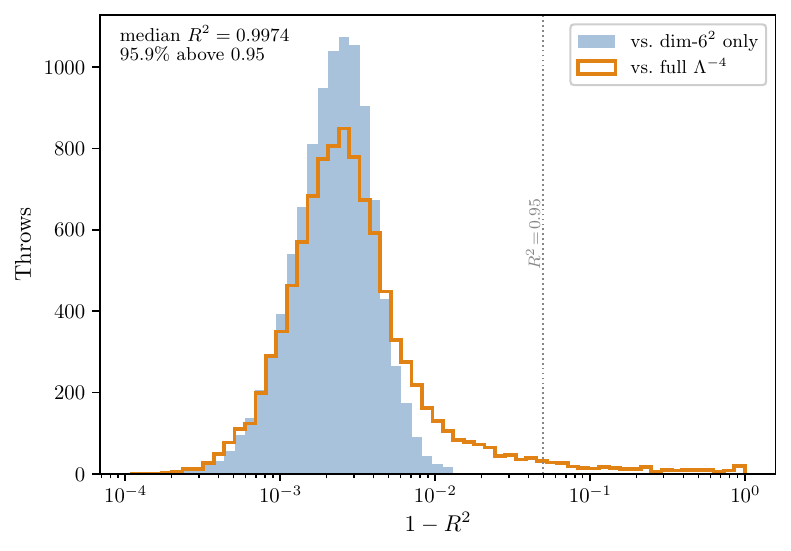}
 \caption{Distribution of $1-R^2$ over $T=10{,}000$ throws for
 VBF $m_{jj}$, on a logarithmic axis, when the $K=2$ representatives
 are fit to the full $\Lambda^{-4}$ truth (orange outline) and, for
 reference, to the dim-6$^2$ piece alone (blue filled). The
 dim-6$^2$ fit is essentially exact, while against the full truth a
 genuine residual tail appears ($95.9\%$ of throws above $0.95$):
 the five-point dim-8 contact produces shapes outside the dim-6$^2$
 span, which the representatives cannot absorb and the
 $\alpha=\sqrt 2$ inflation must cover. The tail extends all the
 way to $R^2 \approx 0$ ($1.6\%$ of throws fall below $R^2=0.05$):
 these are draws in which the sign-indefinite dim-8 interference
 drives the total correction negative, which the non-negative
 quadratic form cannot reproduce.}
 \label{fig:r2_vbf}
\end{figure}

Squaring the amplitude, at the dim-6$^2$ level both the $c_{HQ}^{(3)}$ and $c_{HB}$ self-interferences plus their $c_{HQ}^{(3)}c_{HB}$ cross all sit at $\lambda^{10}$: the two representative directions are parametrically the same order, and the large numerical split in their variance shares is not a $\lambda$ hierarchy. The dimension 8 contact term interferes with the SM and also generates a $\lambda^{10}$ piece. It introduces genuinely new kinematic structure not present in the dim-6$^2$ quadratic form. This structure is directly visible as the residual tail in the $R^2$ comparison of Figure~\ref{fig:r2_vbf}.

The dim-8 piece cannot appear as a representative by construction: the algorithm operates on the dim-6$^2$ kernel $A_b$ alone and has no dim-8 input. That the $\alpha=\sqrt 2$-inflated nuisance nonetheless brackets the dim-8 contribution is precisely a direct validation of the method: it covers a piece it had no a priori knowledge of, and the coverage check confirms it. This is an empirical result, not an a priori shape claim. The contrast with $Vh$, where $p_{T,H}$ extends into low-energy bins that resolve sub-leading directions as new representatives, is that the VBF event selection cuts the soft region, so the leading dim-6$^2$ shape dominates more uniformly across $m_{jj}$.

\subsubsection{Nuisance parameter distributions}
\label{subsec:vbf_dists}

\begin{figure}[H]
 \centering
 \includegraphics[width=0.45\textwidth]{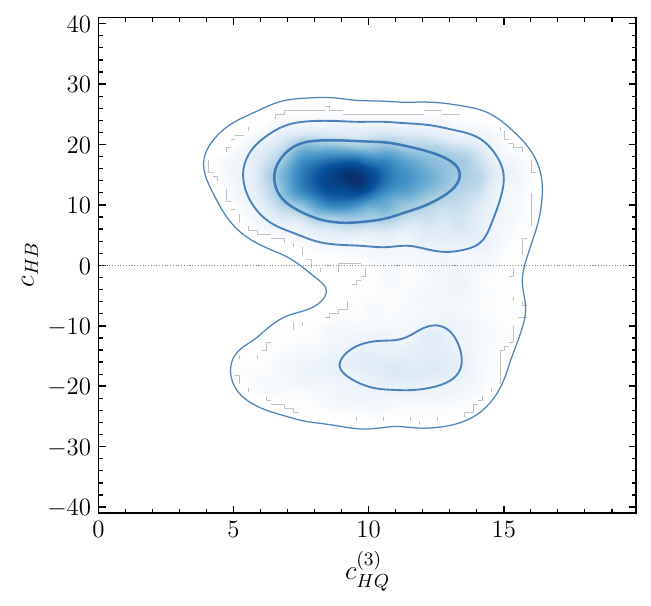}
 \caption{Joint distribution of the two VBF ($m_{jj}$) representatives
 from $T=10{,}000$ throws.}
 \label{fig:vbf_2d}
\end{figure}

The two-representative fit reaches median $R^2 = 0.998$ across the 10 $m_{jj}$ bins, and the pairwise distribution (Figure~\ref{fig:vbf_2d}) shows that the two
representatives $c_{HQ}^{(3)}$ and $c_{HB}$ are essentially uncorrelated
(sample correlation $\approx -0.1$), each spanning an independent
kinematic shape of the $m_{jj}$ spectrum.

\subsubsection{$\Lambda^{-4}$ comparison}
\label{subsec:vbf_lam4}

\begin{figure}[H]
 \centering
 \includegraphics[width=\textwidth]{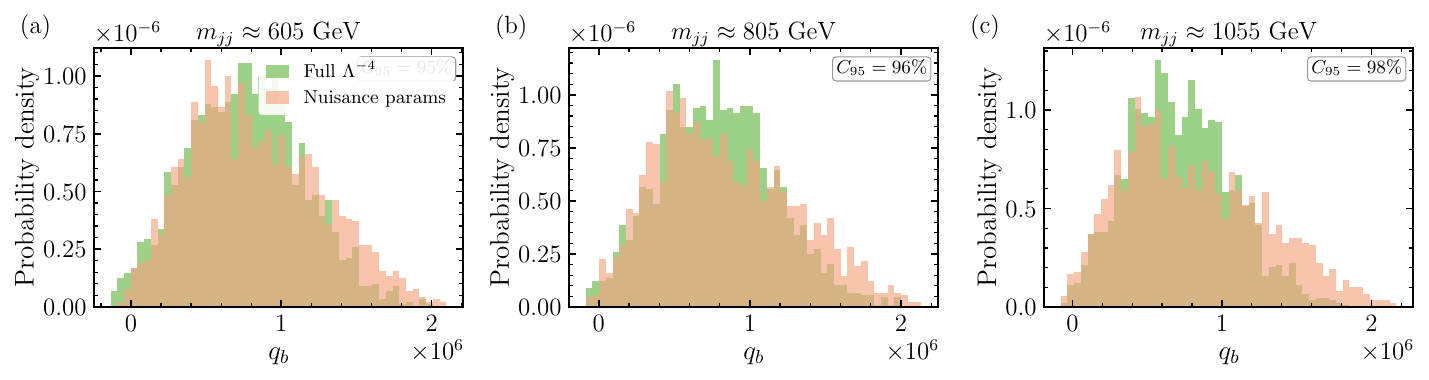}
 \caption{Per-bin comparison of the full $\Lambda^{-4}$ distribution (green) and the
 $K=2$ decorrelated nuisance distribution (orange) for $m_{jj}$ bins in VBF.
 Per-bin $95\%$ coverage $C_{95}$ annotated per panel.}
 \label{fig:vbf_lam4}
\end{figure}

For the $K=2$ comparison (Figure~\ref{fig:vbf_lam4}), the decorrelated monomial prescription with $\alpha=\sqrt{2}$ inflation is sufficient to cover the truth distribution bin by bin throughout the analysis range. The two representatives $c_{HQ}^{(3)}$ and $c_{HB}$ between them span both leading SMEFT directions ($t$-channel $\psi^2H^2D$ and $H^2X^2$ modifications), and the decorrelation step's $\sqrt 2$ inflation absorbs the dim-8 five-point $\mathcal{A}^{(3)}$ contribution without requiring it as a separate representative. Three nuisance parameters suffice for the $m_{jj}$ spectrum even though ten of the dim-8 operators feed a topology the dim-6 simulation never produces.

\subsubsection{Additional VBF observables}
\label{subsec:vbf_other}

We apply the same algorithm to three additional VBF observables: the
Higgs transverse momentum ($B=8$ bins, $200$--$400~\text{GeV}$), the
dijet rapidity separation $\Delta\eta_{jj}$ ($B=12$ bins after a $1\%$
MC statistics cut), and the dijet azimuthal separation $\Delta\phi_{jj}$
($B=5$ bins). The Higgs $p_T$ analysis resolves $K=3$ representatives,
$c_{HQ}^{(3)}$ ($98.6\%$), $c_{H\widetilde{W}}$ ($0.7\%$) and
$c_{H\widetilde{B}}$ ($0.3\%$), with median $R^2=0.999$ and $95\%$
coverage $\geq 0.97$. Notably the two sub-leading directions are the
CP-odd bosonic operators $c_{H\widetilde{W}}, c_{H\widetilde{B}}$,
replacing the CP-even $c_{HB}$ found for $m_{jj}$.
The $\Delta\eta_{jj}$ analysis admits $K=4$ with
representatives $c_{HQ}^{(3)},\,c_{HW},\,c_{H\widetilde{B}},\,c_{H\widetilde{WB}}$,
median $R^2>0.999$ and $95\%$ coverage $\geq 0.9$.
Finally, $\Delta\phi_{jj}$ requires only $K=2$ representatives
($c_{HQ}^{(3)},\,c_{HW}$) because of its limited bin count, with
median $R^2=0.998$ and $95\%$ coverage $\geq 0.98$. Distributions of representative Wilson coefficients and error estimate comparisons to true distributions are presented in \cref{app:figs}. The variation of $K$ across VBF observables reflects how many sub-leading monomial directions produce resolvably distinct bin patterns in the given binning: $\Delta\phi_{jj}$ ($K=2$) is bin-count limited, for $m_{jj}$ the directions beyond the second no longer improve the median $R^2$, $p_{T,H}$ resolves one additional direction ($K=3$), and $\Delta\eta_{jj}$ resolves the most ($K=4$), consistent with the rapidity gap probing kinematic structure that energy-only variables average over.

\section{Summary of results}
\label{sec:summary}

Table~\ref{tab:summary} collects the key results for all processes and
observables studied. For each observable, the table lists the number of
independent kinematic directions $K$, the representative Wilson
coefficients, the fit quality ($R^2$ median), and the minimum
$C_{95}$ across bins of the decorrelated nuisance against the
full $\mathcal{O}(\Lambda^{-4})$ distribution. All results use the
decorrelated monomial prescription
(Section~\ref{subsec:decorr}) with $\alpha = \sqrt{2}$.

\begin{table*}[t]
 \centering
 \small
 \renewcommand{\arraystretch}{1.3}
 \begin{tabular}{llclcc}
 \hline
 \textbf{Process} & \textbf{Observable} & $\boldsymbol{K}$
 & \textbf{Representatives}
 & ${\rm \textbf{med}}~\boldsymbol{R^2}$
 & {\boldmath${{\rm \textbf{min}}~C_{95}}$} \\
 \hline
 DY & $m_{\ell\ell}$ (F+B) & 2 & $c_{LQ}^{(3)},\, c_{HL}^{(3)}$ & 0.999 & 0.993 \\
 DY & $m_{\ell\ell}$ (Fwd) & 2 & $c_{LQ}^{(3)},\, c_{HL}^{(3)}$ & 1.000 & 0.993 \\
 DY & $m_{\ell\ell}$ (Bwd) & 2 & $c_{LQ}^{(3)},\, c_{HL}^{(3)}$ & 1.000 & 0.989 \\
 \hline
 $Vh$ & $p_{T,H}$ & 5 & $c_{HQ}^{(3)},\, c_{HW},\, c_{H\widetilde{W}},\, c_{H\widetilde{B}},\, c_{HL}^{(3)}$ & 0.999 & 0.948 \\
 $Vh$ & $p_{T,\ell}$ & 3 & $c_{HQ}^{(3)},\, c_{HW},\, c_{HD}$ & 1.000 & 0.974 \\
 $Vh$ & $\cos\theta^*$ & 3 & $c_{HQ}^{(3)},\, c_{HW},\, c_{He}$ & 0.998 & 0.989 \\
 $Vh$ & $p_{T,\ell}{\times}\cos\theta^*$ & 4 & $c_{HQ}^{(3)},\, c_{HW},\, c_{H\widetilde{W}},\, c_{He}$ & 0.992 & 0.921 \\
 \hline
 VBF & $m_{jj}$ & 2 & $c_{HQ}^{(3)},\, c_{HB}$ & 0.998 & 0.950 \\
 VBF & $p_{T,H}$ & 3 & $c_{HQ}^{(3)},\, c_{H\widetilde{W}},\, c_{H\widetilde{B}}$ & 0.999 & 0.974 \\
 VBF & $\Delta\eta_{jj}$ & 4 & $c_{HQ}^{(3)},\, c_{HW},\, c_{H\widetilde{B}},\, c_{H\widetilde{WB}}$ & 1.000 & 0.922 \\
 VBF & $\Delta\phi_{jj}$ & 2 & $c_{HQ}^{(3)},\, c_{HW}$ & 0.998 & 0.986 \\
 \hline
 \end{tabular}
 \caption{Summary of nuisance parameter results for all processes and
 observables. $K$ is the required number of representative operators, ${\rm med}~R^2$ the median coefficient
 of determination from the ensemble fit, and min $C_{95}$ is the
 minimum per-bin fraction of full-prior throws falling within the
 $95\%$ credible interval of the decorrelated nuisance
 ($\alpha = \sqrt{2}$). The 2D
 $p_{T,\ell}{\times}\cos\theta^*$ analysis uses a $5\%$ MC
 statistics cut (see text). All other analyses use $1\%$.}
 \label{tab:summary}
\end{table*}

In both $Vh$ and VBF, where the energy-enhanced $\bar qqVh$ contact
vertex is available, $c_{HQ}^{(3)}$ emerges as the leading representative, consistent with the
energy-enhanced $\lambda$-counting of
eqs.~\eqref{eq:vh_qqv}--\eqref{eq:vh_qqvh}. In high-$p_T$ Drell-Yan the
four-fermion contact topology dominates, and just
two representatives $(c_{LQ}^{(3)}, c_{HL}^{(3)})$ suffice ($K=2$), since $c_{HL}^{(3)}$ stands in for high-dimensional flat directions of $s$-channel operators, including $c_{HQ}^{(j)}$ and the right-handed quark and lepton currents, that all produce the same $m_{\ell\ell}$ shape at tree level in this final state. In $Vh$ and VBF the
bosonic operators ($c_{HW}$, $c_{HB}$ and their CP-odd counterparts)
appear as sub-leading directions that resolve additional kinematic
structure at high $p_T$. The number of independent shapes ranges from
$K=2$ to $K=5$, representing a reduction from $N(N{+}1)/2 = 78$ to $105$
monomials to a handful of nuisance parameters.\footnote{Operators that are already tightly constrained by existing measurements (most notably $c_{HQ}^{(3)}$ from LEP $Z\to q\bar q$) still appear as representatives here because the role of the representative list is to span the dim-6$^2$ {shape space}, not to fit Wilson-coefficient values. The NDA prior remains the natural choice for that role independently of whatever bounds individual coefficients carry from other measurements. See also the apparent-circularity discussion in Sec.~\ref{subsec:qb}: the representative list spans the dim-6$^2$ shape space used to build the calibration scan, while the signal Wilson coefficients $\mathbf{c}$ may carry tight external constraints unrelated to that role.}

\subsection{Bin-to-bin correlation of the truncated correction}
\label{sec:bincorr}

The central premise of the paper is a rank statement: the quadratic
correction across $B$ bins lives in a $D_M$-dimensional subspace, with
$D_M \ll B$. The per-bin comparisons of
Sections~\ref{sec:dy}--\ref{sec:vbf} show that the $K$ representative
nuisance reproduces the truth distribution {marginally in each
bin}, but they do not directly probe how the bins move together. Two
bins could each have perfect 1D agreement and still be either
independent (rank $2$) or perfectly redundant (rank $1$). The
marginals cannot tell them apart.

We can test this possibility of incorrect bin correlations directly by computing the correlation
matrix
\begin{equation}
\label{eq:bincorr}
 C_{ab} = \frac{\mathrm{Cov}(q_a, q_b)}{\sqrt{\mathrm{Var}(q_a)\,\mathrm{Var}(q_b)}}
\end{equation}
across throws of the truth ensemble at dim-6$^2$ using all $N$ operators, alongside the same matrix
computed with the bootstrap of fitted tuples (so that
$q_b = \mathbf{c}_{\mathcal{R}}^\top \tilde A_b\,\mathbf{c}_{\mathcal{R}}$
with $\mathbf{c}_{\mathcal{R}}$ drawn from the joint fitted
distribution of Section~\ref{subsec:fitting}). If the
$K$-representative description is faithful, the two matrices and
their eigenvalue spectra should agree.

\begin{figure}[H]
 \centering
 \includegraphics[width=0.7\textwidth]{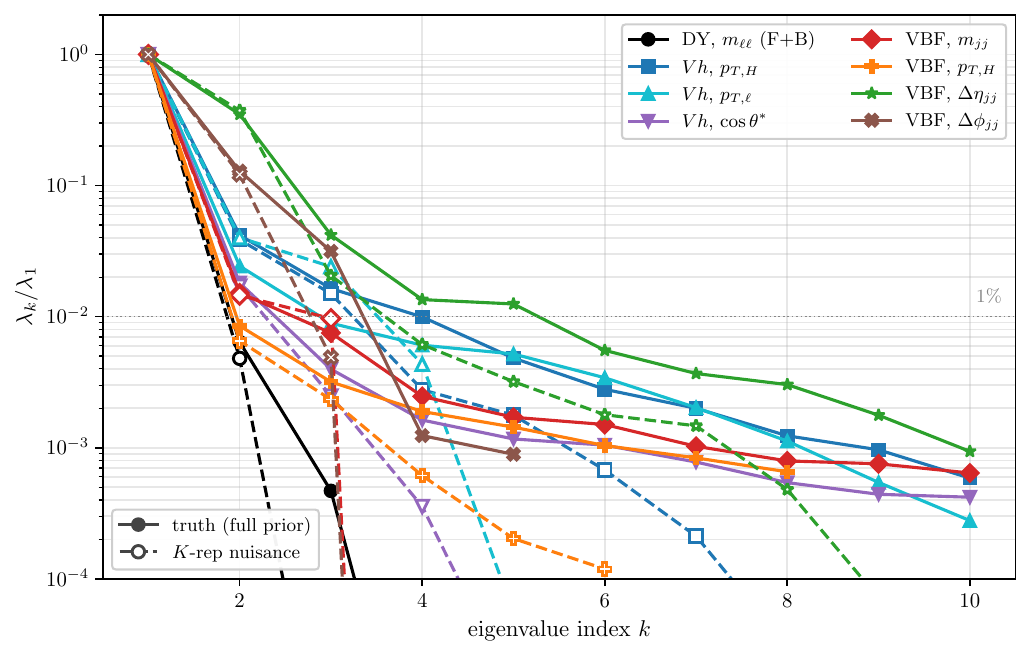}
 \caption{Eigenvalue spectra of the bin-correlation matrix across
 all observables studied in this paper. Filled markers are the
 true dim-6$^2$ distribution using all $N$ operators. Open markers (joined by the same colour) are the
 $K$-representative bootstrap distribution. The spectra fall steeply within the
 first few eigenvalues in every case. The nuisance reproduces the
 leading, variance-dominating eigenvalues and falls off faster than
 the truth in the sub-leading tail, where the eigenvalues are orders
 of magnitude smaller and contribute negligibly to the bin
 covariance. The representative fit therefore reproduces the marginal
 per-bin distributions together with the dominant joint variation
 between bins.}
 \label{fig:bincorr_all}
\end{figure}

Figure~\ref{fig:bincorr_all} compares the eigenvalue spectra across all eight
process--observable combinations of this paper. Across processes the
truth spectra fall by an order of magnitude or more between the
leading and second eigenvalue, and continue to fall steeply
thereafter. The nuisance spectra (open markers) reproduce the leading,
variance-dominating eigenvalues. In the sub-leading tail they fall off
faster than the truth, since the $K$-representative basis spans only the
directions that carry appreciable variance. This is expected and
harmless: the eigenvalues where the two diverge lie orders of magnitude
below the leading ones and contribute negligibly to the bin covariance.
The representative fit therefore captures the marginal per-bin
distributions of Sections~\ref{sec:dy} to \ref{sec:vbf} together with
the dominant {joint} bin-to-bin variation, which is the empirical
content of the rank $D_M$ of $M$.

The bin-correlation diagnostic above probes the joint structure of
the truncated correction in abstract bin-index space. A complementary
view shows the same coverage statement directly in the physical
kinematic variable. Specifically, for each bin we plot the central $95\%$ quantile
band of $q_b$ under the full $\Lambda^{-4}$ truth ensemble and under
the $K$-representative nuisance with $\alpha=\sqrt 2$. This is the band that an analysis using the prescription would assign to the truncated correction as a function of the kinematic variable.

\begin{figure}[H]
 \centering
 \includegraphics[width=0.49\textwidth]{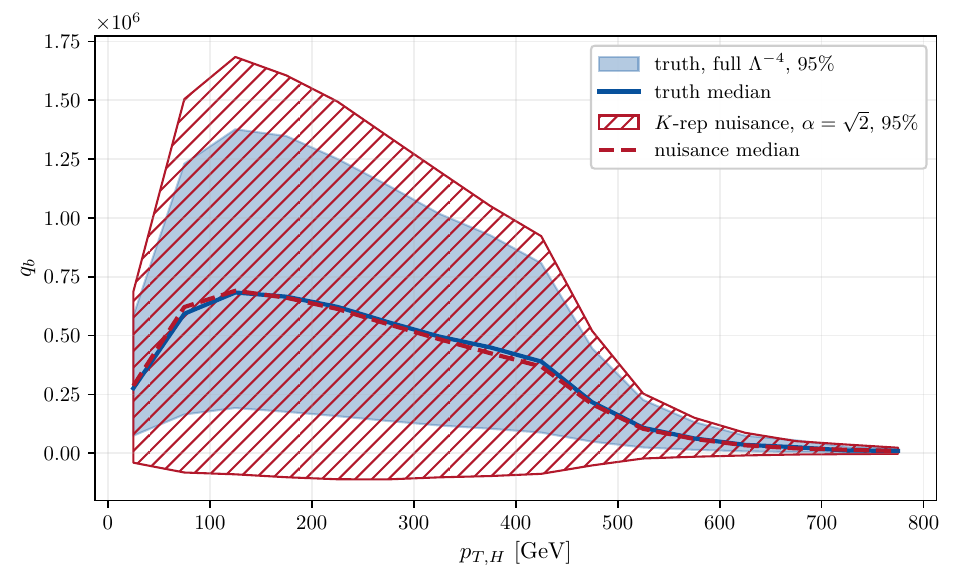}\hfill
 \includegraphics[width=0.49\textwidth]{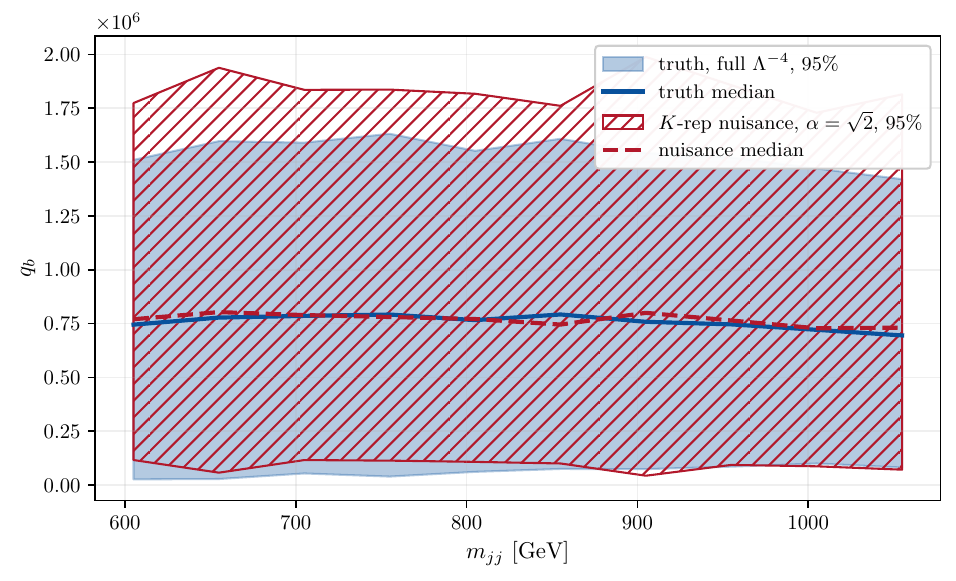}
 \caption{Per-bin 95\% band of the truncated correction $q_b$ as a
 function of the kinematic variable. Filled blue band: full
 $\Lambda^{-4}$ truth distribution, $95\%$ quantile interval. Red hatched band:
 $K$-representative nuisance with the decorrelated prescription
 at $\alpha=\sqrt 2$, $95\%$ quantile interval. Solid and dashed
 lines mark the respective medians. Left: $Vh$ $p_{T,H}$ with
 $K=5$. Right: VBF $m_{jj}$ with $K=2$. The $\alpha=\sqrt 2$
 band covers the truth across the full kinematic range in both
 cases. Note that the nuisance median, built from the dim-6$^2$ kernel alone, coincides with the full $\Lambda^{-4}$ truth median in both panels, showing that the dim-6$^2$ piece is the sole source of positive bias in the full result at $\Lambda^{-4}$. The $\alpha=\sqrt 2$ band gives the roughly symmetric band around this median.}
 \label{fig:envelope}
\end{figure}

Figure~\ref{fig:envelope} confirms in the kinematic-variable view
what the per-bin coverage values reported in
Sections~\ref{sec:vh} and~\ref{sec:vbf} state numerically: the
$\alpha=\sqrt 2$ band covers the truth distribution
bin-by-bin throughout each analysis range. Together with the
rank-$K$ joint structure of Figure~\ref{fig:bincorr_all}, this fixes both the marginal width and
the joint variation of the nuisance in physical terms.

\subsection{Scale parameter dependence}
\label{subsec:alpha}

\begin{figure}[t!]
 \centering
 \includegraphics[width=0.7\textwidth]{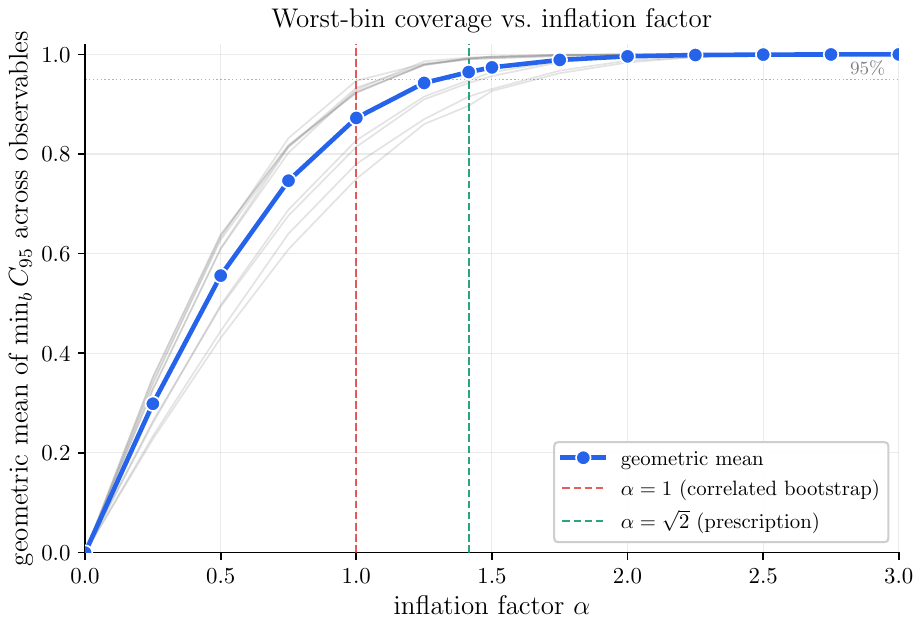}
 \caption{Geometric mean across the ten 1D observables of each
 observable's worst-bin $95\%$ coverage $\min_b C_{95}$, as a
 function of the inflation factor $\alpha$. Faint grey lines
 show the individual observables. The correlated bootstrap
 ($\alpha=1$, red) under-covers the most demanding bins. The
 $\alpha=\sqrt 2$ prescription (green) reaches the knee of the curve
 at a geometric-mean worst-bin coverage of $\approx 0.96$. The truth
 distribution is held fixed across the scan. Only the nuisance side
 is rebuilt at each $\alpha$.}
 \label{fig:cov_vs_alpha}
\end{figure}

The NDA argument of Section~\ref{subsec:decorr} motivates using the scaling factor after decorrelation of
$\alpha=\sqrt 2$. Here, we explore how coverage responds to other possible values for this scaling. Figure~\ref{fig:cov_vs_alpha} scans $\alpha$ from $0$ to $3$ and
plots, for each value, the geometric mean across all ten 1D observables
of each observable's worst-bin $C_{95}$. At $\alpha=1$, the decorrelated bootstrap with no inflation, the geometric mean is $0.87$,
already biased toward under-coverage in the most demanding bins. The
$\alpha=\sqrt 2$ prescription raises it to $\approx 0.96$, after which
the curve flattens and further inflation produces overly conservative errors, undervaluing the sensitivity of the search. The choice
$\alpha=\sqrt 2$ thus sits at the knee of the curve, conservative
without being wasteful. Naturally, choosing to neglect the uncertainty with $\alpha=0$ never correctly covers the actual next-order impact.

\section{Conclusions}
\label{sec:conclusions}

Every EFT analysis needs a principled way of handling the correction that sits at the boundary
between the truncation order it keeps and the next unknown order.
In the SMEFT with Lagrangian defined to $\mathcal{O}(\Lambda^{-2})$ this is the quadratic dim-6$^2$ term $\sum_{ij} A_{b,ij}\,c_i c_j$, where each Wilson coefficient enters from a separate amplitude in the cross section calculation.

The dim-6$^2$ term is fully calculable from the same simulation that produces the signal and is the sole portion of the next-order term which has definite sign, so we recommend including it explicitly in the signal model rather than folding it into the error budget. The calibrated PCA-score nuisances enter each bin through shape weights $W_{b,i}$ from the same kernel $A_b$, so they grow in lockstep with the dim-6$^2$ piece in the signal: when the linear signal at the fitted $\mathbf{c}$ falls below the calibrated nuisance, the profile likelihood absorbs the bin at no cost. A search whose entire kinematic range violates EFT validity therefore returns no constraint, and one spanning valid and invalid regions is dominated by the valid bins. Because the positive-definite dim-6$^2$ term has been explicitly included in the signal, the attenuation is two-sided: each nuisance score is mean-zero with pulls of both signs available, so the argument does not rely on a one-sided ``conservative'' error.

The challenge in using a separately-coefficiented scan of this quadratic kernel to calibrate the residual systematic width has been that the scan ostensibly requires a large number of nuisance parameters with complicated correlations, but we've shown here that this large parameter count is illusory. A much smaller number of operators is able to generate all the needed kinematic shapes for any given analysis. Our algorithm identifies the $K$ independent operators via an SVD of the
monomial feature matrix built from the quadratic kernel, assigns one representative Wilson coefficient to
each, and determines their joint, non-Gaussian distribution through an ensemble of least-squares fits under an
NDA prior drawn independently of the Wilson coefficients carried by the signal model. To conservatively cover any contributions from higher orders
in $1/\Lambda$,
we have introduced a decorrelated monomial prescription that applies
PCA to the fitted monomials and independently bootstraps each
principal component with an $\alpha=\sqrt{2}$ scaling,
whose physical content is the variance addition of the two uncalculable next-order pieces (SM$\times$dim-8 and double insertions), each NDA-comparable to dim-6$^2$. The result is a compact set of $D = K(K{+}1)/2$ mean-zero PCA-score nuisances with empirical marginals that absorb the residual $\mathcal{O}(\Lambda^{-4})$ uncertainty without requiring it to be explicitly computed, though they do not rise to the reliability of a proven-complete parametrisation of the error \`a la Ref.~\cite{Tackmann:2024kci}.

The nuisance parameter reduction is dramatic in all three processes studied:
high-$p_T$ Drell-Yan ($N=14$, $P=105$) collapses to $K=2$ and just $3$ nuisance parameters, $Vh$ production
($N=14$, $P=105$) to $K=3$ to $5$ and at most $15$ parameters, and VBF ($N=12$, $P=78$) to $K=2$ to $4$, with a maximum of $10$ parameters. In every
case the fits achieve median $R^2\geq 0.99$ (above $0.999$ for most
observables) and the calibrated nuisance width
brackets the full $\mathcal{O}(\Lambda^{-4})$ residual at the $\geq 90\%$ level in every bin. The only
inputs required are the quadratic EFT predictions already available
from standard signal simulations, and the simulations needed can be performed once per monomial and then rescaled for all subsequent calculations required by this method.

A natural set of extensions follows immediately. The same machinery
applies to additional processes such as $t\bar{t}H$, dijet, and $WW$
production, and the nuisance sets identified for individual measurements
can be combined across processes within global SMEFT
fits~\cite{Ellis:2020unq, Giani:2023gfq, Celada:2024mcf}. Future work will demonstrate the full likelihood inclusion of these nuisances in a mock signal-extraction analysis, quantifying their impact on Wilson-coefficient bounds at HL-LHC statistics.

\section*{Code availability}

A reference implementation of the algorithm, including a worked
$Vh$ ($p_{T,H}$) example that reproduces the results in
Table~\ref{tab:summary}, is available at
\url{https://github.com/benleo12/smeft_nuisance}. The package takes the
per-bin dim-6$^2$ polynomial in the same form a standard
MadGraph$+$SMEFTSim run produces and is process- and observable-agnostic.
See the repository's \texttt{INPUT\_FORMAT.md} for the CSV specification
and \texttt{examples/} for a complete worked case.

\section*{Acknowledgments}

We would like to thank Joel Walker and Kelci Mohrman for helpful comments. This work was initiated at the Aspen Center for Physics, which is supported by National
Science Foundation grant PHY-2210452. The work of WS is supported by the National Science Foundation under grant no.
PHY2412995. The work of AM is partially supported by the National Science Foundation under grant PHY-2412701. BA acknowledges support in part by the U.S. Department of Energy grants DE-SC1019775 and
DE-SC0026301, and the National Science Foundation grants OAC-2103889, OAC-2411215,
and OAC-2417682. BA's work was performed in part at the Aspen Center for Physics, with support by a grant from the Simons Foundation (1161654,Troyer).

\bibliographystyle{JHEP}
\bibliography{ref}

\appendix
\section{Implementation details}
\label{app:implementation}

\paragraph{Optimiser and regularisation.}
The optimisation in eq.~\eqref{eq:lsq} works directly in the
$K$-dimensional coefficient space, combining a global scan over
directions with local L-BFGS-B descent as described below. The
regularisation
$\lambda_{\rm reg} = 10^{-6}$ is set just large enough for numerical stability when
$\tilde A$ has near-zero eigenvalues and small enough that for typical
draws the unregularised residual is reached to working precision. The
minimum-norm preference it induces breaks otherwise-degenerate residual
landscapes.

\paragraph{Global direction scan and sign convention.}
The quartic objective is non-convex, but its only non-convex freedom is
the direction of $\mathbf{c}_\mathcal{R}$: writing
$\mathbf{c}_\mathcal{R} = r\,\mathbf{u}$ with $|\mathbf{u}|=1$, the
prediction is $r^2 g_b(\mathbf{u})$ with
$g_b(\mathbf{u}) = \mathbf{u}^\top \tilde A_b\, \mathbf{u}$, and the
optimal magnitude along any direction is analytic,
$r^2(\mathbf{u}) = \max\!\big(\langle q, g(\mathbf{u})\rangle, 0\big) /
\|g(\mathbf{u})\|^2$. The fit therefore scans a fixed quasi-random bank
of unit directions (up to $2^{18}$ points, generated once from a
constant seed), scoring each analytically in a single vectorised
operation. The best-scoring directions, together with the positive
eigendirections of the monomial-space linear least-squares solution,
are refined by a short ascent of the smooth, scale-invariant direction
score and then polished with L-BFGS-B on the objective normalised by
$\|q\|^2$, keeping the best result. Every step is deterministic: the
fit carries no random seed and no dependence on initialisation. For
$K=2$ the direction scan is equivalent to the exact one-dimensional
angular solution, and the fitted optimum reproduces it throw by throw
to better than $10^{-8}$ in $R^2$. For $K>2$ no closed-form reference
exists, and we validated the fit against a heavy random-multistart
search on ensembles of throws at $K=4$ and $K=5$, finding agreement to
better than $10^{-9}$ in $R^2$ on every throw tested.
After the fit converges we apply the sign convention $\hat c_{r_1} \ge 0$, flipping the whole tuple if the leading representative came out negative. This leaves every monomial $m_{kl} = \hat c_k\hat c_l$ invariant and only affects the marginal coefficient histograms.

\paragraph{Monte Carlo generation.}
\begin{sloppypar}
Simulations are performed using MadGraph5~\cite{Alwall:2014hca}. For dimension-6 operators we generate a FeynRules~\cite{Christensen:2008py, Degrande:2011ua, Alloul:2013bka} file grabbing the desired operators verbatim from the $U(3)^5$ flavor symmetric SMEFTsim~\cite{Brivio:2017btx}.fr file. Dimension-8 operators are then directly added to this master file. For DY and Vh we run with default run$\_$card settings and parton distribution functions, while for VBF we impose the cuts in Ref.~\cite{Araz:2020zyh}. We generate $20k$ events per dim-6$^2$ coefficient choice for Drell-Yan and VBF, increasing this to $50k$ for $Vh$ to mitigate statistical fluctuations in the $p_{T,\ell}\times\cos\theta^*$ analysis. Mixed-coefficient entries ($c_i c_j$, $i\ne j$) of the monomial matrix $M$ are obtained by subtraction using \cref{eq:subtract}. For simplicity we ignore operator contributions (at both dim-6 and dim-8) to electroweak input parameters. Bins containing less than $1\%$ of the total Monte Carlo events are discarded as outside the large-statistics regime. The threshold is tightened to $5\%$ for the two-dimensional $(p_{T,\ell},\cos\theta^*)$ analysis, where the bin count is much higher and individual bins are correspondingly sparser. The double-insertion test of \cref{subsec:decorr} uses two generations of $pp\to jjh$ in the same model and under the same VBF cuts, one with at most one EFT vertex allowed in the amplitude (\texttt{NP<=1}), which returns the SM, its dim-6 interference and the kernel $A_b$, and one with the squared matrix element kept through $\mathcal O(\Lambda^{-4})$ (\texttt{NP\textasciicircum 2<=2}), which adds the SM interference with the two-vertex amplitude. Every dim-8 coefficient is zero in both, so the two differ by $\Delta_b$ alone. Each is run at $c_{HQ}^{(3)}=0$ and $\pm10^{-5}~\text{GeV}^{-2}$ with $50k$ events, a common seed and an $m_{jj}>480~\text{GeV}$ generation cut, and the second difference $[\sigma(+c)+\sigma(-c)-2\sigma(0)]/(2c^2)$ isolates the piece quadratic in that coefficient bin by bin, free of the linear term. The inclusive ratio of the two generations is $0.874$, against $0.881\pm0.007$ for the mean over the ten analysis bins.
\end{sloppypar}

\paragraph{Ensemble size and cost.}
Typical ensemble sizes are $T=10{,}000$ throws. The direction bank is
projected onto the kernel once per ensemble, after which each throw
costs one vectorised scoring pass plus a handful of L-BFGS-B descents,
tens of milliseconds on a single CPU core for $K \leq 6$. A full
ensemble runs in minutes per process and the per-throw fits are
embarrassingly parallel.

\section{Worked example}
\label{app:mwe}

This appendix walks the algorithm from start to finish on a six-bin,
three-operator toy with rich interference structure. Every quoted
number is reproduced by the accompanying Wolfram script
\verb|mwe.wl|. In the language of the method, the quadratic kernel
below is the calculable piece the representatives fit, and the truth
we validate against is that same quadratic piece. This toy adds no
uncalculable next-order term.

The toy keeps three Wilson coefficients $c_1, c_2, c_3$ and six
kinematic bins. The two amplitudes form two topologies that share the
coefficient $c_2$,
\begin{equation}
\mathcal{M}_{\rm I} \propto c_1 + 2 c_2,
\qquad
\mathcal{M}_{\rm II} \propto c_2 + c_3,
\end{equation}
with per-bin strengths
\begin{equation}
S_{\rm I} = (1, 2, 4, 8, 16, 32),
\qquad
S_{\rm II} = (32, 16, 8, 4, 2, 1),
\end{equation}
two linearly-independent bin shapes spanning a two-dimensional
``shape space'' inside $\mathbb R^6$. The quadratic kernel in bin $b$
(the calculable piece) is the linear combination
\begin{equation}
A_b = S_{\rm I}[b]
\begin{pmatrix} 1 & 2 & 0 \\ 2 & 4 & 0 \\ 0 & 0 & 0 \end{pmatrix}
+ S_{\rm II}[b]
\begin{pmatrix} 0 & 0 & 0 \\ 0 & 1 & 1 \\ 0 & 1 & 1 \end{pmatrix},
\end{equation}
which factorises in topology form as
\begin{equation}
q_b = S_{\rm I}[b]\,(c_1 + 2 c_2)^2 + S_{\rm II}[b]\,(c_2 + c_3)^2.
\end{equation}
The kernel carries off-diagonal entries $A_{b,12}=2\,S_{\rm I}[b]$
between $c_1$ and $c_2$ and $A_{b,23}=S_{\rm II}[b]$ between $c_2$
and $c_3$. The entry $A_{b,13}$ vanishes because $c_1$ and $c_3$
never share an amplitude. This is an example of interference structure
the full processes in the studied examples can exhibit.

\paragraph{Monomial matrix and rank.}
The $B\times P = 6\times 6$ monomial feature matrix from
eq.~\eqref{eq:monomial} reads
\begin{equation}
M \;=\;
\begin{array}{c|cccccc}
 & c_1^2 & c_2^2 & c_3^2 & c_1 c_2 & c_1 c_3 & c_2 c_3 \\ \hline
1 & 1 & 36 & 32 & 4 & 0 & 64 \\
2 & 2 & 24 & 16 & 8 & 0 & 32 \\
3 & 4 & 24 & 8 & 16 & 0 & 16 \\
4 & 8 & 36 & 4 & 32 & 0 & 8 \\
5 & 16 & 66 & 2 & 64 & 0 & 4 \\
6 & 32 & 129 & 1 & 128 & 0 & 2
\end{array}
\end{equation}
The five non-zero columns all live in
$\operatorname{span}\{S_{\rm I}, S_{\rm II}\}$ by inspection
($c_1^2 = S_{\rm I}$, $c_3^2 = S_{\rm II}$, $c_2^2 = 4 S_{\rm I} +
S_{\rm II}$, $c_1 c_2 = 4 S_{\rm I}$, $c_2 c_3 = 2 S_{\rm II}$), so
$M$ has rank $2$ even though there are three operators and five
non-zero monomial columns. The SVD reports the singular values
$\sigma_1 \approx 218.7$, $\sigma_2 \approx 83.4$, and four exact
zeros, which is the empirical rank statement.

\paragraph{Representatives.}
We take the two non-zero SVD directions in order of decreasing singular
value and, for each, score the operators with eq.~\eqref{eq:score}. For
the leading direction ($\sigma_1 \approx 218.7$),
\begin{equation}
\mathrm{score}_1[c_1] \approx 0.47, \quad
\mathrm{score}_1[c_2] \approx 0.97, \quad
\mathrm{score}_1[c_3] \approx 0.01,
\end{equation}
so $c_2$ is selected first: it is touched by both topologies through
$c_2^2$, $c_1 c_2$, and $c_2 c_3$, hence its dominant score. For the
sub-leading direction ($\sigma_2 \approx 83.4$),
\begin{equation}
\mathrm{score}_2[c_1] \approx 0.09, \quad
\mathrm{score}_2[c_2] \approx 0.82, \quad
\mathrm{score}_2[c_3] \approx 0.89,
\end{equation}
and, with $c_2$ already assigned, $c_3$ is selected. The selection
order is thus $c_2 \to c_3$, and the $R^2$ ladder below confirms that no
third representative is needed.

\paragraph{R$^2$ ladder fits coefficients, not monomials.}
For each truth throw $t$, drawn from a uniform prior
$c_i^{(t)}\sim U(-1,1)$ (the toy is invariant under a rescaling of the
prior, so the NDA scale is set to one), the algorithm fits a coefficient vector
$\hat{\mathbf c}_{\mathcal R}^{(t)}\in\mathbb R^K$ by the nonlinear
minimisation
\begin{equation}
\hat{\mathbf c}_{\mathcal R}^{(t)} \;=\; \arg\min_{\mathbf c\,\in\,\mathbb R^K}
\sum_{b=1}^{6}\Bigl(q_b^{(t)} - \mathbf c^\top \tilde A_b\,\mathbf c\Bigr)^2,
\end{equation}
which is quartic in $\mathbf c$. This is the step where the
representation lives in coefficient space, not monomial space. The
median $R^2$ over $T=5000$ throws is
\begin{equation}
\mathrm{med}(R^2)|_{K=1,\{c_2\}} = 0.947,\quad
\mathrm{med}(R^2)|_{K=2,\{c_2,c_3\}} = 1.000,\quad
\mathrm{med}(R^2)|_{K=3,\{c_2,c_3,c_1\}} = 1.000.
\end{equation}
At $K=1$ the fit can only realise the single shape
$4 S_{\rm I} + S_{\rm II}$ from $\hat c_2^2\,\tilde A_{b,22}$. It
absorbs the topology-I content but is starved of the topology-II
direction. Adding $c_3$ at $K=2$ supplies the second shape, and
since the rank of the truth is two, the fit becomes exact. The
$K=3$ step adds the operator $c_1$, but $c_1$ enters only in the
combination $(c_1+2c_2)$ which $\hat c_2$ already represents by
rescaling. No new shape is unlocked and the median $R^2$ does not
move, so $c_1$ is rejected and the algorithm terminates at
$K=2$ with representatives $\{c_2, c_3\}$.

\paragraph{Decorrelation of the fitted ensemble.}
With $K$ fixed, the algorithm switches to monomial space. From the
ensemble of fitted coefficient tuples
$\{(\hat c_2^{(t)},\hat c_3^{(t)})\}_{t=1}^T$ we form the
$K(K{+}1)/2 = 3$ monomials
\begin{equation}
m_{22}^{(t)} = \bigl(\hat c_2^{(t)}\bigr)^2, \quad
m_{23}^{(t)} = \hat c_2^{(t)}\,\hat c_3^{(t)}, \quad
m_{33}^{(t)} = \bigl(\hat c_3^{(t)}\bigr)^2.
\end{equation}
By construction these three numbers, viewed as a vector in
$\mathbb R^3$, lie on the two-dimensional surface $m_{23}^2 = m_{22}\,m_{33}$,
a curved two-dimensional sheet inside the three-dimensional monomial
space. The reason is that they are entries of the outer product
$\hat{\mathbf c}\hat{\mathbf c}^\top$, which has rank one whatever
$\hat{\mathbf c}$ is.

We then centre the monomial ensemble, $m_{kl}^{(t)} \to
m_{kl}^{(t)} - \bar m_{kl}$, and take the PCA. The PCA returns three
principal directions, of which two carry the variance and one is
nearly null in expectation. Concretely the script reports the PCA
sigmas
$\boldsymbol\sigma_{\rm PC} \approx (0.479,\ 0.321,\ 0)$. This zero in the third direction reflects the fact that the toy's monomial matrix has rank 2. In a more involved case the third monomial direction could carry genuine variance while $K=2$ representatives still suffice for the fit. Thus, our algorithm still provides only two nontrivial nuisance parameters, despite the intermediate investigation of a third potential direction.

In a case where an independent third shape is generated by cross-terms of the two representative Wilson coefficients, independently bootstrapping the principal-component scores, the decorrelation step of Section~\ref{subsec:decorr}, would break the
two-dimensional constraint: the resulting monomial draws would no longer satisfy
$m_{23}^2 = m_{22}\,m_{33}$, and the cloud of error-estimate points in monomial space would spill off of the curved
sheet (this is the analogue of Figure~\ref{fig:decorr_cartoon}c in the body).

Each bootstrap draw is then scaled by
$\alpha=\sqrt 2$, which doubles the variance of every PCA direction.
The decorrelated, $\alpha$-inflated nuisance observable in bin $b$ is
\begin{equation}
q^{\rm dec}_b \;=\; \sum_{k \le l} M_{b,(kl)}\, \tilde m_{kl},
\end{equation}
with $M$ the $K=2$ sub-monomial matrix. Crucially in a truly rank-3 case $q^{\rm dec}_b$ would be
allowed to take negative values, since the bootstrapped
$\tilde m_{kl}$ would no longer come from a valid outer product.

\paragraph{Coverage.}
The per-bin $95\%$ interval of the decorrelated nuisance distribution
is compared bin by bin against the truth ensemble of the
opening. Note that this ``truth ensemble'' differs from those in the physics examples: there the truth included uncalculable next-order pieces beyond the quadratic input, whereas this toy adds no such piece. The script reports
\begin{equation}
\begin{aligned}
\text{coverage at } \alpha = 1: \quad & (0.950,\, 0.965,\, 0.968,\, 0.965,\, 0.968,\, 0.969), \\
\text{coverage at } \alpha = \sqrt 2:\quad & (0.998,\, 0.997,\, 0.993,\, 0.995,\, 0.997,\, 0.997).
\end{aligned}
\end{equation}
The uninflated bootstrap already sits at $95\%$, as it should for an
ensemble drawn from the same prior as the truth, and the
$\alpha=\sqrt 2$ inflation lifts the worst bin to $99.3\%$. This is
the headroom that would absorb the uncalculable next-order pieces the
quadratic input does not see, none of which are added in this toy.

The example exhibits the three claims of the paper in miniature.
First, the rank statement is empirical: $N=3$ operators produced a
rank-2 monomial matrix, so the $\Lambda^{-2}$ truncation lives in a
two-dimensional shape space whatever the prior. Second, the
$K$-representative reparametrisation is faithful in the sense of
$R^2 = 1$ at the validated rank, because the $K(K{+}1)/2$ monomials of
$K=2$ representatives are enough to express any element of that
shape space. Third, the PCA-and-bootstrap decorrelation introduces independent variation along directions the dim-6$^2$-only fit cannot populate, and the $\sqrt 2$ inflation opens headroom along those directions. The toy's truth contains no uncalculable next-order piece, so what it demonstrates is the headroom itself, the worst-bin coverage rising from $0.95$ at $\alpha=1$ to $0.993$ at $\alpha=\sqrt 2$. In the physics examples that headroom is what absorbs the SM$\times$dim-8 and double-insertion terms. The same three ingredients drive
the Drell-Yan, $Zh$, and VBF analyses in the body.

\section{Supplementary figures}
\label{app:figs}

Here we collect figures from alternative choices of observable for the example processes discussed in \cref{sec:examples}.

\subsection*{Drell-Yan observables}
\begin{figure}[H]
 \centering
 \includegraphics[width=0.5\textwidth]{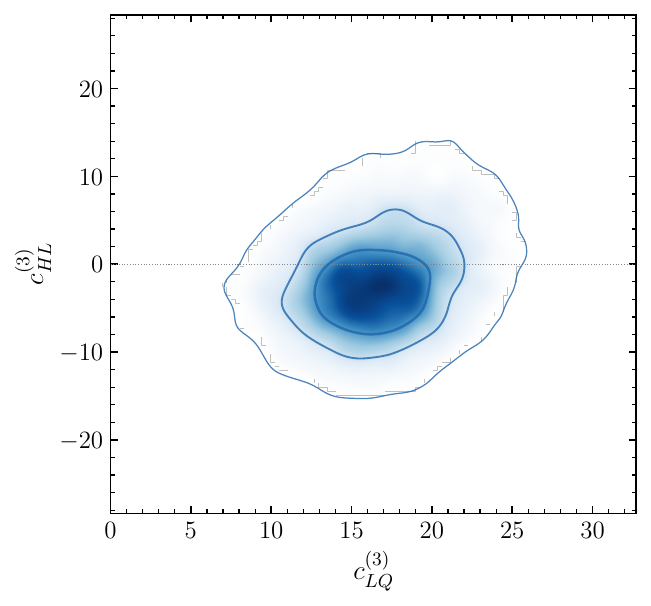}
 \caption{Joint distribution of the two forward-hemisphere representatives
 $c_{LQ}^{(3)}$ and $c_{HL}^{(3)}$ from $T=10{,}000$ throws.}
 \label{fig:dy_fwd_2d}
\end{figure}

\begin{figure}[H]
 \centering
 \includegraphics[width=\textwidth]{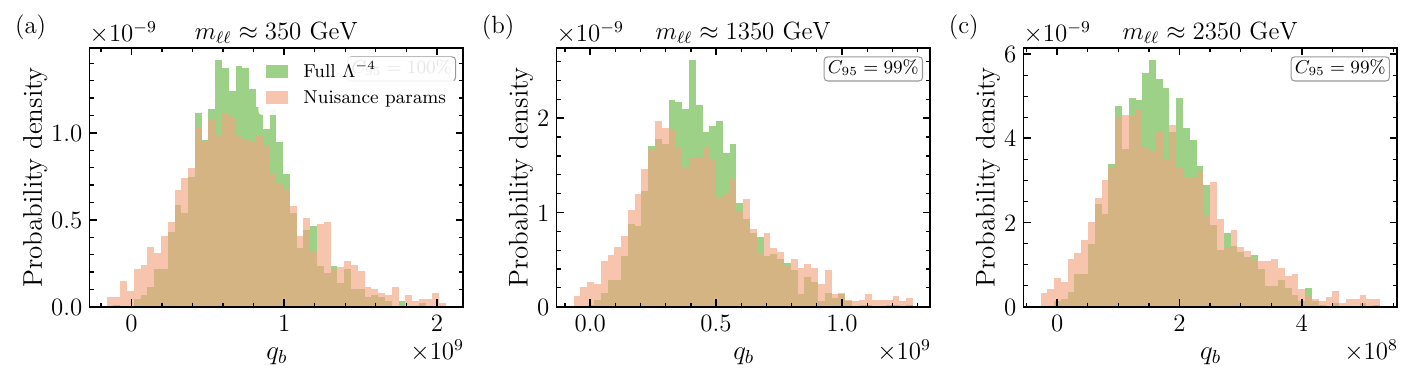}
 \caption{Per-bin $\Lambda^{-4}$ comparison for the forward
 $m_{\ell\ell}$ distribution. Per-bin $95\%$ coverage $C_{95}$ annotated per panel.}
 \label{fig:dy_fwd_lam4}
\end{figure}

\begin{figure}[H]
 \centering
 \includegraphics[width=0.5\textwidth]{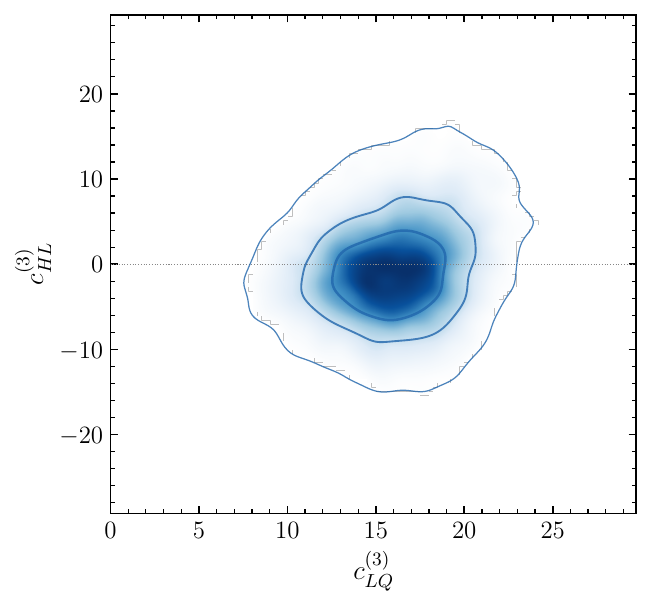}
 \caption{Joint distribution of the two backward-hemisphere representatives
 $c_{LQ}^{(3)}$ and $c_{HL}^{(3)}$ from $T=10{,}000$ throws.}
 \label{fig:dy_bwd_2d}
\end{figure}

\begin{figure}[H]
 \centering
 \includegraphics[width=\textwidth]{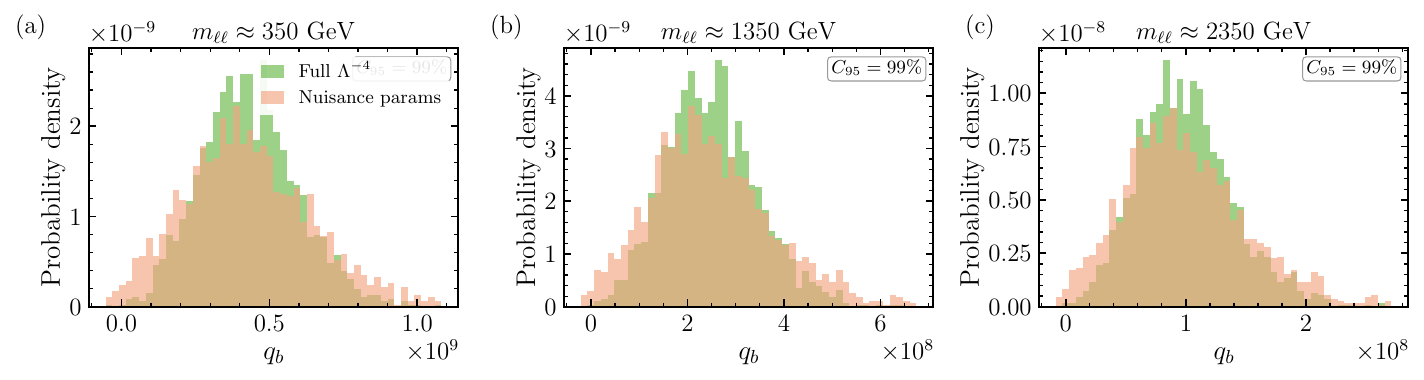}
 \caption{Per-bin $\Lambda^{-4}$ comparison for the backward
 $m_{\ell\ell}$ distribution. Per-bin $95\%$ coverage $C_{95}$ annotated per panel.}
 \label{fig:dy_bwd_lam4}
\end{figure}

\subsection*{$Vh$ observables}

\begin{figure}[H]
 \centering
 \includegraphics[width=\textwidth]{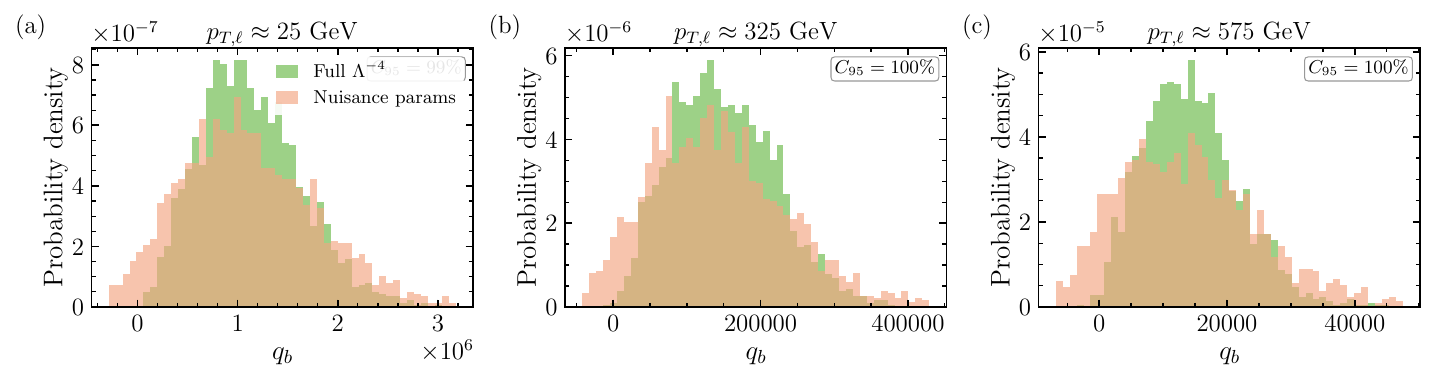}
 \caption{Per-bin $\Lambda^{-4}$ comparison for the lepton $p_T$ observable
 in $Vh$ production ($K=3$). Per-bin $95\%$ coverage $C_{95}$ annotated per panel.}
 \label{fig:vh_lpt_lam4}
\end{figure}

\begin{figure}[H]
 \centering
 \subfloat{\includegraphics[width=0.32\textwidth]{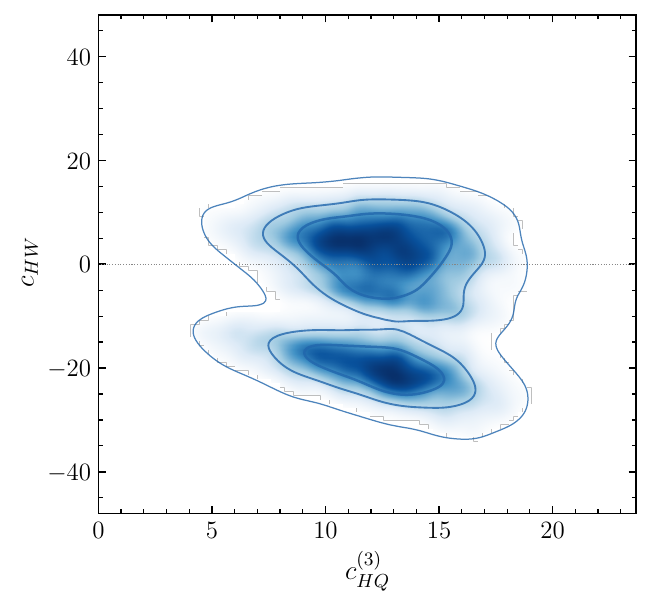}}\hfill
 \subfloat{\includegraphics[width=0.32\textwidth]{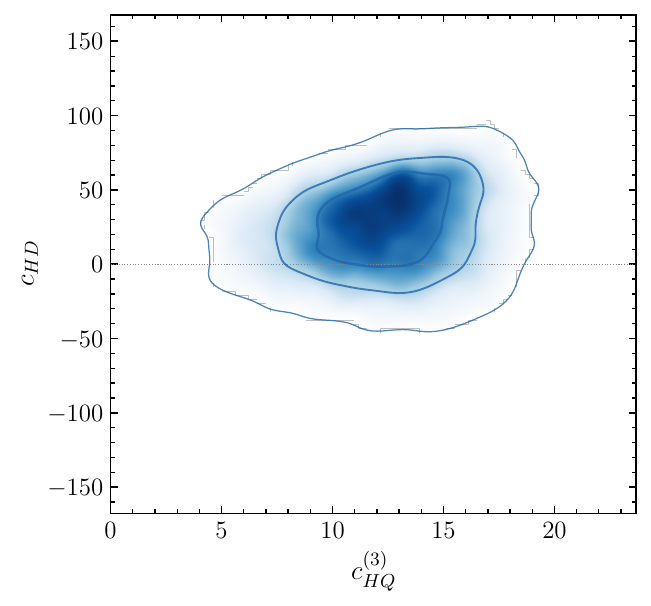}}\hfill
 \subfloat{\includegraphics[width=0.32\textwidth]{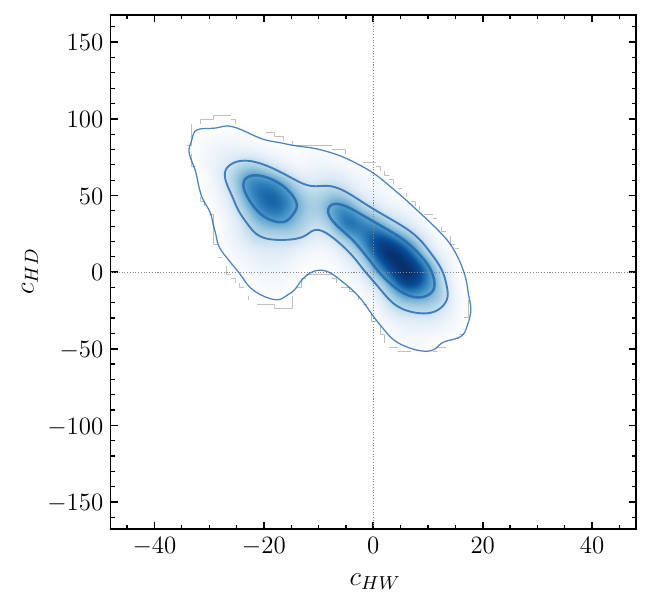}}
 \caption{Pairwise joint distributions of the three lepton $p_T$
 representatives from $T=10{,}000$ throws.}
 \label{fig:vh_lpt_2d}
\end{figure}

\begin{figure}[H]
 \centering
 \includegraphics[width=\textwidth]{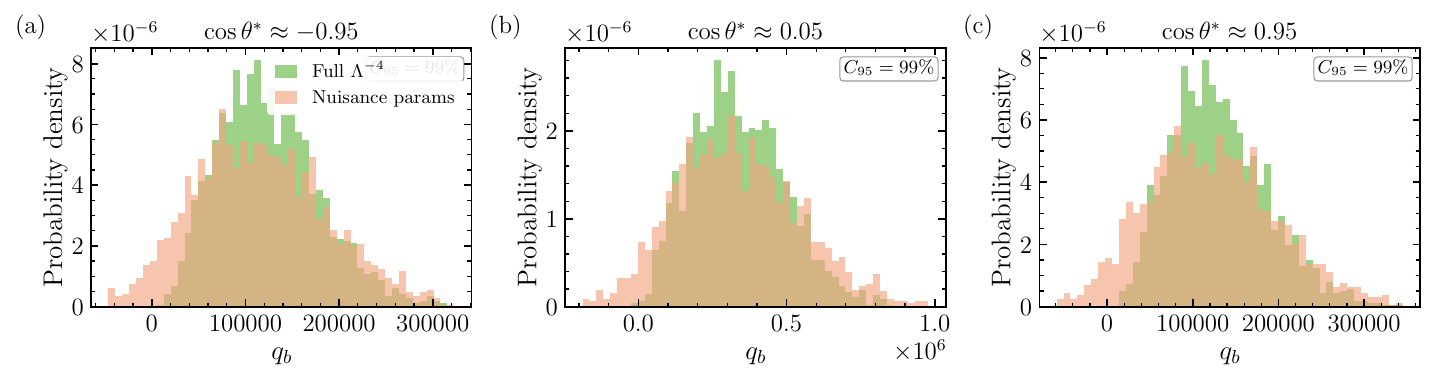}
 \caption{Per-bin $\Lambda^{-4}$ comparison for $\cos\theta^*$
 in $Vh$ production ($K=3$). Per-bin $95\%$ coverage $C_{95}$ annotated per panel.}
 \label{fig:vh_costheta_lam4}
\end{figure}

\begin{figure}[H]
 \centering
 \subfloat{\includegraphics[width=0.32\textwidth]{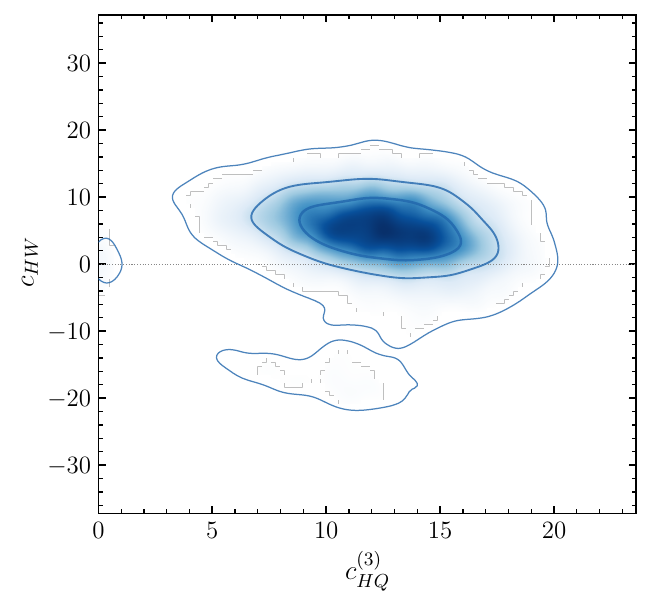}}\hfill
 \subfloat{\includegraphics[width=0.32\textwidth]{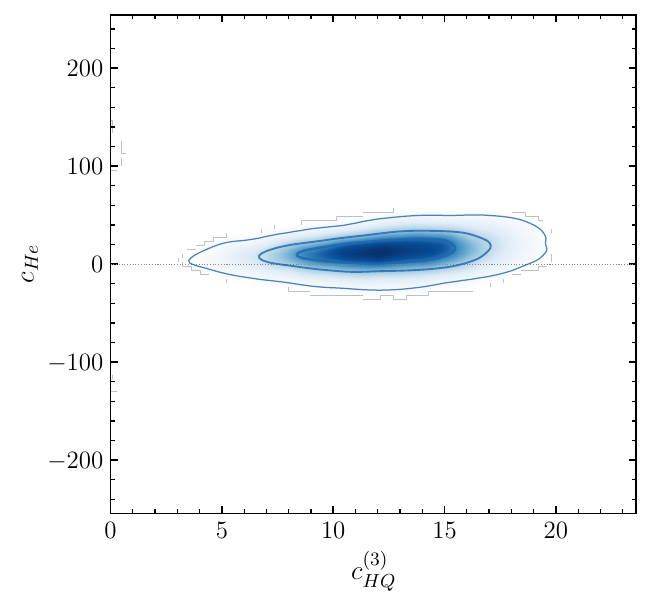}}\hfill
 \subfloat{\includegraphics[width=0.32\textwidth]{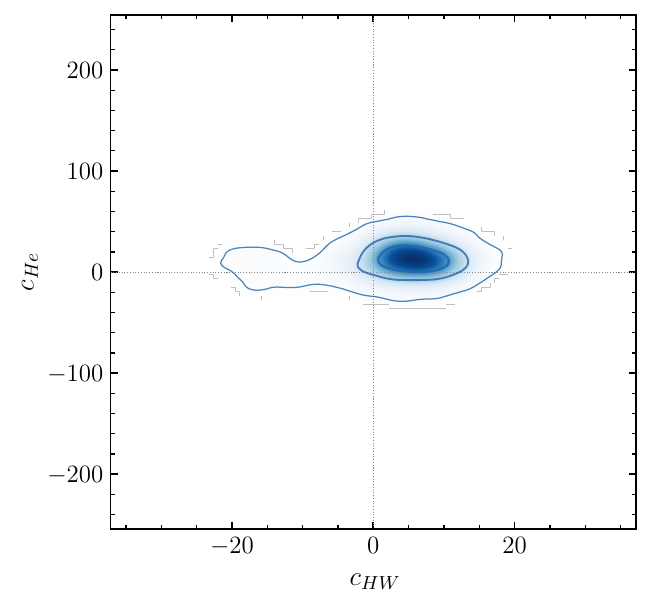}}
 \caption{Pairwise joint distributions of the three $\cos\theta^*$
 representatives from $T=10{,}000$ throws.}
 \label{fig:vh_costheta_2d}
\end{figure}

\begin{figure}[H]
 \centering
 \subfloat{\includegraphics[width=0.32\textwidth]{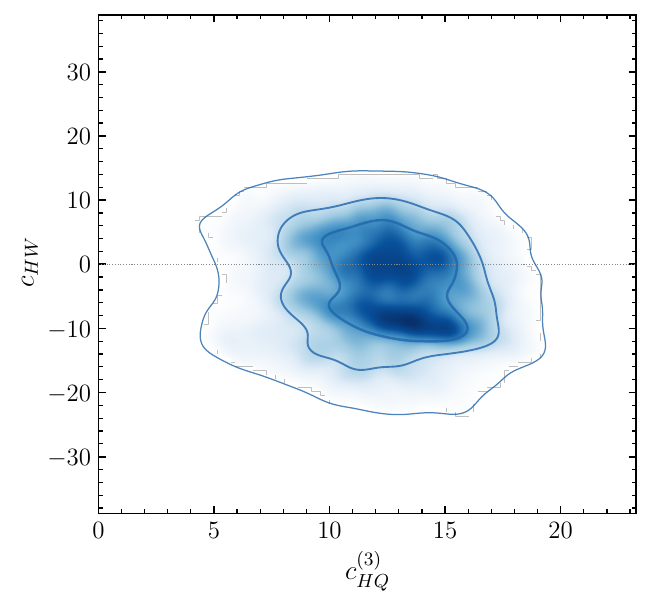}}\hfill
 \subfloat{\includegraphics[width=0.32\textwidth]{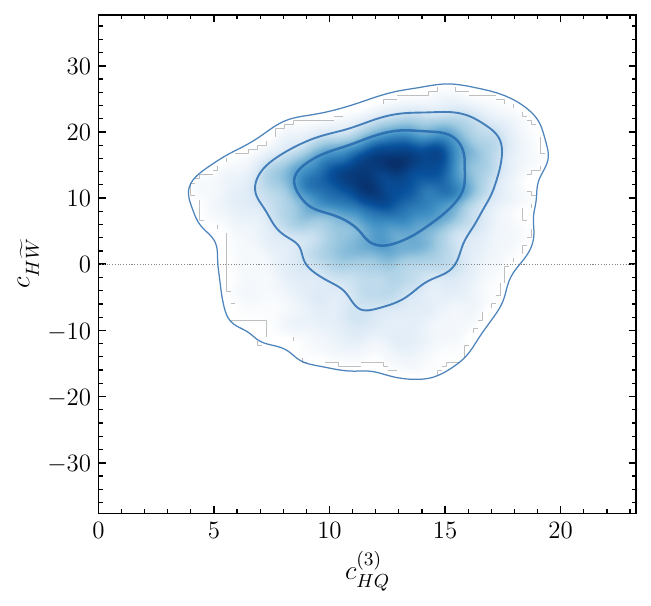}}\hfill
 \subfloat{\includegraphics[width=0.32\textwidth]{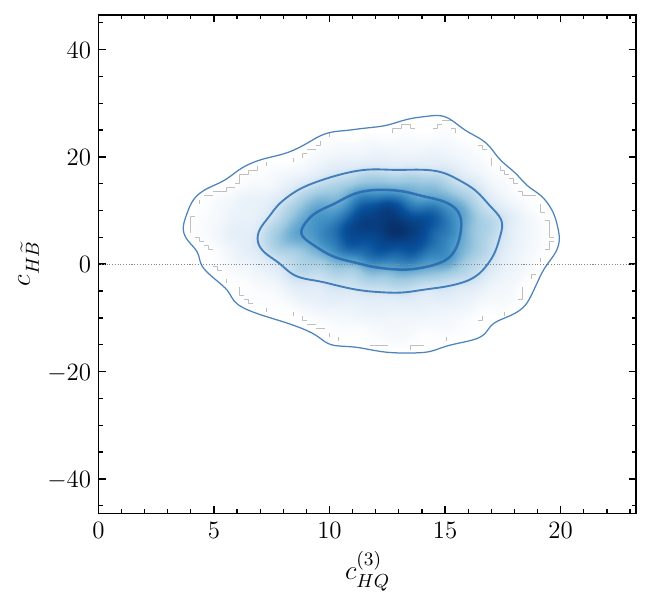}}\\
 \subfloat{\includegraphics[width=0.32\textwidth]{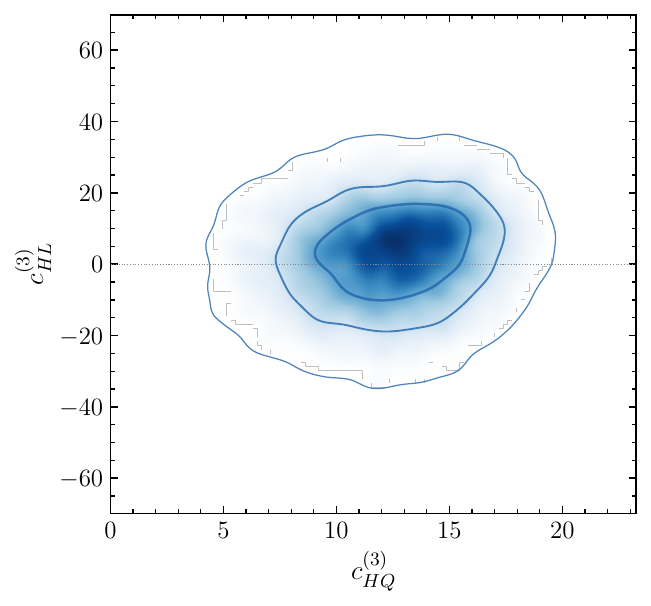}}\hfill
 \subfloat{\includegraphics[width=0.32\textwidth]{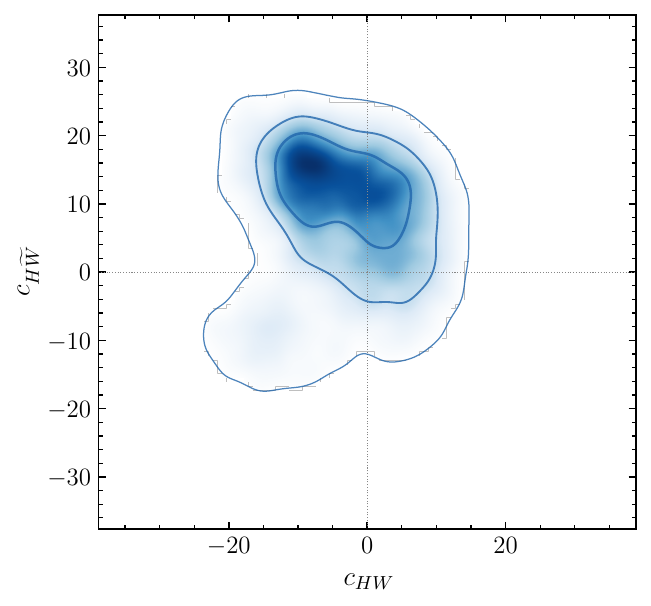}}\hfill
 \subfloat{\includegraphics[width=0.32\textwidth]{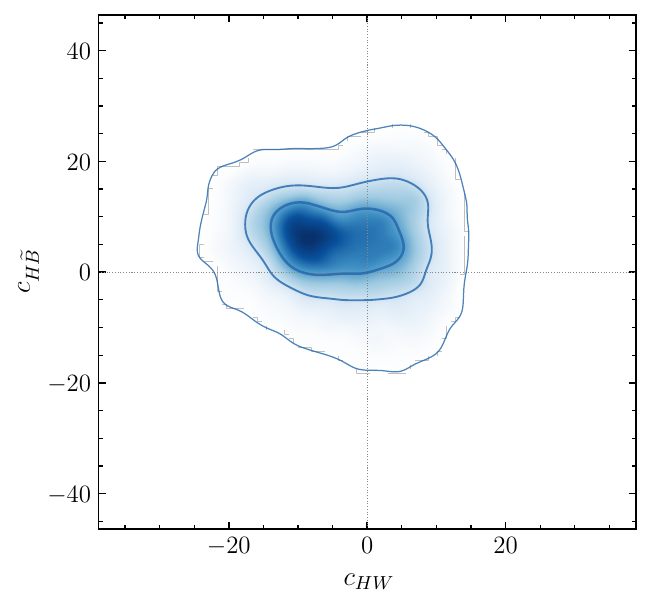}}\\
 \subfloat{\includegraphics[width=0.32\textwidth]{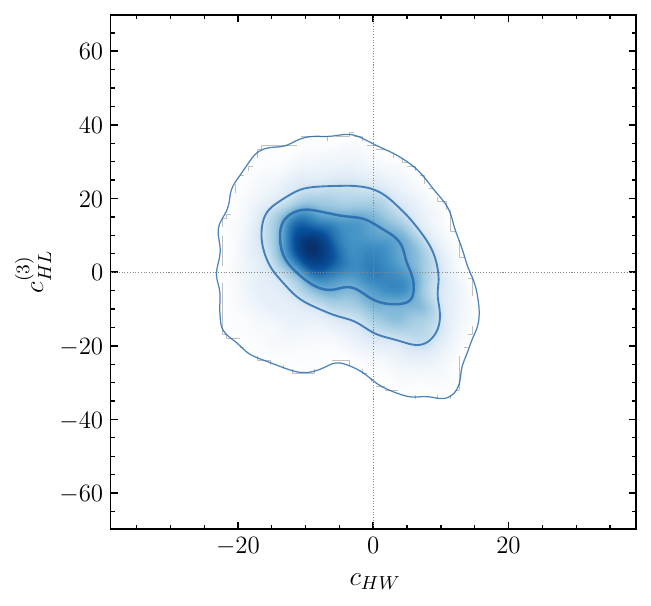}}\hfill
 \subfloat{\includegraphics[width=0.32\textwidth]{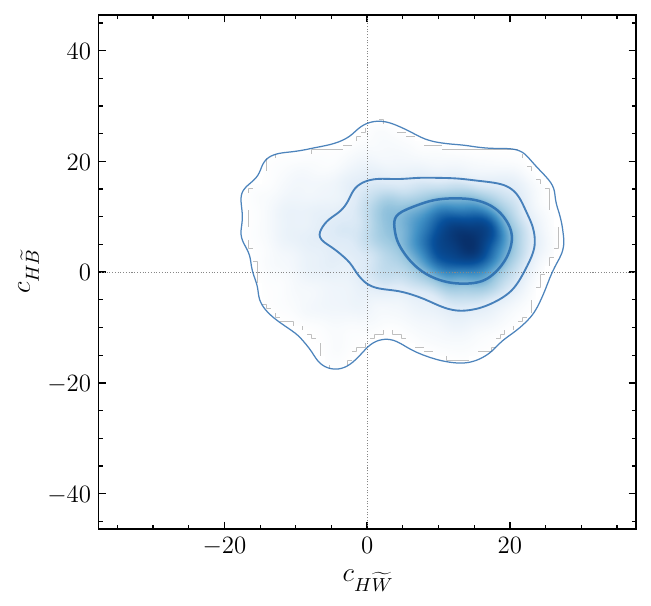}}\hfill
 \subfloat{\includegraphics[width=0.32\textwidth]{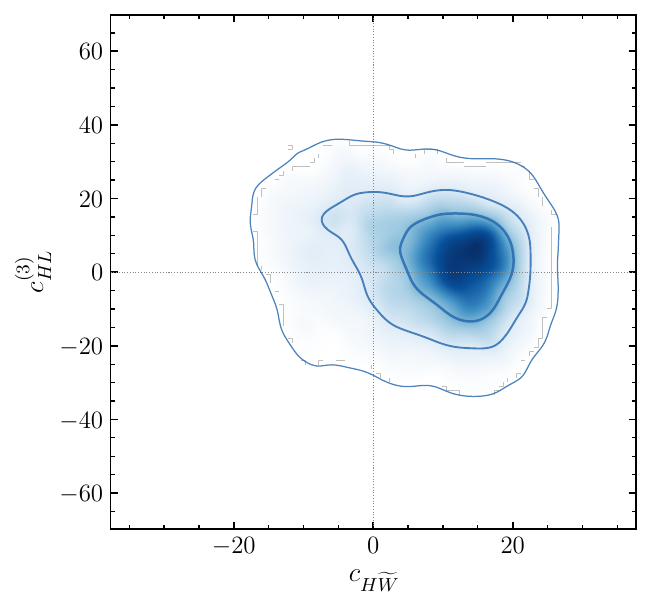}}\\
 \subfloat{\includegraphics[width=0.32\textwidth]{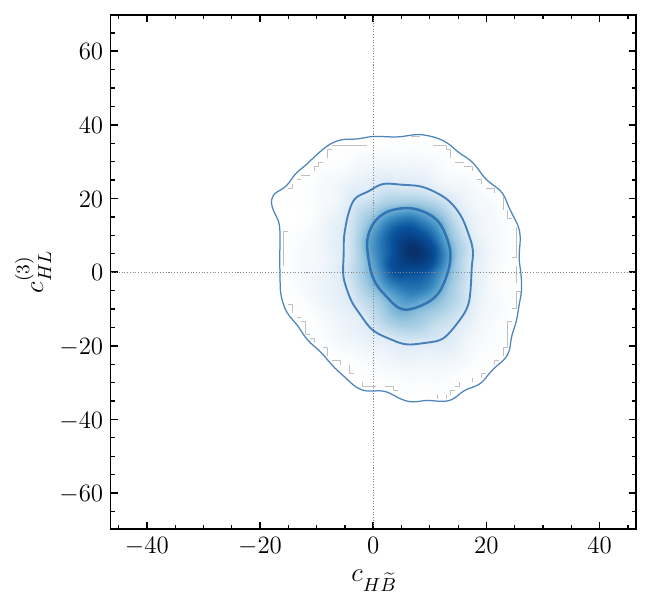}}
 \caption{Pairwise joint distributions of the five $Vh$ ($p_{T,H}$)
 representatives from $T=10{,}000$ throws.
 Blue: KDE density contours at $38\%$, $68\%$, and $95\%$ enclosed probability.}
 \label{fig:vh_theory_2d}
\end{figure}

\subsection*{VBF Higgs observables}

\begin{figure}[H]
 \centering
 \includegraphics[width=\textwidth]{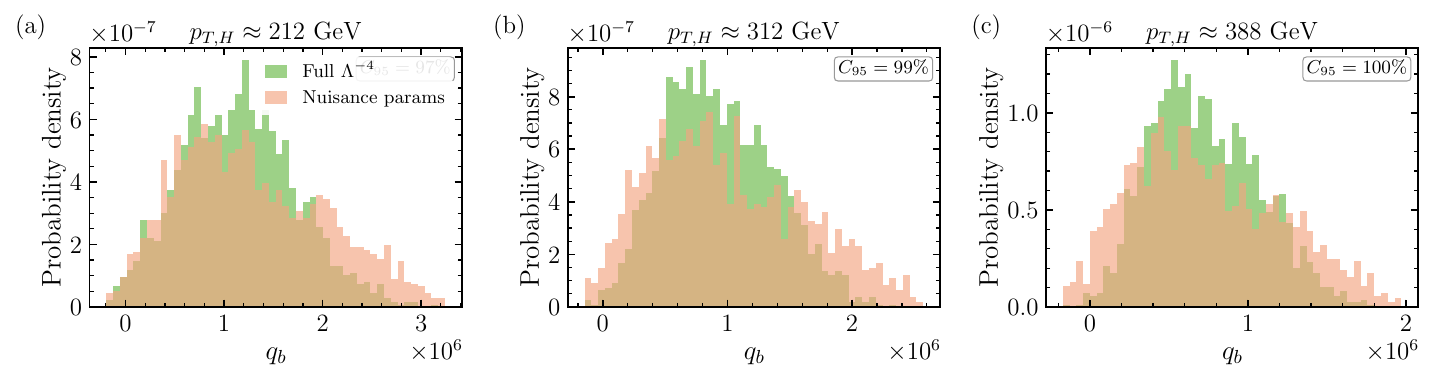}
 \caption{Per-bin $\Lambda^{-4}$ comparison for the VBF Higgs $p_T$
 observable ($K=3$). Per-bin $95\%$ coverage $C_{95}$ annotated per panel.}
 \label{fig:vbf_hpt_lam4}
\end{figure}

\begin{figure}[H]
 \centering
 \subfloat{\includegraphics[width=0.32\textwidth]{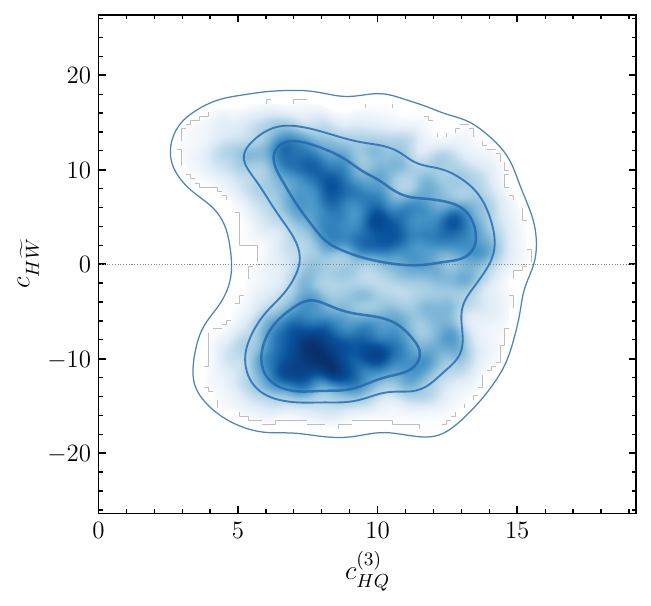}}\hfill
 \subfloat{\includegraphics[width=0.32\textwidth]{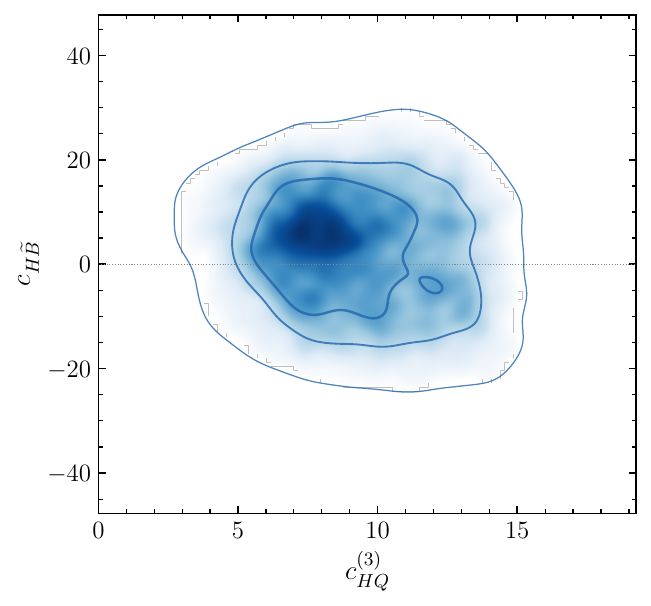}}\hfill
 \subfloat{\includegraphics[width=0.32\textwidth]{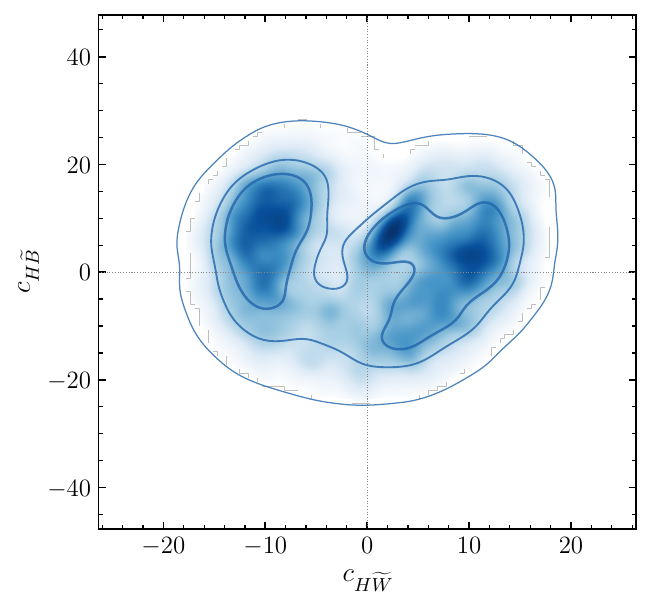}}
 \caption{Pairwise joint distributions of the three VBF Higgs $p_T$
 representatives from $T=10{,}000$ throws.}
 \label{fig:vbf_hpt_2d}
\end{figure}

\begin{figure}[H]
 \centering
 \includegraphics[width=\textwidth]{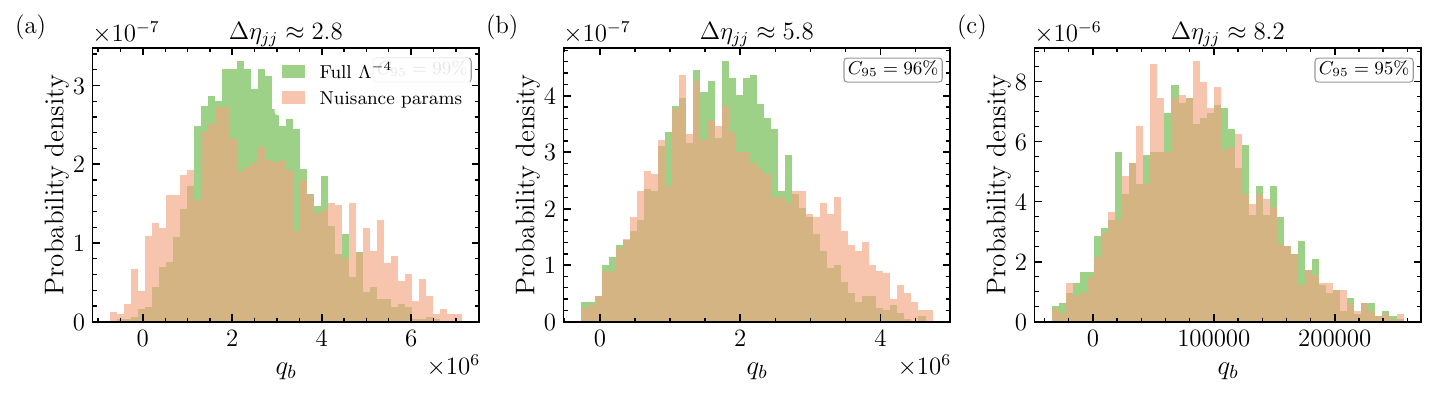}
 \caption{Per-bin $\Lambda^{-4}$ comparison for VBF $\Delta\eta_{jj}$
 ($K=4$). Per-bin $95\%$ coverage $C_{95}$ annotated per panel.}
 \label{fig:vbf_deta_lam4}
\end{figure}

\begin{figure}[H]
 \centering
 \subfloat{\includegraphics[width=0.32\textwidth]{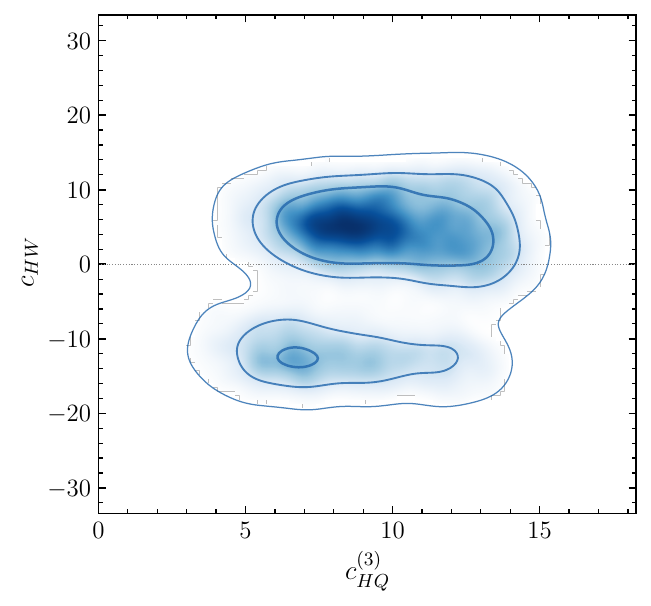}}\hfill
 \subfloat{\includegraphics[width=0.32\textwidth]{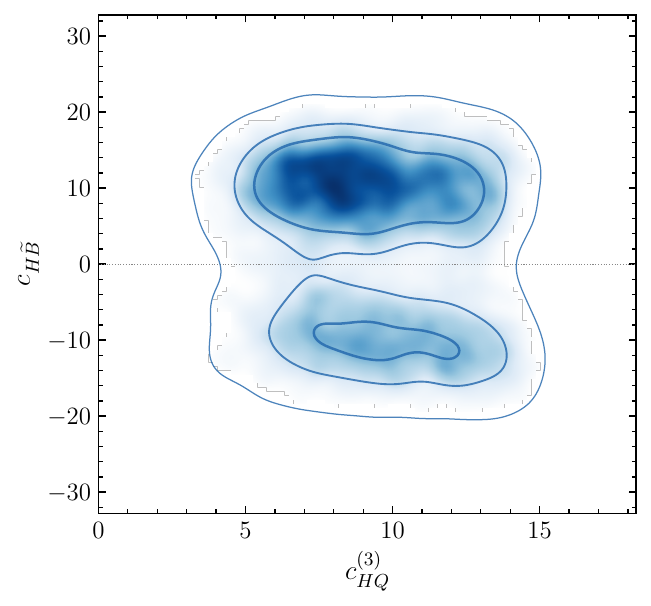}}\hfill
 \subfloat{\includegraphics[width=0.32\textwidth]{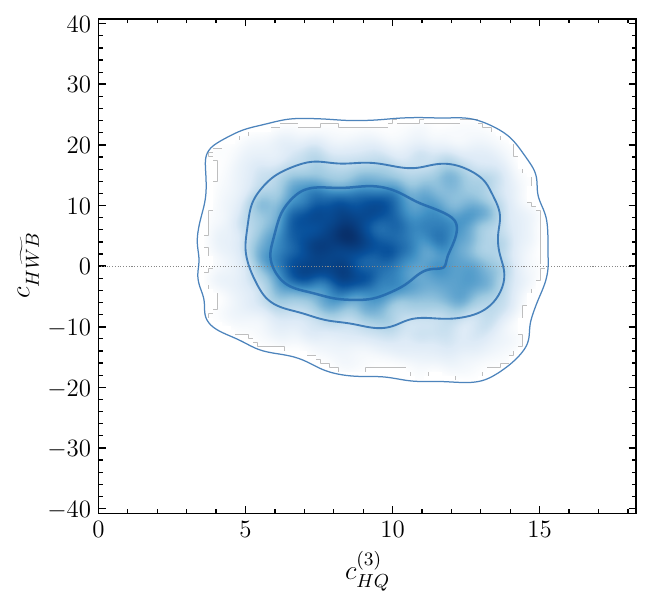}}\\
 \subfloat{\includegraphics[width=0.32\textwidth]{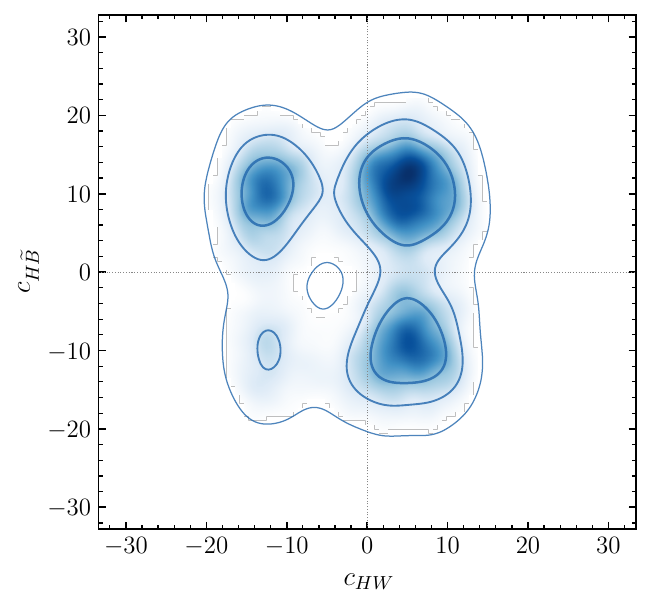}}\hfill
 \subfloat{\includegraphics[width=0.32\textwidth]{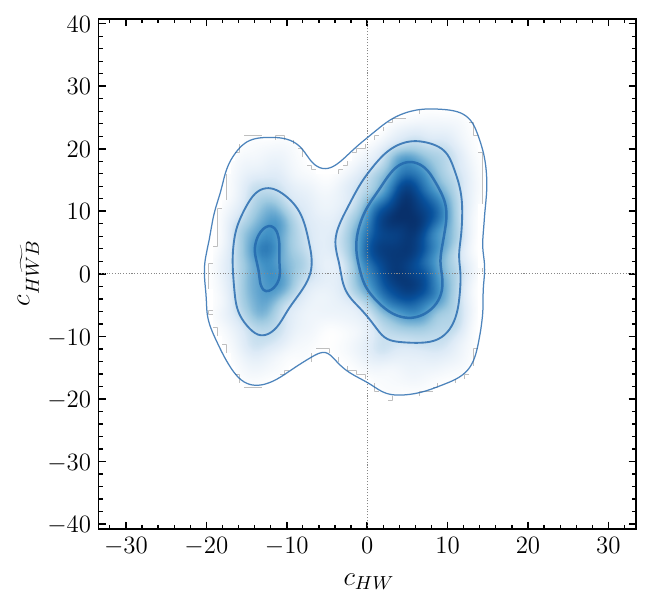}}\hfill
 \subfloat{\includegraphics[width=0.32\textwidth]{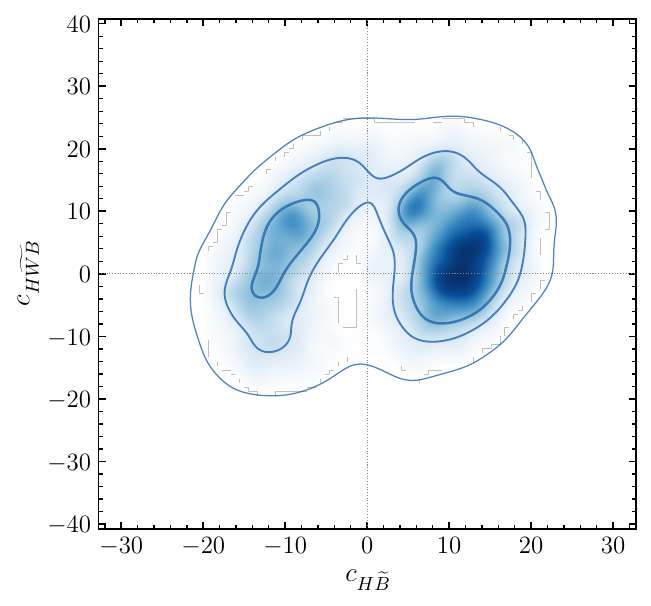}}
 \caption{Pairwise joint distributions of the four VBF $\Delta\eta_{jj}$
 representatives from $T=10{,}000$ throws.}
 \label{fig:vbf_deta_2d}
\end{figure}

\begin{figure}[H]
 \centering
 \includegraphics[width=\textwidth]{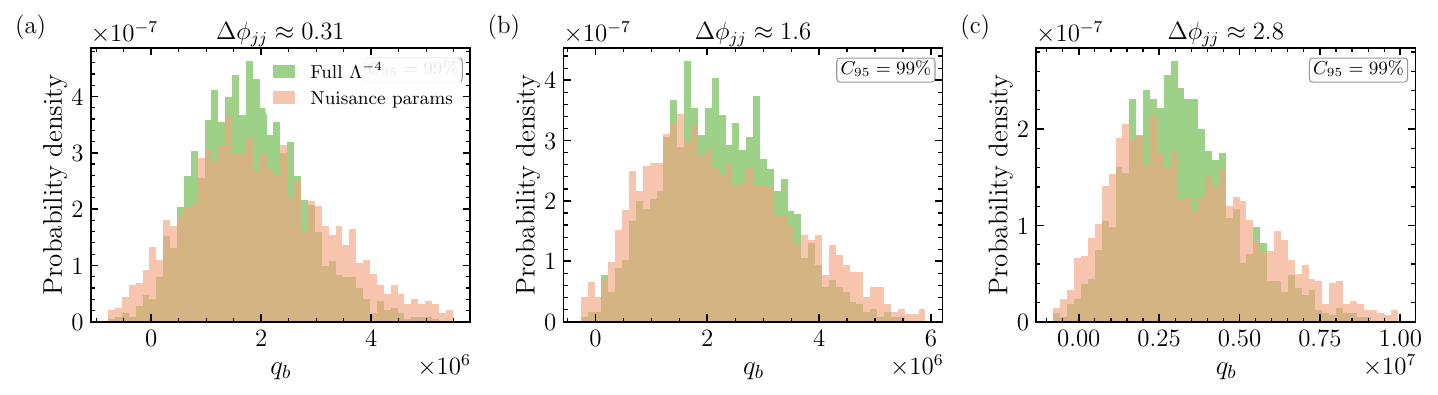}
 \caption{Per-bin $\Lambda^{-4}$ comparison for VBF $\Delta\phi_{jj}$
 ($K=2$). Per-bin $95\%$ coverage $C_{95}$ annotated per panel.}
 \label{fig:vbf_dphi_lam4}
\end{figure}

\begin{figure}[H]
 \centering
 \includegraphics[width=0.45\textwidth]{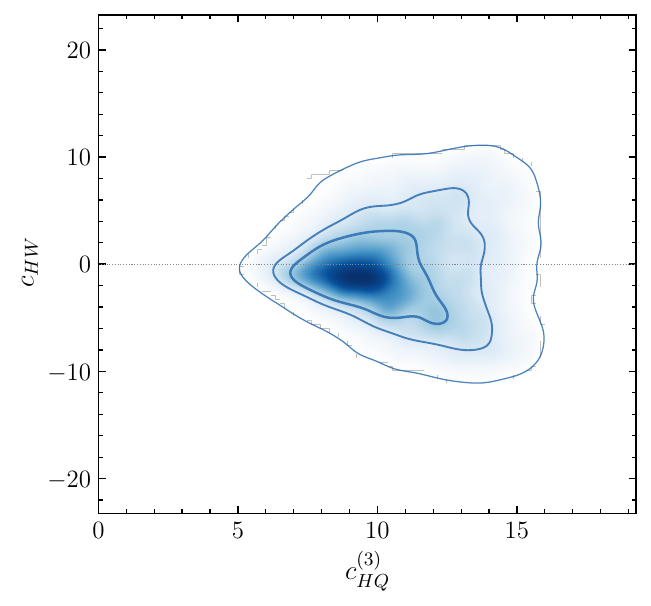}
 \caption{Joint distribution of the two VBF $\Delta\phi_{jj}$
 representatives from $T=10{,}000$ throws.}
 \label{fig:vbf_dphi_2d}
\end{figure}

\end{document}